\pdfoutput=1
\documentclass[acmsmall, authorversion]{acmart}
\AtBeginDocument{%
  \providecommand\BibTeX{{%
    \normalfont B\kern-0.5em{\scshape i\kern-0.25em b}\kern-0.8em\TeX}}}
\setcopyright{acmlicensed}
\acmJournal{POMACS}
\acmYear{2019} \acmVolume{3} \acmNumber{3} \acmArticle{58} \acmMonth{12} \acmPrice{15.00}\acmDOI{10.1145/3366706}
\usepackage[utf8]{inputenc} 
\usepackage[T1]{fontenc}
\usepackage{url}
\usepackage{hyperref}       
\usepackage{booktabs}       
\usepackage{amsfonts}       
\usepackage{nicefrac}       
\usepackage{microtype}      
\usepackage{ifthen}
\usepackage[linesnumbered,ruled]{algorithm2e}
\usepackage{amsthm, amssymb}
\usepackage{enumitem}
\usepackage{nccmath}
\usepackage{graphicx}
\usepackage{cleveref}
\usepackage{subfig}
\usepackage{wrapfig}
\usepackage[font=small]{caption}
\usepackage[colorinlistoftodos,prependcaption]{todonotes}

\newtheorem{lem}{Lemma}
\newtheorem{thm}{Theorem}
\newtheorem{defn}{Definition}
\newtheorem{coro}{Corollary}

\newtheorem{rem}{Remark}
\newtheorem{assum}{Assumption}

\newcommand{\E}[1]{\mathbb{E}\left[{#1}\right]}

\graphicspath{{Figures/}}

\crefname{equation}{}{}
\Crefname{equation}{}{}
\crefname{thm}{theorem}{theorems}
\Crefname{thm}{Theorem}{Theorems}
\crefname{clm}{claim}{claims}
\Crefname{clm}{Claim}{Claims}
\Crefname{coro}{Corollary}{Corollaries}
\Crefname{lem}{Lemma}{Lemmas}
\Crefname{sec}{Section}{Sections}
\crefname{app}{appendix}{appendices}
\Crefname{app}{Appendix}{Appendices}
\crefname{prop}{proposition}{propositions}
\Crefname{prop}{Proposition}{Propositions}
\Crefname{propty}{Property}{Properties}
\crefname{figure}{fig.}{figures}
\Crefname{figure}{Fig.}{Figures}
\crefname{defn}{definition}{definitions}
\Crefname{defn}{Definition}{Definitions}
\crefname{fact}{fact}{facts}
\Crefname{fact}{Fact}{Facts}
\crefname{appendix}{appendix}{appendices}
\Crefname{appendix}{Appendix}{Appendices}
\crefname{algo}{algorithm}{algorithms}
\Crefname{algo}{Algorithm}{Algorithms}
\crefname{algorithm}{algorithm}{algorithms}
\Crefname{algorithm}{Algorithm}{Algorithms}
\crefname{tbl}{table}{table}
\Crefname{tbl}{Table}{Table}
\crefname{table}{table}{table}
\Crefname{table}{Table}{Table}
\crefname{algorithm}{algorithm}{algorithms}
\Crefname{algorithm}{Algorithm}{Algorithms}

\crefname{conj}{conjecture}{conjectures}
\Crefname{conj}{Conjecture}{Conjectures}
\crefname{obs}{observation}{observations}
\Crefname{obs}{Observation}{Observations}

\newcommand{\mat}{\mathbf{A}}
\newcommand{\encmat}{\mathbf{A_e}}
\newcommand{\vect}{\mathbf{x}}
\newcommand{\res}{\mathbf{b}}
\newcommand{\encres}{\mathbf{b_e}}
\newcommand{\rows}{m}
\newcommand{\encrows}{m_e}
\newcommand{\randdecrows}{M'}
\newcommand{\decrows}{m_d}
\newcommand{\cols}{n}
\newcommand{\proc}{p}
\newcommand{\strag}{\alpha}
\newcommand{\numrep}{r}
\newcommand{\mdsmat}{\mathbf{F}}
\newcommand{\mdsnum}{k}
\newcommand{\fcdeg}{d}
\newcommand{\fcpc}{c}
\newcommand{\fcpdel}{\delta}

\newcommand{\RS}{\rho}
\newcommand{\numcomp}{C}
\newcommand{\workcomp}{B}

\newcommand{\runtime}{T}
\newcommand{\worktime}{Y}
\newcommand{\queuetime}{Z}
\newcommand{\shifttime}{\tau}
\newcommand{\smalltime}{t}
\newcommand{\exprate}{\mu}

\begin{document}
\title{Rateless Codes for Near-Perfect Load Balancing in Distributed Matrix-Vector Multiplication}
\author{Ankur Mallick}
\authornote{Correspondence Author}
\affiliation{%
  \institution{Carnegie Mellon University}
  \city{Pittsburgh}
  \state{PA}
}
\email{amallic1@andrew.cmu.edu}
\author{Malhar Chaudhari}
\authornote{Work done while at CMU}
\affiliation{%
  \institution{Oracle Corporation}
  \city{Redwood City}
  \state{CA}
}
\email{malharchaudhari@gmail.com}
\author{Utsav Sheth}
\authornotemark[2]
\affiliation{%
  \institution{Automation Anywhere}
  \city{San Jose}
  \state{CA}
}
\email{utsavsheth1994@gmail.com}
\author{Ganesh Palanikumar}
\authornotemark[2]
\affiliation{%
  \institution{Apple Inc.}
  \city{Cupertino}
  \state{CA}
}
\email{ganeshpkumar93@gmail.com}
\author{Gauri Joshi}
\affiliation{%
  \institution{Carnegie Mellon University}
  \city{Pittsburgh}
  \state{PA}
}
\email{gaurij@andrew.cmu.edu}

\begin{abstract}
Large-scale machine learning and data mining applications require computer systems to perform massive matrix-vector and matrix-matrix multiplication operations that need to be parallelized across multiple nodes. The presence of straggling nodes -- computing nodes that unpredictably slowdown or fail -- is a major bottleneck in such distributed computations. Ideal load balancing strategies that dynamically allocate more tasks to faster nodes require knowledge or monitoring of node speeds as well as the ability to quickly move data. Recently proposed fixed-rate erasure coding strategies can handle unpredictable node slowdown, but they ignore partial work done by straggling nodes thus resulting in a lot of redundant computation. We propose a \emph{rateless fountain coding} strategy that achieves the best of both worlds -- we prove that its latency is asymptotically equal to ideal load balancing, and it performs asymptotically zero redundant computations. Our idea is to create linear combinations of the $m$ rows of the matrix and assign these encoded rows to different worker nodes. The original matrix-vector product can be decoded as soon as slightly more than $m$ row-vector products are collectively finished by the nodes. We conduct experiments in three computing environments: local parallel computing, Amazon EC2, and Amazon Lambda, which show that rateless coding gives as much as $3\times$ speed-up over uncoded schemes.
\end{abstract}

\maketitle

\section{Introduction}
\label{sec:introduction}

Matrix-vector multiplications form the core of a plethora of scientific computing and machine learning applications that include solving partial differential equations \cite{ames2014numerical}, forward and back propagation in neural networks \cite{dally2015high}, computing the PageRank of graphs \cite{page1999pagerank} etc. In the age of Big Data, most of these applications involve multiplying extremely large matrices and vectors and the computations cannot be performed efficiently on a single machine. This has motivated the development of several algorithms \cite{kumar1994introduction}, \cite{fox1987matrix} that seek to speed-up matrix-vector multiplication by distributing the computation across multiple computing nodes. The individual nodes (the \emph{workers}) perform their respective tasks in parallel while a central node (the \emph{master}) aggregates the output of all these workers to complete the computation. 

\noindent \textbf{The Problem of Stragglers.} Unfortunately, large-scale distributed computation jobs are often bottlenecked by tasks that are run on unpredictably slow or unresponsive workers \emph{called stragglers} \cite{dean2013tail}. Since the job is complete only when all its parallel tasks are executed, this problem is aggravated for jobs with a large number of parallel tasks. Even a tiny probability of a node slowing down and becoming a straggler can cause a big increase in the expected latency of the job. As pointed out in \cite[Table 1]{dean2013tail}, the latency of executing many parallel tasks could be significantly larger ($140$ ms) than the median latency of a single task ($1$ ms). Straggling of nodes is widely observed in cloud infrastructure \cite{dean2013tail} and it is the norm rather than an exception.
\subsection{Previous Solution Approaches} 
\textbf{Load Balancing Strategies.} An obvious solution to overcome the bottleneck of waiting for slow nodes is to move tasks from busy or slow nodes to idle or fast nodes. Such work stealing or dynamic load balancing strategies are often implemented in shared and distributed memory settings \cite{dinan2007dynamic,dinan2009scalable,harlap2016addressing}. This approach involves establishing a protocol for continually monitoring workers and moving tasks from slow to fast workers. It entails considerable centralized control over the computing environment, which may not be feasible in cloud systems where the nodes can unpredictably slow down due to background processes, network outages etc. There may also be concerns regarding data privacy, and the communication cost of moving data between nodes in a distributed system spread over a large geographical area. Thus it is desirable to develop a principled and easy-to-implement straggler-mitigation approach that does not involve moving data between workers.

\noindent \textbf{Task Replication.} Existing systems like MapReduce \cite{dean2008mapreduce} and Spark \cite{zaharia2010spark} generally deal with the problem of stragglers by launching replicas of straggling tasks, which are referred to as \emph{back-up} tasks. This strategy of task replication has many variants such as \cite{ananthanarayanan2010reining,ananthanarayanan2013effective}, and has been theoretically analyzed in \cite{wang2014efficient, wang2015using, wang2019efficient} where schemes for adding redundant copies based on the tail of the runtime distribution at the workers are proposed. In the area of queueing theory there is a line of interesting recent works analyzing the effect of task replication on queueing delays in multi-server systems \cite{gardner2016s&x, gardner2015reducing, gardner2015analyzing, joshi2018synergy, joshi2017boosting, sun_shroff_2016, sun2015provably}. For distributed matrix-vector multiplication, which is the focus of this work, a simple replication strategy is to divide matrix $\mat$ into $\proc/\numrep$ (where $\numrep$ divides the number of workers $\proc$) sub-matrices and replicate each sub-matrix at $\numrep$ workers. Then the master waits for the fastest worker from each set of $\numrep$ to finish multiplying its sub-matrix with the vector $\vect$ in order to recover the overall result $\res = \mat \vect$.

\noindent \textbf{Erasure Coded Matrix-vector Multiplication.} From a coding-theoretic perspective, task replication is a special case of more general \emph{erasure codes} that overcome loss or erasure of data and recover the message from a subset of the transmitted bits. Erasure codes were first employed to overcome stragglers in the context of fast content download from distributed storage \cite{huang2012codes, joshi2012coding, joshi2014delay}. A file that is divided into $k$ chunks and encoded using a $(p,k)$ maximum-distance-separable (MDS) code (for example a Reed-Solomon code), can be recovered by downloading any $k$ out of $p$ encoded chunks. Queueing models to analyze the latency of coded content download jobs were proposed and analyzed in \cite{joshi2014delay, joshi2015queues, shah2016when, shah2017mds, joshi2017efficient}.

Unlike distributed storage, erasure coding of computing jobs is not straightforward. A job with $n$ parallel tasks needs to be designed such that the execution of any $k$ out of $n$ tasks is sufficient to complete the job. However, this is possible for linear computations such as matrix-vector multiplication. The usage of codes to provide error-resilience in computation has its origins in works on algorithmic fault tolerance \cite{huang1984algorithm}. Recent works such as \cite{lee2017speeding,li2016unified,dutta2016short} have employed Maximum Distance Separable (MDS) codes to speed up the computation of matrix vector products in a distributed setting. For example, suppose that we want to multiply a matrix $\mat$ with vector $\vect$ using $3$ worker nodes and a $(3,2)$ MDS code. Then we split $\mat$ along rows into two matrices $\mat_1$ and $\mat_2$ such that $\mat = [\mat_1^T \,\,\, \mat_2^T]^T$. The worker nodes store matrices $\mat_1$, $\mat_2$ and $\mat_1 + \mat_2$ respectively, and each node multiplies its matrix with $\vect$. Results from any two worker nodes are sufficient to obtain $\mat \vect$, and thus the system is tolerant to $1$ straggling node.

\noindent \textbf{Erasure Coding for Other Linear Computations.} A natural generalization of matrix-vector multiplication is matrix-matrix multiplication, considered in \cite{yu2017polynomial, dutta2018optimal, wang2018coded}. There are also works dealing with coded gradient descent \cite{tandon2017gradient,halbawi2017improving,li2017near}, coded convolution \cite{dutta2017coded}, coded Fourier Transform \cite{yu2017coded}, Page Rank \cite{yang2017coded}, and coded distributed optimization \cite{karakus2017encoded} -- in general any distributed \emph{linear} computation can essentially be expressed as a matrix multiplication/addition operation and can be made straggler-proof using erasure codes. Most of these works use MDS codes as the core idea and modify it for the specific computation in question. In this work, we focus on the original problem of coded distributed matrix-vector multiplication, but we consider a rateless-coded approach in contrast to MDS codes. We expect that the underlying principles of our work can be extended to matrix-matrix multiplication and other linear computations like gradient descent in the future.
\subsection{Rateless Coding Approach and its Benefits}
\label{sec:rateless_benefits}
The replication or MDS coding strategies used for matrix-vector multiplication are fixed-rate strategies, that is, they fix a redundancy rate $k/\proc$ when encoding matrix $\mat$, use the results of the fastest $k$ out of $\proc$ worker nodes. The key drawbacks of this approach are that: 1) it cannot perform load balancing within the fastest $k$ nodes and account for variabilities in their speeds, and 2) it discards all partial work done by the $p-k$ straggling workers. We address both these issues by proposing the use of \emph{rateless fountain codes}, specifically Luby Transform (LT) codes \cite{luby2002lt,shokrollahi2006raptor,joshi2010fountain} which are known to be scalable and form the basis for practical erasure coding schemes in wireless communication standards \cite{shokrollahi2011raptor}.

The rateless coded matrix-vector multiplication algorithm generates $\encrows = \alpha m$ ($\alpha > 1$) coded linear combinations of the $\rows$ rows of matrix $\mat$ and distributes them equally across $\proc$ worker nodes. Each of these linear combinations is generated by choosing $d$ of the $\rows$ rows uniformly at random and adding them. For example, if $d=2$, and we choose rows $\mathbf{a}_1$ and $\mathbf{a}_3$ of $\mat$, then the encoded row is $\mathbf{a}_1 + \mathbf{a}_3$, as shown in \Cref{fig:encoding_graph}. The value $d$, referred to as the degree of the linear combination is an i.i.d. realization of a carefully chosen degree distribution $\rho(d)$. For LT codes, $\rho(d)$ is the Robust Soliton distribution (given in \eqref{eqn:robust_soliton} below). Each worker receives $\encrows/\proc$ encoded rows of matrix $\mat$ and a copy of the vector $\vect$. It computes row-vector products, for example $(\mathbf{a}_1 + \mathbf{a}_3)^T \vect = b_1 + b_3$ for the encoded row $(\mathbf{a}_1 + \mathbf{a}_3)$, and sends them back to the master node. Due to the carefully chosen degree distribution $\rho(d)$, the master node can use an iterative peeling decoder \cite{luby2002lt} (illustrated in \Cref{fig:decoding_graph}) to recover each element of the product vector $\res =\mat\vect$ with a low decoding complexity of $\mathcal{O}(\log\rows)$. Overall, it needs to wait for any $\randdecrows = \rows(1+ \epsilon)$ row-vector products to be completed across \emph{all} the nodes, where $\epsilon$ is a small overhead; $\epsilon \rightarrow 0$ as $\rows \rightarrow \infty$).

Rateless codes offer the following key benefits over previously proposed coding techniques based on MDS codes.
\begin{enumerate}[wide,labelwidth=!, labelindent=0pt]
\item \textbf{Near-Ideal Load Balancing.} In order to adjust to varying speeds of worker nodes and minimize the overall time to complete the multiplication $\mat \vect$, one can use an \emph{ideal load-balancing scheme that dynamically assigns one row-vector product computation task to each worker node as soon as the node finishes its current task}. Thus, faster nodes complete more tasks than slower nodes, and the final product $\res =\mat \vect$ is obtained when the $\proc$ nodes collectively finish $\rows$ row-vector products. Our rateless coding strategy achieves nearly the same load balancing benefit without the communication overhead of dynamically allocating the tasks one row-vector product at a time. In our strategy, the nodes need to collectively finish $\randdecrows = \rows(1+\epsilon)$ row-vector products, for small $\epsilon$ that goes to zero as $\rows \rightarrow \infty$. In contrast, MDS coding strategies do not adjust to different degrees of node slowdown; they use the results from $\mdsnum$ nodes, and ignore the remaining $\proc-\mdsnum$ nodes. As a result rateless codes achieve a much lower delay than MDS coding strategies.
\item \textbf{Negligible Redundant Computation.} A major drawback of MDS coding is that if there is no straggling, the worker nodes collectively perform $\rows \proc/\mdsnum$ row-vector products, instead of $\rows$. With the rateless coding strategy, the nodes collectively perform a maximum of $\randdecrows =   \rows (1+\epsilon)$ row-vector products where, the overhead $\epsilon$ goes to zero as $m$, the number of rows in the matrix $\mat$ increases. 
\item \textbf{Maximum straggler tolerance.} A $(\proc,\mdsnum)$ MDS coded distributed computation is robust to $\proc-\mdsnum$ straggling nodes, for $\mdsnum \in [1, 2, \dots \proc]$. Reducing $\mdsnum$ increases straggler tolerance but also adds more redundant computation. The rateless coding scheme can tolerate up to $\proc-1$ stragglers, with negligible redundant computation overhead.
\item \textbf{Low Decoding Complexity.}
One may argue that MDS coding approaches can also use partial computations and achieve near-perfect load balancing if we construct an $(\encrows,\rows)$ MDS code (for a given amount of redundancy $\encrows/\rows$) to encode a $\rows\times\cols$ matrix. The decoding complexity of such a code is $\mathcal{O}(\rows^3)$ which is unacceptable for large $\rows$ in practical implementations. Rateless codes offer a low decoding complexity: $O(\rows \log \rows)$ for LT codes \cite{luby2002lt}, and $O(\rows)$ for Raptor codes \cite{shokrollahi2006raptor}.
\end{enumerate}
\noindent \textbf{Difference from \cite{severinson2017block,wang2018coded} on LT-coded Matrix-vector Multiplication.} The use of Luby Transform (LT) codes for matrix-vector multiplication has been recently proposed in \cite{severinson2017block,wang2018coded}. \emph{However, these works do not utilize the `rateless' property of LT codes and instead use them in a fixed-rate setting.} For example, the algorithm in \cite{severinson2017block} generates $\rows_e$ LT-coded rows from an $m$-row matrix using LT codes, and it allocates each row to $\eta q$ workers for some $\frac{1}{\proc} \leq \eta \leq 1$. Each worker completes the entire set of row-vector product tasks assigned to them, and the master waits for the fastest $q$ workers to finish. Partial computations performed by slow workers are discarded. The scheme proposed in \cite{wang2018coded} also uses LT codes in this fixed-rate setting and focuses on using the sparsity of LT codes to reduce the decoding complexity of coded matrix multiplication. Thus, although these works use LT codes, they are similar in spirit to fixed-rate MDS-coding approaches.

To the best of our knowledge, our work is the first to exploit the \emph{rateless} nature of LT codes to perform load-balancing in distributed matrix computations and utilize all the partial work done by slow workers. We also provide the first theoretical analysis of the latency achieved by this strategy with ideal load balancing and show that it asymptotically achieves near-perfect latency and computation cost. Previous works \cite{severinson2017block, wang2018coded} do not present such analyses. Moreover, we present extensive experimental results on $3$ different computing environments: local parallel computing, distributed computing on Amazon EC2 and serverless computing on Amazon Lambda.
\begin{figure}
    \centering
    \includegraphics[width=0.4\textwidth]{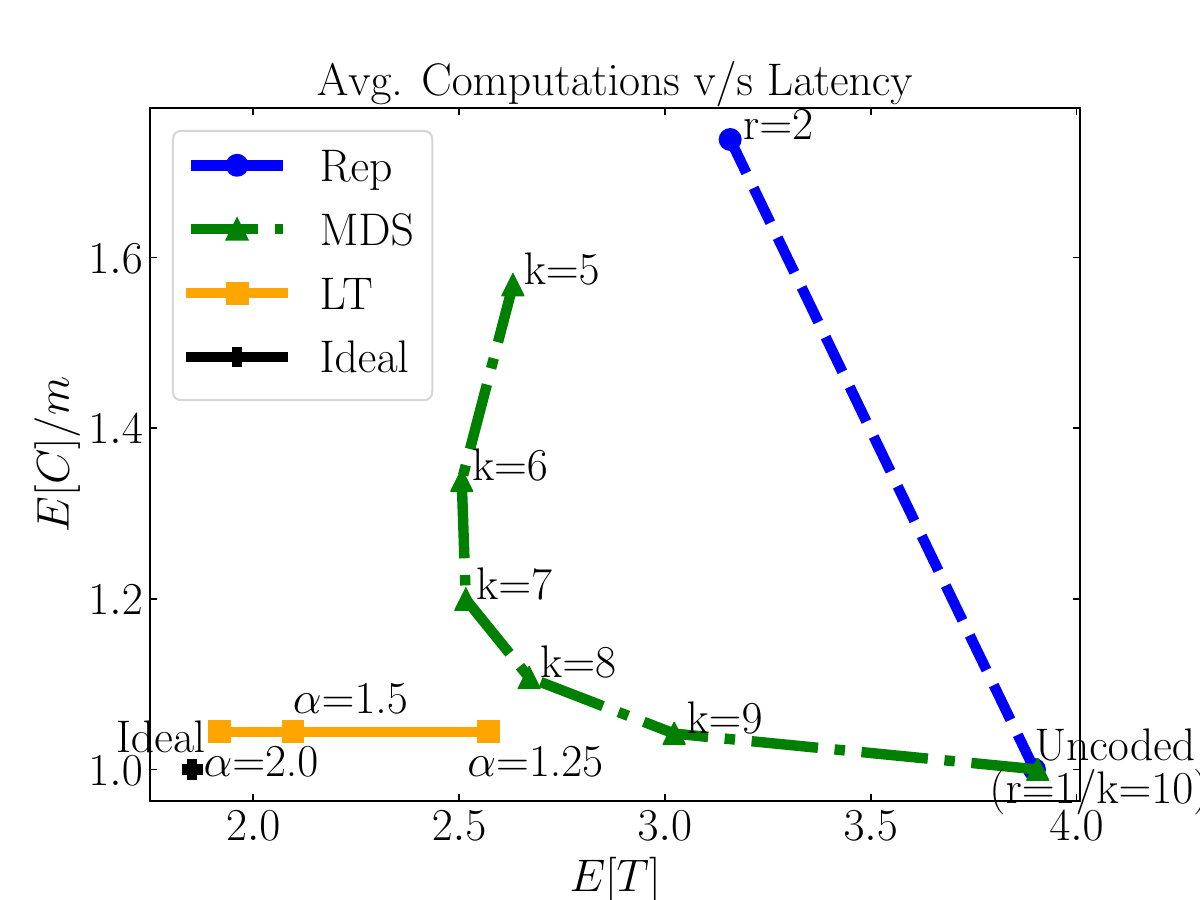}
    \caption{The expected latency ($\mathbb{E}[\runtime]$) of the LT-Coded approach smoothly decays on adding redundancy (increasing $\alpha$) and approaches the Ideal approach without any increase in computational overhead ($\mathbb{E}[\numcomp]/\rows$). Previous approaches - Replication (Rep) and MDS coding not only have a higher latency for the same task but also perform far more redundant computations. The simulation parameters are $\rows = 10000$, number of worker nodes $\proc = 10$ and delay model parameters $\exprate = 1.0, \shifttime = 0.001$.}
    \label{fig:latcompsim}
    \vspace{-0.5cm}
\end{figure}
\subsection{Main Theoretical and Experimental Results}
Besides proposing the rateless coding strategy, one of the main contributions of our work is to theoretically analyze and compare it with ideal load balancing. In particular, we consider two performance metrics: 1) latency $\runtime$, which is the time until $\res = \mat \vect$ can be recovered by the master, and 2) number of computations $\numcomp$, which is the number of row-vector product tasks completed by the $\proc$ workers until $\res$ can be recovered. We consider a simple delay model where worker $i$ has an initial delay of $X_i$ after which it spends a constant time $\tau$ per row-vector product task. 

\noindent \textbf{Comparison with Ideal Load Balancing.} In the ideal load balancing strategy, the $\rows$ row-vector product tasks (which comprise the job of multiplying the $\rows \times \cols$ size $\mat$ with vector $\vect$) are kept in a central queue at the master and dynamically allocated to idle workers one task at a time. The job is complete when $\rows$ tasks are collectively finished by the workers. The rateless coding strategy differs from this ideal policy in two ways due to which its latency is larger: 1) each worker gets $m_e/\proc = \alpha m/\proc$ encoded rows and thus a fast worker may run out of rows before the master is able to recover $\res = \mat \vect$ and 2) the workers collectively need to finish $m(1+\epsilon)$ tasks where $\epsilon$ is a small overhead that diminishes as $\rows \rightarrow \infty$. Our main theoretical result stated in the following (informal) theorem compares the two latencies.
\begin{thm}
\label{thm:T_LT_informal}
The latency $\runtime_{\text{LT}}$ and computations $\numcomp_{\text{LT}}$ of  our LT coded distributed matrix-vector multiplication strategy in computing the product of a $\rows\times\cols$ matrix $\mat$ with a $\cols \times 1$ vector $\vect$ satisfy the following for large $\rows$:
\begin{align}
    \Pr(\runtime_{\text{LT}} > \runtime_{\text{ideal}}) &= \proc\exp \left(-\frac{\exprate\shifttime\rows(\strag-1)}{\proc^2}\right) \\
    \frac{\mathbb{E}[\runtime_{\text{LT}}]- \mathbb{E}[\runtime_{\text{ideal}}]}{\mathbb{E}[\runtime_{\text{ideal}}]} &=  O \left(\exp(-\frac{\shifttime\rows(\strag-1)}{\proc^2} )\right) \\
    \frac{\mathbb{E}[\numcomp_{\text{LT}}]}{\mathbb{E}[\numcomp_{\text{ideal}}]} &= \frac{m(1+\epsilon)}{m} \text{ where } \epsilon \rightarrow 0 \text{ as } m \rightarrow \infty.
\end{align}
where $\encrows = \alpha m$ (for $\alpha > 1$) is the number of encoded rows, the initial delay at each worker is $X_i \sim \exp(\exprate)$ and $\tau$ is the time taken to complete each row-vector product task. Due to the inherent design of LT codes, the overhead $\epsilon \rightarrow 0$ as $m \rightarrow \infty$.
\end{thm}

This results shows that as long as the number of encoded rows $\encrows$ is sufficiently larger than $\rows$, despite not performing dynamic task assignment, the rateless coding strategy can seamlessly adapt to varying initial delays at the workers. Its runtime $\runtime_{\text{LT}}$ and $\numcomp_{\text{LT}}$ asymptotically converge to the ideal strategy. The exact results are derived in \Cref{thm:probltlatency} and \Cref{thm:ltlatency}.

\noindent \textbf{Comparison with MDS and Replication Strategies.} Unlike our rateless coding strategy, MDS-coded and replication-based strategies give strictly worse latency and cost than the ideal scheme and the gap does not go to zero. In \Cref{sec:theoretical} we analyze the expected latency and computations of these strategies. \Cref{fig:latcompsim} shows simulation plots of the latency-computation trade-off of these strategies clearly demonstrating the superiority of using rateless LT codes.
\begin{figure*}[t]
\centering
\subfloat[Uncoded\label{fig:lb_uncoded}]{\includegraphics[width=0.4\linewidth]{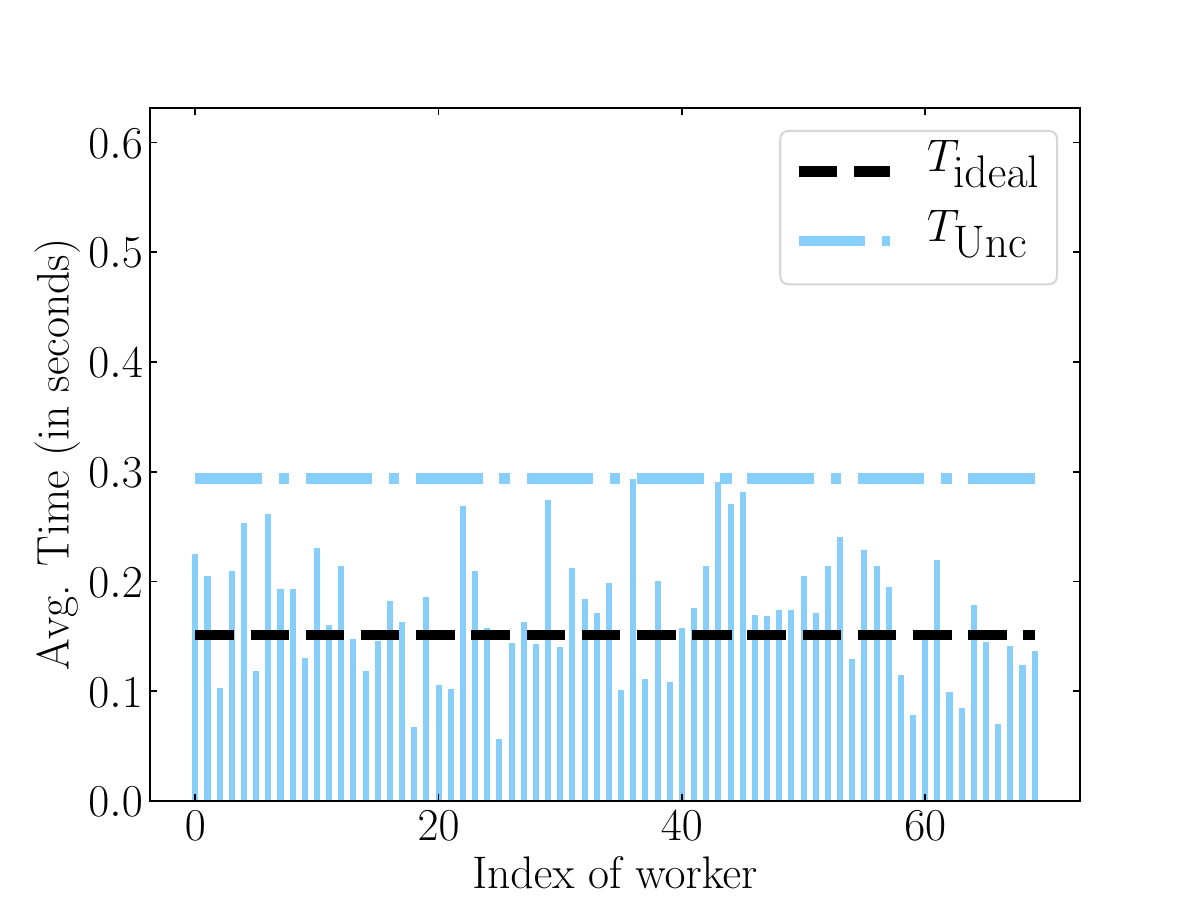}}
  \subfloat[2-Replication\label{fig:lb_rep}]{\includegraphics[width=0.4\linewidth]{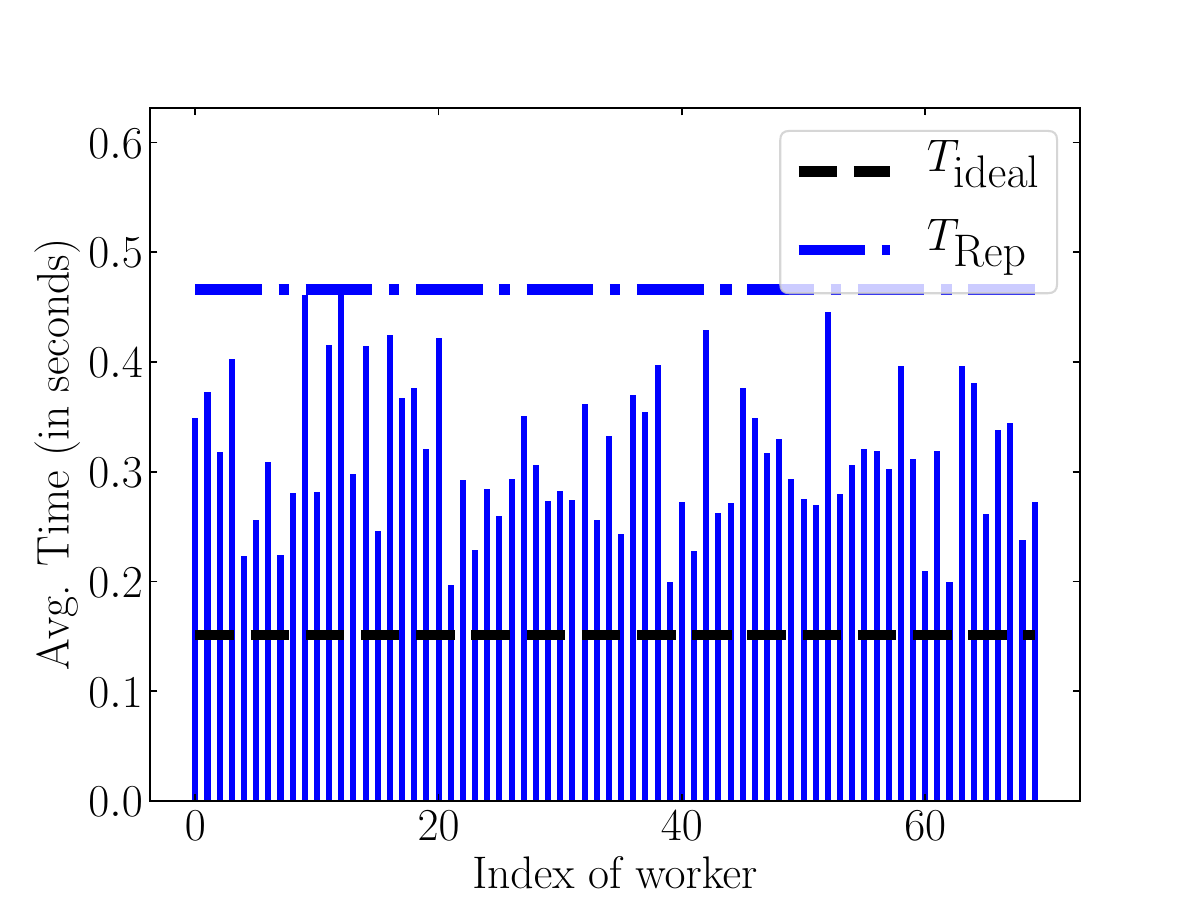}}\\
\subfloat[MDS\label{fig:lb_mds}]{\includegraphics[width=0.4\linewidth]{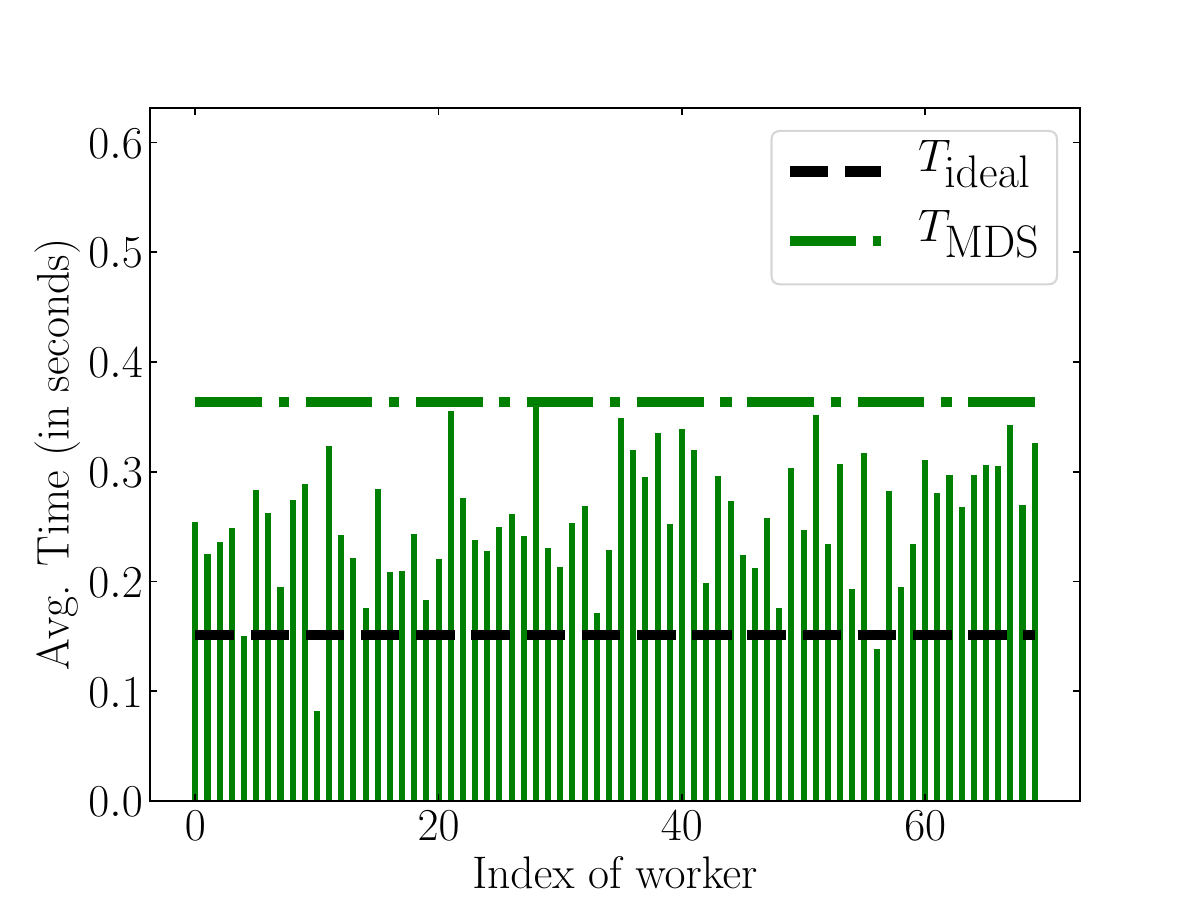}}
  \subfloat[LT\label{fig:lb_lt}]{\includegraphics[width=0.4\linewidth]{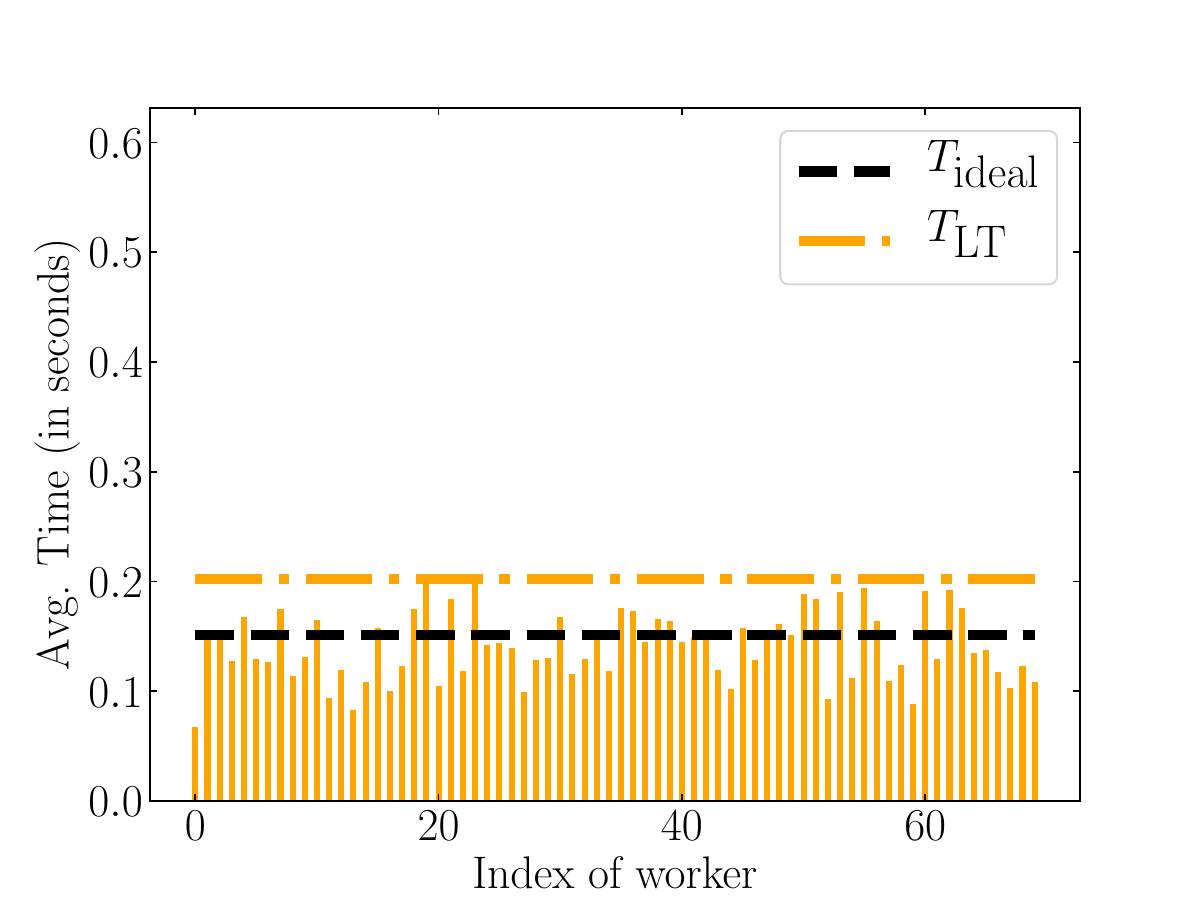}}\\
    \caption{Comparison of load balancing across different matrix-vector multiplication approaches. The rows of a $11760 \times 9216$ matrix $\mat$ are encoded and distributed among $70$ EC2 workers. The height of the bar plot for each worker indicates the time spent by the worker computing row-vector products either until it finishes its assigned tasks or is terminated by the master because the final matrix-vector product $\mat \vect$ has been successfully decoded. The dash-dot line indicates the overall latency (time at which matrix-vector product $\mat \vect$ can be successfully decoded) in each case, and the black dashed line is the latency of ideal load balancing. The LT coded approach exhibits near-ideal load balancing, and has lower latency than other approaches.} \vspace{-1.0em}
    \label{fig:loadbalance}
\end{figure*}

\noindent \textbf{Experimental Results}. \Cref{fig:loadbalance} shows the results of implementing our rateless coded strategy for a real distributed matrix-vector multiplication task on a cluster of $70$ EC2 \cite{amazon_ec2} workers deployed using Kubernetes \cite{Kubernetes}. The computation involves multiplying a $11760 \times 9216$ matrix $\mat$ extracted from the STL-10 dataset \cite{coates2011analysis} with vectors extracted from the same dataset, and is implemented using Dask \cite{Dask} a popular framework for parallel computing in Python. The proposed rateless coded strategy significantly outperforms the uncoded ($3\times-$speedup) and MDS coded ($2\times-$speedup) approaches. The plots in \Cref{fig:loadbalance} also show that the variability in individual worker times is significantly lower for our rateless coded strategy (\Cref{fig:lb_lt}) than for other approaches as fast nodes perform more tasks than slow nodes under our approach leading to much better load balancing. The latency of each approach is also compared to $\runtime_{\text{ideal}}$, the latency of the ideal load-balancing strategy, approximated in this case as the minimum time required by the workers to compute $11760$ encoded row-vector products in total. Observe that $\runtime_{\text{LT}}$ is closest to $\runtime_{\text{ideal}}$. We also obtain similar improvements with LT coding in parallel computing using Python's Multiprocessing library \cite{Multiprocessing} library on a single machine, and in serverless computing on AWS Lambda \cite{amazon_lambda} as described in \Cref{sec:experiments}.
\subsection{Organization}
The rest of the paper is organized as follows. \Cref{sec:prob_formu} presents the system model, performance criteria and comparison benchmarks. \Cref{sec:rateless} describes our rateless fountain coding strategy for distributed matrix-vector multiplication. \Cref{sec:theoretical} shows theoretical analyses and a latency-cost comparison of rateless coding with other strategies. \Cref{sec:queueing} extends these results to the queueing setting where vectors $\vect_1,\vect_2,\ldots$ that need to be multiplied with matrix $\mat$ arrive at rate $\lambda$. Experimental results are presented in \Cref{sec:experiments}. All proofs are deferred to the Appendix.
\section{Problem Formulation}
\label{sec:prob_formu}
\subsection{System Model}
\label{sec:model}
\begin{figure*}[t]
\centering
 \includegraphics[width=0.8\textwidth]{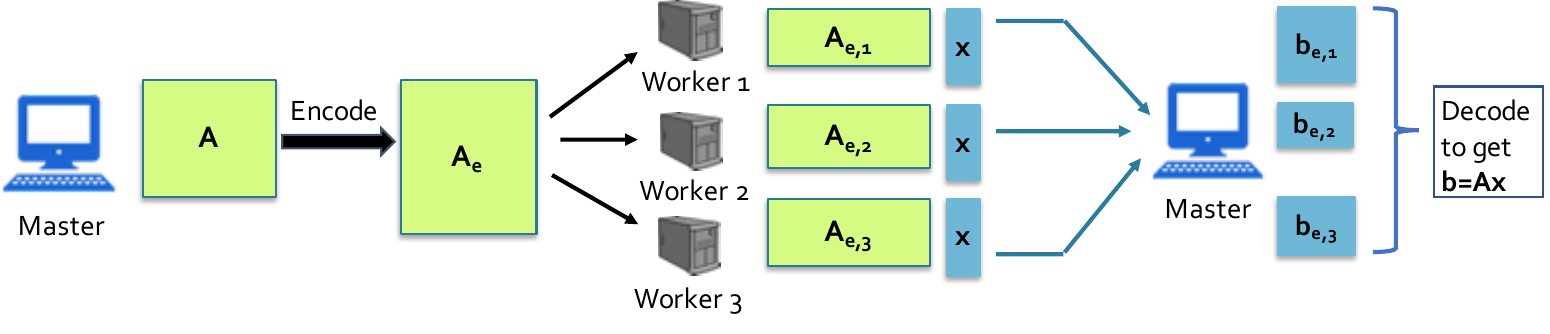}
 \caption{The system model for coded distributed matrix vector multiplication with a master-worker framework. The master generates the encoded matrix $\encmat$ by applying a coding scheme to the rows of $\mat$. Worker $i$ stores a submatrix of $\encmat$ denoted by $\encmat_{,i}$ and sends encoded row-vector products $\encres_{,i}$ to the master ($i=1,\ldots, \proc$). Different $\encres_{,i}$'s may have different sizes. The master decodes the encoded row-vector products in $\encres_{,i}$, $i=1,\ldots, \proc$ to recover $\res = \mat\vect$}.\vspace{-1.5em}
\label{fig:sysmodel}
\end{figure*}
Consider the problem of multiplying a $\rows \times \cols$ matrix $\mat$ with a $\cols \times 1$ vector $\vect$ using $\proc$ worker nodes and a master node as shown in \Cref{fig:sysmodel}. The worker nodes can only communicate with the master, and cannot directly communicate with other workers. The goal is to compute the result $\res = \mat\vect$ in a distributed fashion and mitigate the effect of unpredictable node slowdown or straggling. The rows of $\mat$ are encoded using an error correcting code to give the $\encrows \times \cols$ encoded matrix $\encmat$, where $\encrows \geq \rows$. We denote the amount of redundancy added by the parameter $\strag = \encrows/\rows$.  Matrix $\encmat$ is split along its rows to give $\proc$ submatrices $\encmat_{,1},\ldots,\encmat_{,\proc}$ of equal size such that worker $i$ stores submatrix $\encmat_{,i}$. To compute the matrix-vector product $\res=\mat\vect$, the vector $\vect$ is communicated to the workers such that Worker $i$ is tasked with computing the product $\encmat_{,i} \vect$.

To complete the assigned task, each worker needs to compute a sequence of row vector products of the form $\mathbf{a}_{\mathbf{e},j}\vect$ where $\mathbf{a}_{\mathbf{e},j}$ is the $j^{\text{th}}$ row of $\encmat$. The time taken by a worker node to finish computing one or more row-vector products may be random due to variability in the node speed or variability in the amount of computation assigned to it.
The master node aggregates the computations of all, or a subset of, the workers into the vector $\encres$, which is then decoded to give the final result $\res=\mat\vect$. If $\encres$ is not decodable, the master waits until workers compute more row-vector products. 
\subsection{Performance Criteria}
\label{sec:perf_metrics}
We use the following metrics to compare different distributed matrix-vector multiplication schemes via theoretical analysis and associated simulations (\Cref{sec:theoretical}), and experiments  in parallel, distributed, and serverless environments (\Cref{sec:experiments}).
\begin{defn}[Latency ($\runtime$)]
\label{defn:latency}
The latency $\runtime$ is the time required by the system to complete enough computations so that $\res=\mat\vect$ can be successfully decoded from worker computations aggregated in $\encres$.
\end{defn}
\begin{defn}[Computations ($\numcomp$)]
\label{defn:comp}
The number of computations $\numcomp$ is defined as the total number of row-vector products $\mathbf{a}_{\mathbf{e},j}\vect$ performed collectively by the worker nodes until $\res = \mat \vect$ is decoded.
\end{defn}
For any strategy we always have $\numcomp\geq\rows$ where $\rows$ is the number of rows of $\mat$ or the number of elements in $\res$.
\subsection{Benchmarks for Comparison}
\label{sec:benchmarks}
\begin{figure*}[t]
 \centering
 \subfloat[Ideal]{\includegraphics[width=0.27\linewidth]{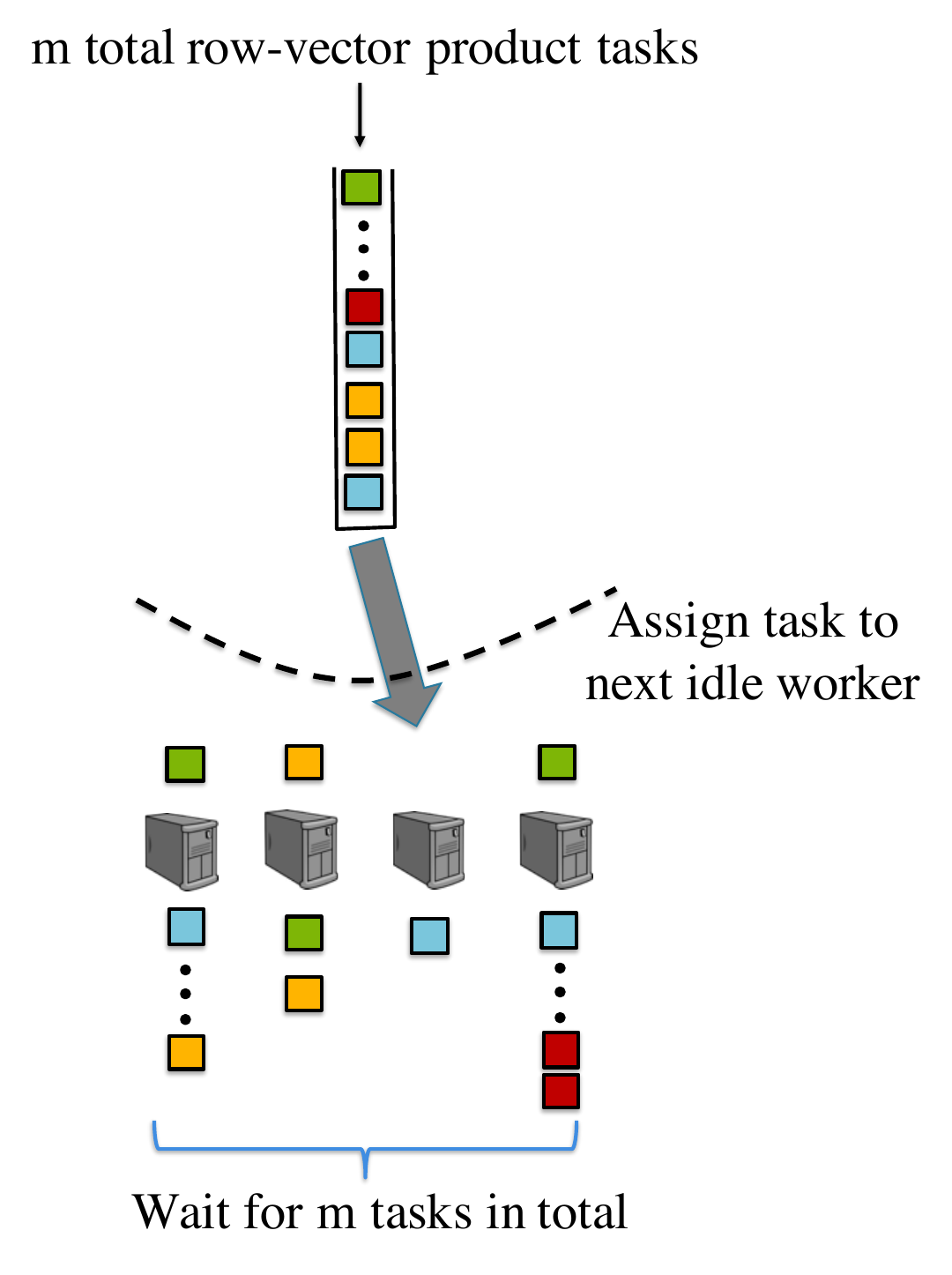}}
 \subfloat[Replication]{\includegraphics[width=0.24\linewidth]{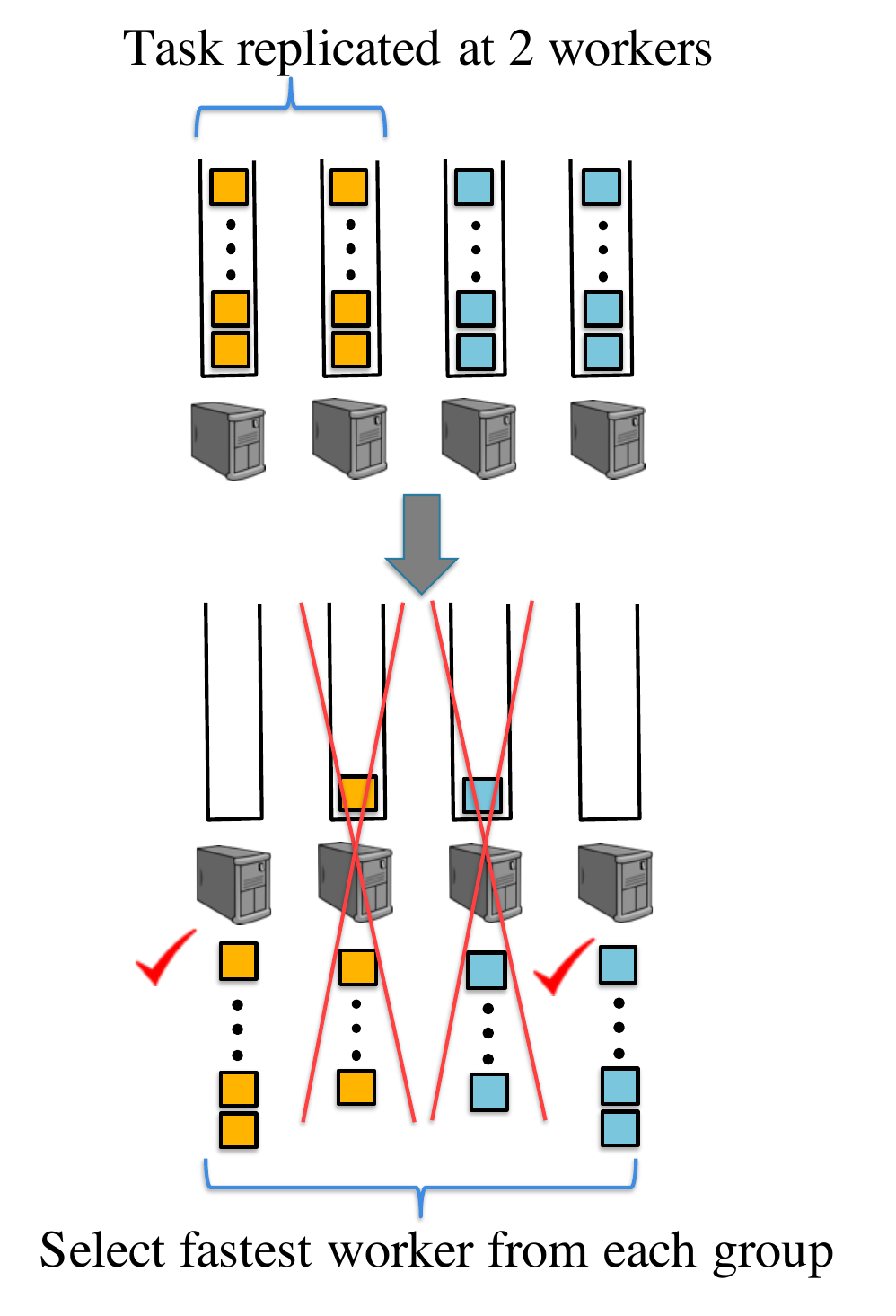}}
  \subfloat[MDS Coded]{\includegraphics[width=0.22\linewidth]{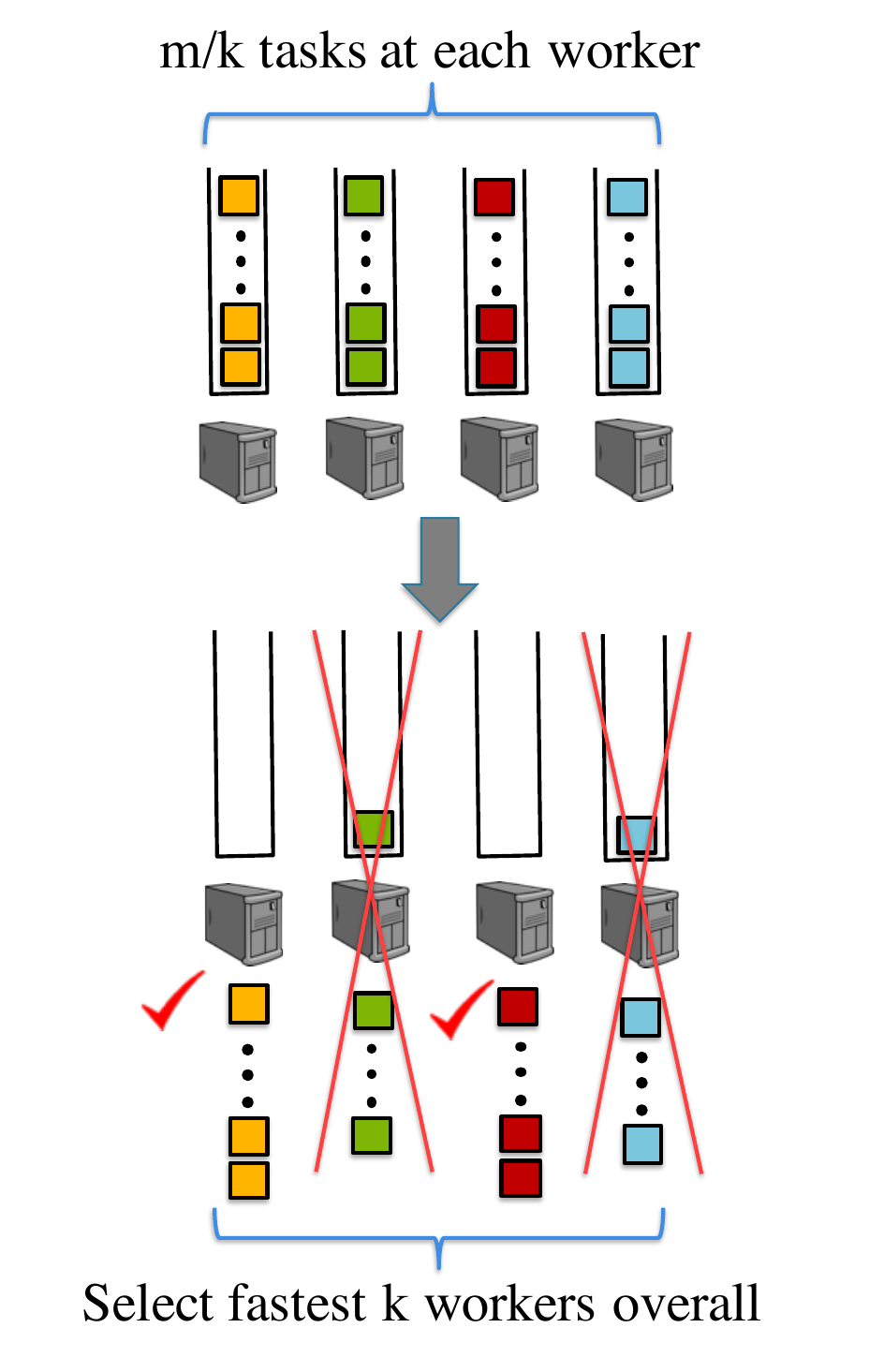}}
  \subfloat[Rateless Coded]{\includegraphics[width=0.24\linewidth]{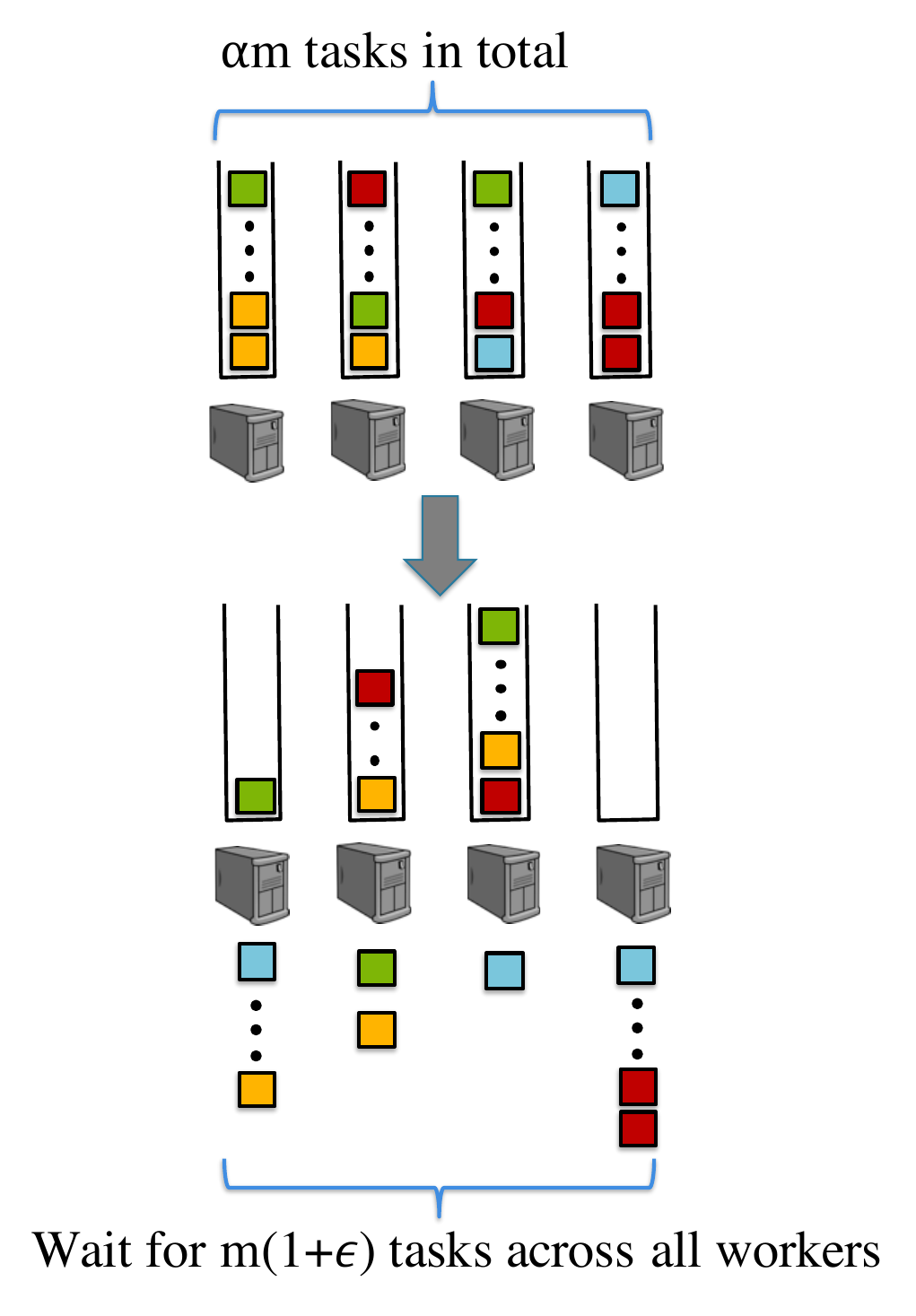}}\\
  \caption{Each square represents one row-vector product task out of a total of $\rows$ tasks to be completed by the $\proc$ workers. In the ideal scheme we have a central queue of $\rows$ tasks and each worker is assigned a new task as soon as it becomes idle until all $\rows$ tasks are completed. In the replication scheme, the master waits for the fastest worker for each sub-matrix. With MDS coding, the master needs to wait for $\mdsnum$ out of $\proc$ workers, but each worker has to complete $\rows/\mdsnum$ tasks. The rateless coded strategy requires waiting for only $\rows(1+\epsilon)$ tasks across \emph{all} workers.\label{fig:computingstrategies}}\vspace{-1.0em}
\end{figure*}
We compare the performance of the proposed rateless coded strategy with three benchmarks: ideal load balancing, $r$-replication, and the $(p,k)$ MDS-coded strategy, which are described formally below. \Cref{fig:computingstrategies} illustrates the differences in the way row-vector product tasks are assigned to and collected from workers in each strategy.

\textbf{Ideal Load Balancing.} The multiplication of the $\rows \times n$ matrix $\mat$ with the $n \times 1 $ vector $\vect$ can be treated as a job with $\rows$ tasks, where each task corresponding to one row-vector product. In the ideal load balancing strategy, the master node maintains a central queue of these $m$ tasks. It dynamically assigns one task to each of the $\proc$ workers as soon as a worker finishes its previous task. The matrix-vector multiplication is complete when exactly $\rows$ tasks are collectively finished by the workers. This strategy seamlessly adapts to varying worker speeds without performing any redundant computation ($C=m$); hence it gives the optimal latency-computation trade-off. This strategy may be impractical due to the constant communication between the master and the worker nodes. Nevertheless, it serves as a good theoretical benchmark for comparison with the rateless, replication and MDS strategies.

\textbf{The $\numrep-$Replication Strategy.} A simple distributed multiplication strategy is to split $\mat$ along its rows into $\proc/\numrep$ submatrices $\mat_{1}, \ldots, \mat_{\proc/\numrep}$, with $\numrep\rows/\proc$ rows each (assume that $\proc/\numrep$ divides $\rows$) and multiply each submatrix with $\vect$ in parallel on $\numrep$ distinct worker nodes. The master collects the results from the fastest of the $\numrep$ nodes that have been assigned the task of computing the product $\mat_{i}\vect$ for all $i$. The computed products are aggregated into the $\rows \times 1$ vector $\res$. \emph{Setting $r=1$ corresponds to the naive or uncoded strategy where $\mat$ is split into $p$ sub-matrices and each worker node computes the corresponding submatrix-vector product.} While this approach performs the least number of computations it is susceptible to straggling nodes or node failures. Increasing the number of replicas provides greater straggler tolerance at the cost of redundant computations. Real distributed computing frameworks like MapReduce \cite{dean2008mapreduce} and Spark \cite{zaharia2010spark} often use $\numrep=2$ i.e. each computation is assigned to $2$ different worker nodes for added reliability and straggler tolerance.

\textbf{The $(\proc,\mdsnum)$ MDS Coded Strategy.}
\label{sec:mdscoded}
Recent works like \cite{lee2017speeding,li2016unified} have applied MDS coding to overcome the problem of stragglers in the uncoded strategy. The strategy involves pre-multiplying $\mat$ at the central node with a suitable encoding matrix $\mdsmat$ denoting the MDS codes. For encoding using a $(\proc, \mdsnum)$ MDS code, the matrix $\mat$ is split along its rows into $\mdsnum$ matrices $\mat_{1}, \ldots, \mat_{\mdsnum}$, each having $\rows/\mdsnum$ rows. The MDS code adds $\proc-\mdsnum$ redundant matrices $\mat_{\mdsnum+1}, \ldots, \mat_{\proc}$ which are independent linear combinations of the matrices $\mat_{1}, \ldots, \mat_{\mdsnum}$. Worker $i$ computes the product $\mat_i\vect$. Thus the system is robust to $\proc-\mdsnum$ stragglers. However this strategy adds a significant computation overhead. When none of the nodes are slow, the system performs $\rows \proc/\mdsnum$ row-vector products (as opposed to $\rows$ row-vector products in the uncoded case).
\section{Proposed Rateless Coded Strategy}
\label{sec:rateless}
We describe how rateless codes, specifically LT codes \cite{luby2002lt}, can be applied to perform coded matrix vector multiplication, and then propose a distributed implementation of this scheme for straggler mitigation in computing the matrix-vector product $\res=\mat\vect$ using the master-worker framework of \Cref{sec:model}. 
\begin{figure*}[t]
 \centering
 \subfloat[Encoding Graph\label{fig:encoding_graph}]{\includegraphics[width=0.3\linewidth]{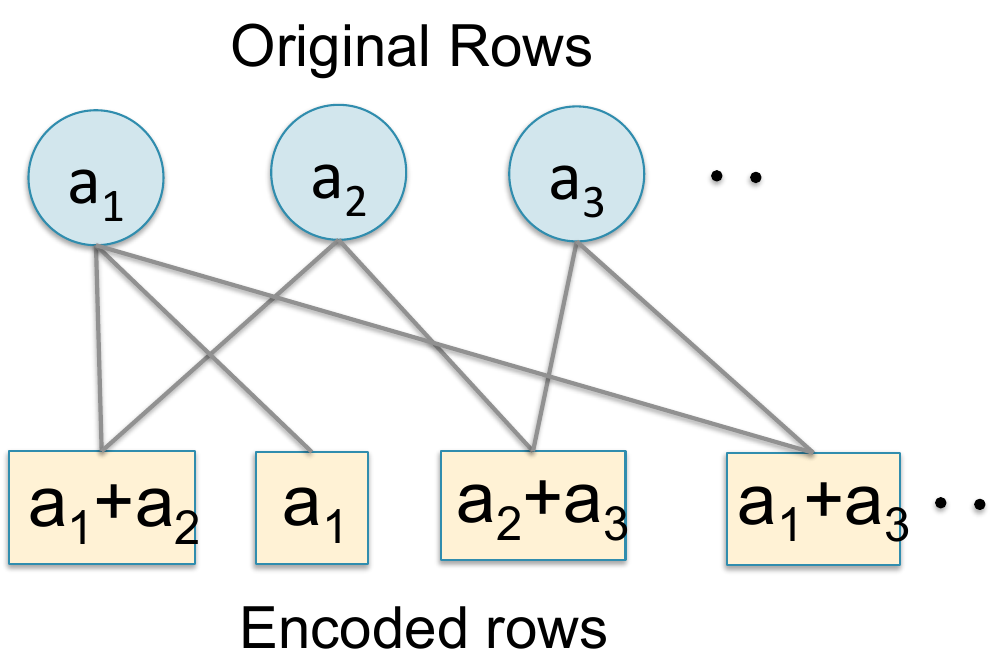}}
  \subfloat[Decoding Graph\label{fig:decoding_graph}]{\includegraphics[width=0.6\linewidth]{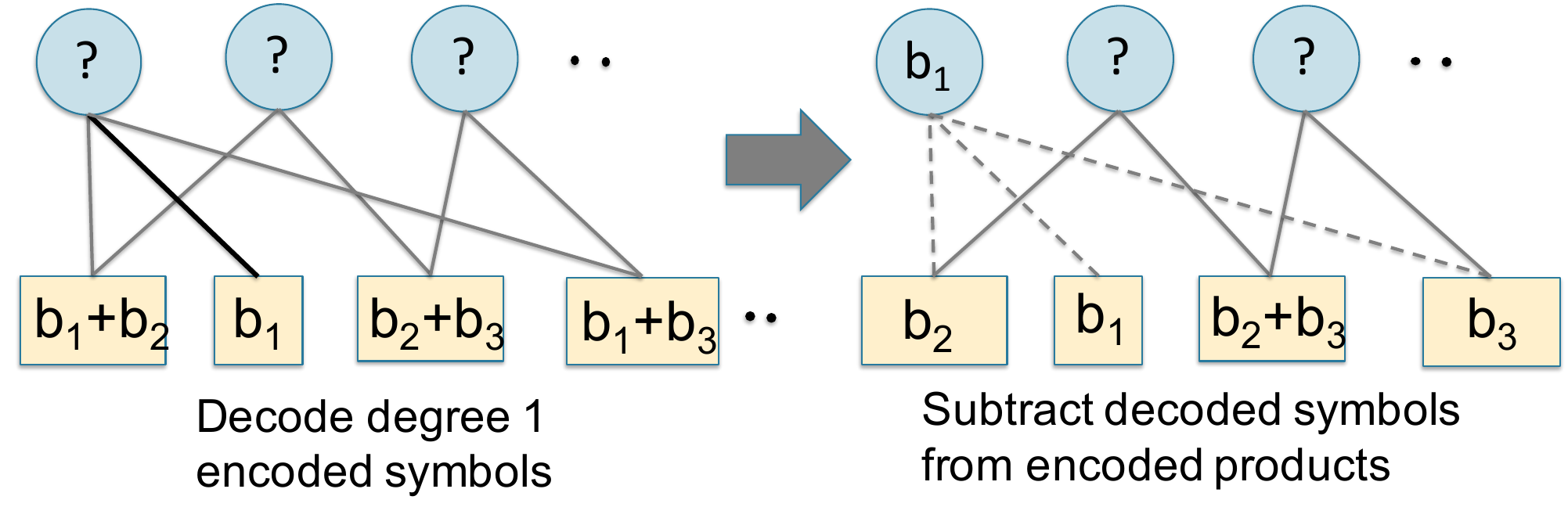}}
  \caption{(a) Bipartite graph representation of the encoding of the rows $\mathbf{a}_1, \mathbf{a}_2, \dots \mathbf{a}_m$ of matrix $\mat$. Each encoded row is the sum of $d$ rows of $\mat$ chosen uniformly at random, where $d$ is drawn from the Robust Soliton degree distribution given by \eqref{eqn:robust_soliton}. (b) In each step of the iterative decoding process, a single degree one encoded symbol is decoded directly, and is subtracted from all sums in which it participates.}\vspace{-1.0em}
\end{figure*}
\subsection{LT-Coded Matrix-vector Multiplication}
\label{sec:LTmatvect}
Luby Transform (LT) codes proposed in \cite{luby2002lt} are a class of erasure codes that can be used to generate a limitless number of encoded symbols from a finite set of source symbols. We apply LT codes to matrix-vector multiplication by treating the  $\rows$ rows of the matrix $\mat$ as source symbols. Each encoded symbol is the sum of $\fcdeg$ source symbols chosen uniformly at random from the matrix rows. Thus if $\mathcal{S}_\fcdeg \subseteq \{1,2, \dots \rows\}$ is the set of $\fcdeg$ row indices, the corresponding encoded row is $\mathbf{a}_{\mathbf{e}}=\sum_{i\in \mathcal{S}_\fcdeg}\mathbf{a}_i$. 

The number of original rows in each encoded row, or the degree $\fcdeg$, is chosen according to the Robust Soliton degree distribution
\begin{align}
\RS(\fcdeg) = \begin{cases} \frac{R}{\fcdeg\rows} + \frac{1}{\rows}&  \text{for } \fcdeg=1\\
\frac{R}{\fcdeg\rows} + \frac{1}{\rows(\rows-1)}&  \text{for } \fcdeg=2,\dots, \rows/R-1 \\
\frac{R \ln (R/\delta)}{\rows} + \frac{1}{\rows(\rows-1)}&  \text{for } \fcdeg = \rows/R \\
\frac{1}{\rows(\rows-1)}& \text{for } \fcdeg = \rows/R+1, \dots, \rows \end{cases} \label{eqn:robust_soliton}
\end{align}
where $R = \fcpc\log(\rows/\fcpdel)\sqrt{\rows}$ for some $\fcpc>0$ and $\delta \in [0,1]$, with $\fcpc$ and $\delta$ being design parameters. Some guidelines for choosing $\fcpc$ and $\fcpdel$ can be found in \cite{mackay2003information}. The probability of choosing $\fcdeg = \fcdeg_0$ is equal to $\rho(\fcdeg_0)/\sum_{i=1}^{m}\rho(i)$. Once the degree $\fcdeg$ is chosen, encoding  is performed by choosing $\fcdeg$ source symbols uniformly at random (this determines $\mathcal{S}_{\fcdeg}$) and adding them to generate an encoded symbol. The encoding process is illustrated in \Cref{fig:encoding_graph}.

Once the rows of the encoded matrix $\encmat$ are generated, we can compute the encoded matrix vector product $\encres = \encmat\vect$. To decode the desired matrix vector product $\res=\mat\vect$ from a subset of $\randdecrows$ symbols of $\encres$ we use the the \emph{iterative peeling decoder} described in \cite{luby2002lt,shokrollahi2006raptor,shokrollahi2011raptor}. If $\res = [b_1, b_2, \dots b_m]$, the decoder may receive symbols $b_1 + b_2 + b_3$, $b_2 + b_4$, $b_3$, $b_4$, and so on since each row of $\encmat$ is a sum of some rows of $\mat$. Decoding is performed in an iterative fashion. In each iteration, the decoder finds a degree one encoded symbol, covers the corresponding source symbol, and subtracts the symbol from all other encoded symbols connected to that source symbols. This decoding process is illustrated in \Cref{fig:decoding_graph}.

Since the encoding uses a random bipartite graph, the number of symbols required to decode the $\rows$ source symbols successfully is a random variable $\randdecrows$ which we call the decoding threshold,
\begin{defn}[Decoding Threshold ($\randdecrows$)]
\label{defn:decrows}
The decoding threshold $\randdecrows$ is the number of encoded symbols required to decode a set of $\rows$ source symbols using the rateless coding strategy.
\end{defn}
For the Robust Soliton distribution, \cite{luby2002lt} gives the following high probability bound on $M'$.
\begin{lem}[Theorems 12 and 17 in \cite{luby2002lt}]
\label{lem:LTcomp}
The original set of $\rows$ source symbols can be recovered from a set of any $M' = \rows+\mathcal{O}(\sqrt{\rows}\ln^2(\rows/\fcpdel))$ with probability at least $1-\fcpdel$.
\end{lem}
\begin{rem}
\normalfont
While $\mat$ can be encoded using any random linear code to ensure successful decoding of $\res$ from $\rows$ symbols of $\encres$ with a high probability, the key benefit of using LT codes is the low decoding complexity owing to the careful design of the Robust Soliton distribution. The complexity of LT decoding is $\mathcal{O}(\rows\ln\rows)$ while for any other random linear code it would be $\mathcal{O}(\rows^3)$ which is unacceptable for large $\rows$. (see \Cref{sec:LTProps})
\end{rem}
\subsection{Distributed Implementation}
The $\rows\times\cols$ matrix $\mat$ is encoded to generate an $\encrows\times\cols$ encoded matrix $\encmat$ where $\encrows=\strag\rows$. Each row of $\encmat$ is the sum of a random subset of rows of $\mat$ as described in \Cref{sec:LTmatvect}. The knowledge of the mapping between the rows of $\mat$ and the rows of $\encmat$ is crucial for successful decoding as illustrated in \Cref{fig:encoding_graph,fig:decoding_graph}. Hence this mapping is stored at the master. The encoding step can be treated as a pre-processing step in that it is only performed initially.

The $\strag\rows$ rows of the encoded matrix are distributed equally among the $\proc$ worker nodes as illustrated in \Cref{fig:sysmodel}. To multiply $\mat$ with a vector $\vect$, the master communicates $\vect$ to the workers. Each worker multiplies $\vect$ with each row of $\encmat$ stored in its memory and returns the product (a scalar) to the master. The master collects row-vector products of the form $\textbf{a}_{\textbf{e},j}\vect$ (elements of $\encres$) from the workers until it has enough elements to be able to recover $\res$. If a worker node completes all the $\alpha \rows/\proc$ row-vector products assigned to it before the master is able to decode $\res$, it will remain idle, while the master collects more row-vector products from other workers.

Once the master has collected a sufficient number of coded row-vector products from the workers it can recover the desired matrix vector product $\res=\mat\vect$ from the subset of the elements of $\encres = \encmat\vect$ that it has collected using the iterative peeling decoder. Once the master decodes all elements of the product vector $\res = \mat \vect$, it sends a \emph{done} signal to all workers nodes to stop their local computation.

The following modifications can make the current implementation even more efficient in real systems:
\begin{enumerate}[wide, labelwidth=!, labelindent=0pt]
    \item \textbf{Blockwise Communication: } To truly monitor the partial work done by each worker the master needs to receive each encoded row-vector product $\mathbf{a}_{\mathbf{e},j}\vect$ from the workers. However this imposes a large communication overhead which may increase latency in a slow network. To prevent this, in our distributed computing experiments, we communicate submatrix-vector products $\encmat_{i}^{j}\vect$ where $\encmat_{i}^{j}$ is the $j^{\text{th}}$ part of the encoded submatrix $\encmat_{i}$ stored at worker $i$, and each part corresponds to approximately $10\%$ of the total rows of the submatrix. Note that if $\mat$ is very large then it will not be feasible for worker $i$ to read the entire submatrix $\encmat_{i}$ from memory at once. As a result $\encmat_{i}\vect$ needs to be computed in parts for \emph{any} coding scheme.
    \item \textbf{Using Raptor Codes: } Despite their ease of implementation and fast decoding, LT codes \cite{luby2002lt} are sub-optimal in practice due to the overhead of $\randdecrows - \rows$ extra symbols required to decode the original $\rows$ source symbols. In our experiments we observe that for a matrix $\mat$ with $\rows=11760$ rows, we need to wait for $12500$ encoded row-vector products to decode $\res = \mat\vect$ with $99\%$ probability. Advanced rateless codes like Raptor Codes \cite{shokrollahi2006raptor} can decode $\rows$ source symbols from $\rows(1+\epsilon)$ symbols for any \emph{constant} $\epsilon$ even for \emph{finite values} of $\rows$. Since Raptor Codes are the rateless codes used in practical wireless standards \cite{shokrollahi2011raptor} we expect them to be used in practical implementations of our coded distributed matrix vector multiplication strategy to improve efficiency.
    \item \textbf{Using Systematic Rateless Codes: } We can entirely avoid decoding (in the absence of significant straggling) by using Systematic LT/Raptor Codes \cite{shokrollahi2006raptor} where the $\rows$ source rows $\mathbf{a}_1, \mathbf{a}_2, \dots \mathbf{a}_m$ form a subset of the encoded rows in $\encmat$. The overall scheme can be designed so that each worker first computes the row-vector products corresponding to the systematic symbols $\mathbf{a}_1, \mathbf{a}_2, \dots \mathbf{a}_m$ and then computes other encoded products (in the event of node slowdown). This would preclude the need for decoding if there is no/little straggling thereby reducing the overall latency.
\end{enumerate}

\section{Performance Analysis}
\label{sec:theoretical}
In this section we theoretically analyze the performance of LT coding and the three benchmark strategies --- ideal load balancing, $(\proc,\mdsnum)$-MDS, and $\numrep-$Replication --- in terms of latency (\Cref{defn:latency}) and computations (\Cref{defn:comp}). Our results are summarized in Table 1 and the proofs of the theoretical results are contained in \Cref{sec:Proofs_Delay}. We begin by describing our delay model.
\subsection{Delay Model}
We assume that worker $i$ requires time $\worktime_i$ to perform $\workcomp_i$ row-vector product computations where
\begin{equation}
\label{eq:delaymodel}
\worktime_i = X_i + \shifttime\workcomp_i, \quad \text{for all } i=1,\dots, \proc
\end{equation}
Thus, the delay involves the sum of two components: 1) a random variable $X_i$ that includes initial setup time at the worker before it actually begins performing the computations, and 2) a shift that is linear in the number of computations performed at the worker. This delay model is motivated by the observations of \cite{dean2013tail} where it is noted that the variability in latency arises largely from delays due to background tasks running at worker nodes and that once a request actually begins execution, the variability is considerably lower. When $X_i$ is exponentially distributed with rate $\exprate$, the time taken by worker $i$ to perform $b$ computations is distributed as
\begin{equation}
\Pr(\worktime_i\leq\smalltime) = 1 - \exp(-\exprate(\smalltime - \shifttime b)).
\end{equation}
While this follows the shifted exponential delay models used in \cite{lee2017speeding}, \cite{dutta2016short} and \cite{dutta2017coded}, the key difference is that the shift is parameterized by the number of computations at each worker. We believe this is a more realistic model as it captures the effect of increasing the amount of computations on the delay -- if a worker is assigned more computations, there is larger delay. Moreover, unlike previous works, the decay rate $\mu$ of the exponential part of the delay does not change with the number of computations performed by that worker. \Cref{fig:delay_model} illustrates our delay model.

Also, in our analysis, we use $X_{\mdsnum:\proc}$ to denote the $\mdsnum^{\text{th}}$ order statistic i.e. the $\mdsnum^{\text{th}}$ smallest of $\proc$ random variables $X_1,\ldots,X_\proc$ and we define $U_{l} = X_{l+1:p} - X_{l:p}$, $l=1,\ldots,\proc-1$ as the difference of consecutive order statistics. We also use the notation $H_{j} = \sum_{v=1}^{j} 1/v$, for the $j^{\text{th}}$ Harmonic number.
\subsection{Ideal Load Balancing Strategy}
\label{sec:idealanalysis}
\begin{figure}[t]
\centering
  \includegraphics[width=0.4\textwidth]{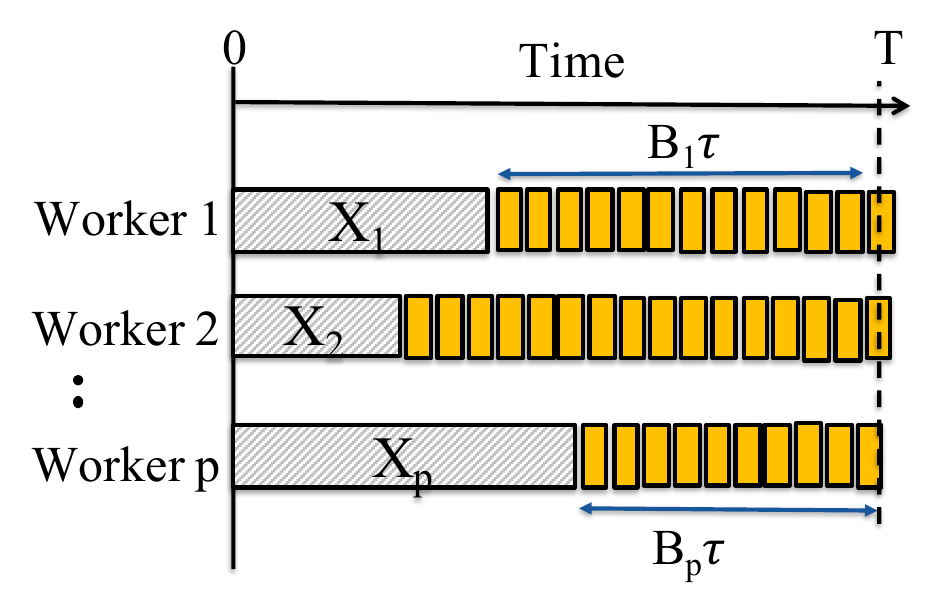}
  \caption{Worker $i$ has a random initial delay $X_i$, after which it completes row-vector product tasks (denoted by the small rectangles), taking time $\tau$ per task. The latency $\runtime$ is the time until enough tasks have been completed for the product $\res = \mat\vect$ to be recovered. \label{fig:delay_model}}
\end{figure}
\begin{table}[t]
\begin{center}
\begin{tabular}{ |c|c|c|l| } 
 \hline
 \textbf{Strategy} &  \textbf{Latency} &  \textbf{\# of Comp} &  \textbf{Complexity}  \\ 
 \hline
 Ideal & $\frac{\shifttime \rows}{\proc} + \frac{1}{\mu}$ & $\rows$ & $O(\rows)$\\ 
 \hline
LT (large $\alpha$) & $\frac{\shifttime \rows(1+\epsilon)}{\proc} + \frac{1}{\mu}$ & $\rows(1+\epsilon)$ & $O(\rows \log \rows)$\\ 
\hline
 $r$-Replication & $\frac{\shifttime \numrep\rows}{\proc} + \frac{1}{\numrep\mu}\log \frac{\proc}{r}$ & $\numrep\rows$ & $O(\rows)$ \\
 \hline
 $(p,k)$ MDS & $\frac{\shifttime \rows}{\mdsnum} + \frac{1}{\mu}\log \frac{\proc}{\proc-\mdsnum}$ & $\rows \proc/\mdsnum$ & $O(\rows \mdsnum + \mdsnum^3)$  \\ 
  \hline
\end{tabular}
\end{center}
\caption{Comparison of different strategies to multiply a $\rows \times \cols$ matrix $\mat$ with vector $\vect$ using $\proc$ worker nodes. The latency values are approximate, and number of computations values are for the case when none of the nodes slowdown.}
\label{tbl:comp}
\vspace{-1.5em}
\end{table} 
Recall that in the ideal load balancing strategy, we have a central queue at the master and tasks being allocated to a worker as soon as it becomes idle (either immediately after the initial delay or after it completes the current task) as illustrated in \Cref{fig:delay_model}. Thus it computes exactly $C = \rows$ row-vector products in total when an $\rows \times \cols$ matrix is multiplied with $\cols \times 1$ vector and performs zero redundant computations. \Cref{thm:idealoptimal} below proves the optimality of ideal load balancing in terms of latency and \Cref{lem:ideallatency} and \Cref{coro:ideallatency} give bounds on the expected latency.
\begin{thm}[Optimality of Ideal Load Balancing]
\label{thm:idealoptimal}
For any distributed matrix-vector multiplication scheme, for the delay model of \Cref{eq:delaymodel}, the latency $\runtime$ is no less than the latency of ideal load balancing, denoted by $\runtime_{\text{ideal}}$. In other words, for any scheme,
\begin{align}
\runtime \geq \runtime_{\text{ideal}}.
\end{align}
\end{thm}
\begin{lem}[Latency of Ideal Load Balancing]
\label{lem:ideallatency}
The latency for the ideal load balancing strategy with $\proc$ workers has the following upper and lower bounds.
\begin{align}
 \frac{\shifttime \rows}{\proc} + X_{1:\proc} \leq \mathbb{E}[\runtime_{\text{ideal}}] &\leq \frac{\shifttime\rows}{\proc}+ \frac{1}{\proc}\sum_{i=1}^{\proc} X_i + \shifttime. \label{eqn:ideal_bnds}
\end{align}
\end{lem}
\begin{coro}
\label{coro:ideallatency}
The expected latency for the ideal strategy with $X_i \sim \exp(\exprate)$ for all workers $i=1,\ldots,\proc$, has the following upper and lower bounds.
\begin{align}
\frac{\shifttime\rows}{\proc} + \frac{1}{\proc\exprate} \leq \mathbb{E}[\runtime_{\text{ideal}}] &\leq \frac{\shifttime\rows}{\proc}+ \frac{1}{\exprate} + \shifttime. \label{eqn:expideal_bnds}
\end{align}
\end{coro}
Note that the ideal load balancing scheme is not exactly realizable in practice. Approaches like work stealing \cite{dinan2007dynamic,harlap2016addressing} can potentially approximate this strategy by physically moving tasks from busy workers to idle workers. However implementing such approaches may not be feasible in all settings, for e.g.\ when the communication latency between workers is too large, or the data is restricted to lie on a particular worker due to privacy concerns. In this work we aim to show that it is possible to \emph{algorithmically} achieve near-ideal latency performance for matrix-vector multiplication by using the rateless coded computing strategy described in \Cref{sec:rateless} which does not require physically moving data between workers. 
\subsection{Rateless Coded Strategy}
\label{sec:ltanalysis}
We make the following assumption for analyzing the latency of the proposed rateless coded strategy.
\begin{assum}
The decoding threshold $\randdecrows$ (\Cref{defn:decrows}) of the LT coded strategy satisfies $\randdecrows \simeq \rows$. 
\end{assum}
We believe the above assumption is reasonable because the problem of distributed matrix vector multiplication arises only when $\rows$ (the number of rows of $\mat$) is large and the high probability bound of \Cref{lem:LTcomp} can be used to show that $\mathbb{E}[M'] = \rows(1+\epsilon)$, where $\epsilon \rightarrow 0$ as $\rows \rightarrow \infty$. Note that this assumption is only to facilitate a better \emph{theoretical} comparison between the LT coded and ideal strategies. In our experiments in \Cref{sec:experiments} we choose a value of $\randdecrows$ according to \Cref{lem:LTcomp} that is slightly larger than $\rows$ and ensures that the original matrix-vector product $\res = \mat\vect$ can be recovered with high $(>99\%)$ probability.
\begin{rem}
\normalfont
\label{rem:ltlatency}
The Rateless coded computing strategy described in \Cref{sec:rateless} is identical to the ideal load balancing strategy described above for large values of $\rows$ and infinite redundancy i.e. $\strag = \encrows/\rows \rightarrow \infty$. This is because both the rateless coded strategy and ideal load balancing strategies are based on collecting a pre-determined number of computations across all workers by greedily picking the next available task at each worker.
\end{rem}
However in practice, we cannot set $\strag = \encrows/\rows \rightarrow \infty$ owing to limitations in computation power and memory at workers. Instead the amount of redundancy in the LT coded strategy is fixed initially by choosing the number of encoded rows $\encrows = \strag\rows/\proc$ for some $\strag >1$. Computations are divided equally among the $\proc$ workers and thus each worker can perform a maximum of $\strag\rows/\proc$ computations. 
\begin{thm}[Rateless v/s Ideal]
\label{thm:probltlatency}
The latency of the proposed rateless coded strategy, $\runtime_{\text{LT}}$ decreases on increasing $\strag$ and approaches the latency of the ideal strategy $\runtime_{\text{ideal}}$. This is quantified by the following probabilistic upper bound:
\begin{align}
\Pr(\runtime_{\text{LT}} > \runtime_{\text{ideal}}) &\leq \sum_{j=2}^{\proc}\Pr\left(\sum_{l=1}^{j-1}U_{l} \geq \frac{\shifttime\rows(\strag - 1)}{\proc - 1}\right), \label{eq:probltupper}
\end{align}
where $U_{l} = X_{l+1:p} - X_{l:p}$.
\end{thm}
\begin{rem}
\normalfont
The effect of straggling is captured through the term $U_{l} = X_{l+1:p} - X_{l:p}$ in the above expression. High straggling, implies a high variability in the initial worker delays $X_i$ due to which $U_{l}$ is large and the probability of $\runtime_{\text{LT}} > \runtime_{\text{ideal}}$ is also higher. 
\end{rem}
Thus as $\strag$ (and consequently $\encrows$) increases, $\runtime_{\text{LT}}$ is equal to $\runtime_{\text{ideal}}$ with a high probability. A cleaner result is obtained for the case when $X_i \sim \exp(\exprate)$, as given below.
\begin{coro}
\label{coro:probltlatency}
If $X_i \sim \exp(\exprate)$ for all workers $i=1,\ldots,\proc$ then
\begin{align}
\Pr(\runtime_{\text{LT}} > \runtime_{\text{ideal}}) &\leq \proc\exp \left(-\frac{\exprate\shifttime\rows(\strag-1)}{\proc^2}\right).
\end{align}
\end{coro}
We also derive an upper bound on the difference between the expected latencies of the rateless and the ideal strategies. We only state the result for $X_i \sim \exp(\exprate)$ over here and defer the (more complicated) general result and its proof to \Cref{sec:Proofs_Delay}.
\begin{thm}[Latency of the Rateless Coded strategy]
\label{thm:ltlatency}
If $X_i \sim \exp(\exprate)$ for all workers $i=1,\ldots,\proc$ then
\begin{align}
&\mathbb{E}[\runtime_{\text{LT}}] - \mathbb{E}[\runtime_{\text{ideal}}] \leq \left(\shifttime\strag\rows\proc^2 + \frac{\proc^2}{\exprate} + \shifttime\proc \right)\exp \left(-\frac{\exprate\shifttime\rows(\strag-1)}{\proc^2} \right).
\end{align}
\end{thm}
The second term decays exponentially and dominates the first term, which is a polynomial. The rate of decay increases as the amount of redundancy $\alpha$ increases. In other words, $\mathbb{E}[\runtime_{\text{LT}}]$ approaches $\mathbb{E}[\runtime_{\text{ideal}}]$ exponentially fast as redundancy is increased.
\begin{rem}
\normalfont
Another important advantage of the rateless coded strategy is that the number of computations performed by the workers, $\numcomp_{\text{LT}}$, is always equal to $\randdecrows$ (the decoding threshold, defined in \Cref{defn:decrows}). and does not increase on increasing  redundancy (increasing $\strag$) unlike for the MDS and Replication strategies. Moreover 
since $\mathbb{E}[\randdecrows] = \rows(1+\epsilon)$ and $\epsilon\rightarrow 0$ as $\rows\rightarrow\infty$, $\mathbb{E}[\numcomp_{\text{LT}]}$ asymptotically approaches the minimum number of computations ($\rows$) required to recover a $\rows-$dimensional matrix-vector product. 
\end{rem}
In the following subsections (and in \Cref{sec:additional_theory}) we show that the latency of the MDS and Replication strategies is much larger than that of ideal load balancing and does not converge to $\runtime_{\text{ideal}}$ on increasing redundancy. We also show that the number of computations performed by both replication and MDS coding in computing $\res = \mat\vect$ is much larger than $\rows$.
\subsection{MDS Coded Strategy}
\label{sec:mdslatency}
Recall that for the $(\proc,\mdsnum)$ MDS coded strategy, the encoded submatrices $\encmat_1$,$\ldots$,$\encmat_\proc$ are generated by applying a $(\proc,\mdsnum)$ MDS Code to submatrices $\mat_1,\ldots,\mat_\mdsnum$. The master then waits for the fastest $\mdsnum$ workers to complete all the tasks assigned to them. 
\begin{lem}[Latency of the MDS Coded Strategy]
\label{thm:mdslatency}
The latency of the $(\proc,\mdsnum)$ MDS-coded strategy, $\runtime_{\text{MDS}}$, is given by
\begin{align}
\runtime_{\text{MDS}} &= X_{\mdsnum:\proc} + \shifttime\frac{\rows}{\mdsnum}.
\end{align}
\end{lem}
\begin{coro}
\label{coro:mdslatency}
The expected latency of the $(\proc,\mdsnum)$ MDS-coded strategies with $X_i \sim \exp(\exprate)$ for all workers $i=1,\ldots,\proc$ is
\begin{align}
\mathbb{E}[\runtime_{\text{MDS}}] &= \frac{\shifttime \rows}{\mdsnum} + \frac{1}{\mu}(H_\proc - H_{\proc-\mdsnum}) \simeq \frac{\shifttime \rows}{\mdsnum} + \frac{1}{\mu} \log\frac{\proc}{\proc - \mdsnum}. \label{eqn:mds_latency}
\end{align}
\end{coro}
Observe that in \eqref{eqn:mds_latency} above, adding redundancy (reducing $\mdsnum$) leads to an increase in the first term (more computation at each node) and decrease in the second term (less delay due to stragglers). Thus, straggler mitigation comes at the cost of additional computation at the workers which might even lead to an increase in latency. This is in contrast to \Cref{thm:probltlatency,thm:ltlatency} which indicates that the expected latency of the rateless coded strategy always decreases on adding redundancy (increasing $\strag$). Moreover, the presence of the log-factor in the second term causes $\runtime_{\text{MDS}}$ to always be larger than $\runtime_{\text{ideal}}$ since there is no log-factor in the term containing $1/\exprate$ in the upper bound on $\runtime_{\text{ideal}}$ (\Cref{lem:ideallatency}).

We now analyze the number of computations performed by the $(\proc,\mdsnum)$ MDS coding. The following result shows that with a high probability, the number of computations performed by the MDS Coded strategy is very close to the worst-case number of computations ($\rows\proc/\mdsnum$) i.e.\ when all the workers perform all the tasks assigned to them in time $\runtime_{\text{MDS}}$.
\begin{lem}[Tail of Computations for MDS Coding]
\label{thm:mdscomptail}
The tail of the number of computations of the MDS coded strategy, $\numcomp_{\text{MDS}}$, with $\proc$ workers and the delay model of \Cref{eq:delaymodel} is bounded as
\begin{align}
\Pr(\numcomp_{\text{MDS}}\geq\frac{\rows\proc}{\mdsnum}-\numcomp_0) \geq 1-\Pr\left( \sum_{l=\mdsnum}^{\proc-1}U_{l} \geq\frac{\shifttime C_0}{\proc-\mdsnum}-\shifttime \right).
\end{align}
When $X_i \sim \exp(\exprate)\  \forall i=1,\ldots,\proc$ this reduces to
\begin{align}
\Pr(\numcomp_{\text{MDS}}\geq\frac{\rows\proc}{\mdsnum}-\numcomp_0) \geq 1-\exp \left(-\exprate \left(\frac{\shifttime\numcomp_0}{(\proc-\mdsnum)^{2}}-\frac{\shifttime}{\proc-\mdsnum}\right)\right).
\end{align}
\end{lem}
Even for a small value of $\numcomp_0$ in the above expression, $\Pr(\numcomp_{\text{MDS}}\geq\frac{\rows\proc}{\mdsnum}-\numcomp_0)$ can be very large. Thus the overhead $\numcomp_{\text{MDS}}-\rows$ is quite large (we only need $\rows$ computations in the uncoded case to reconstruct the $\rows$-dimensional matrix-vector product). 
\subsection{Replication Strategy}
\label{sec:replatency}
\begin{figure*}[t]
  \centering
   \subfloat[Latency Tail]{\includegraphics[width=0.33\linewidth]{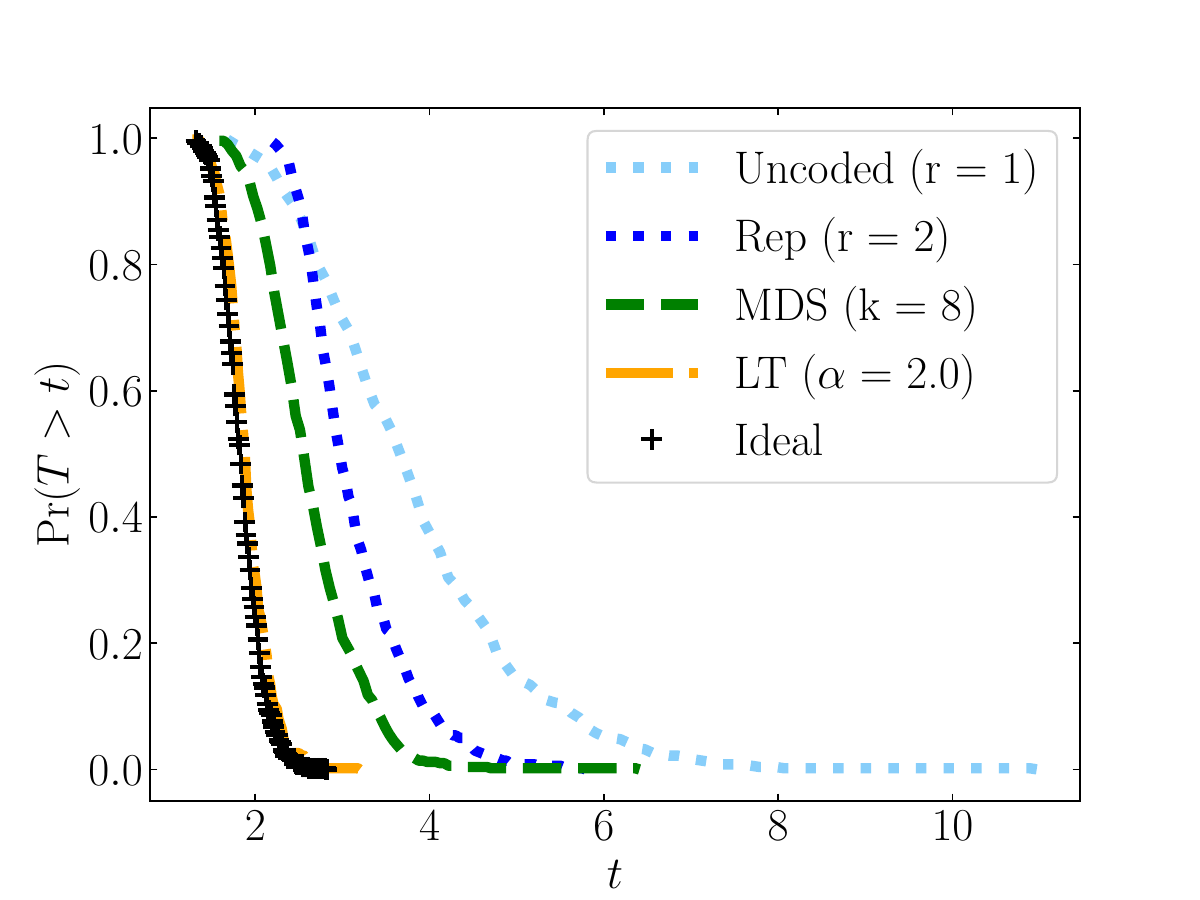}\label{fig:latsim}}
  \subfloat[Computation Tail]{\includegraphics[width=0.33\linewidth]{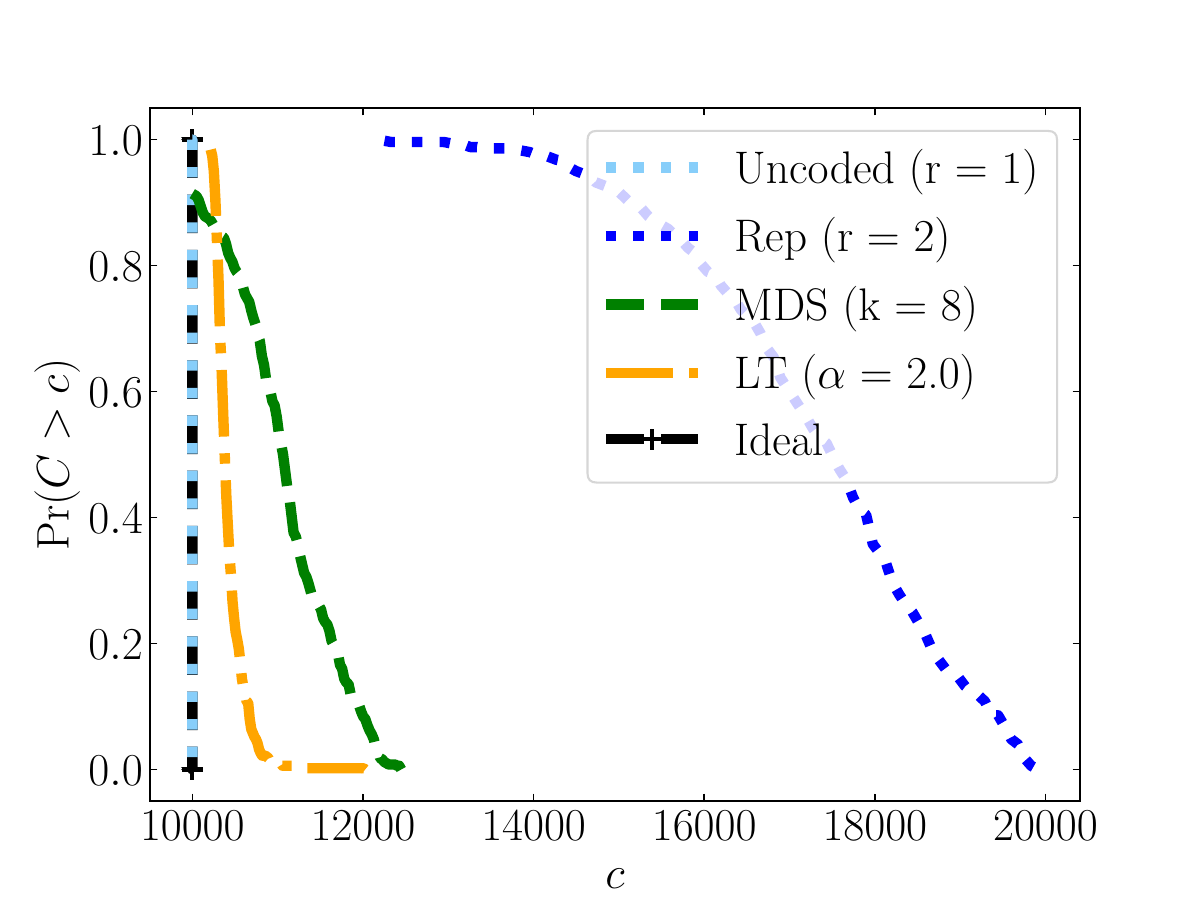}\label{fig:compsim}}
   \subfloat[Mean Response Time]{\includegraphics[width=0.33\linewidth]{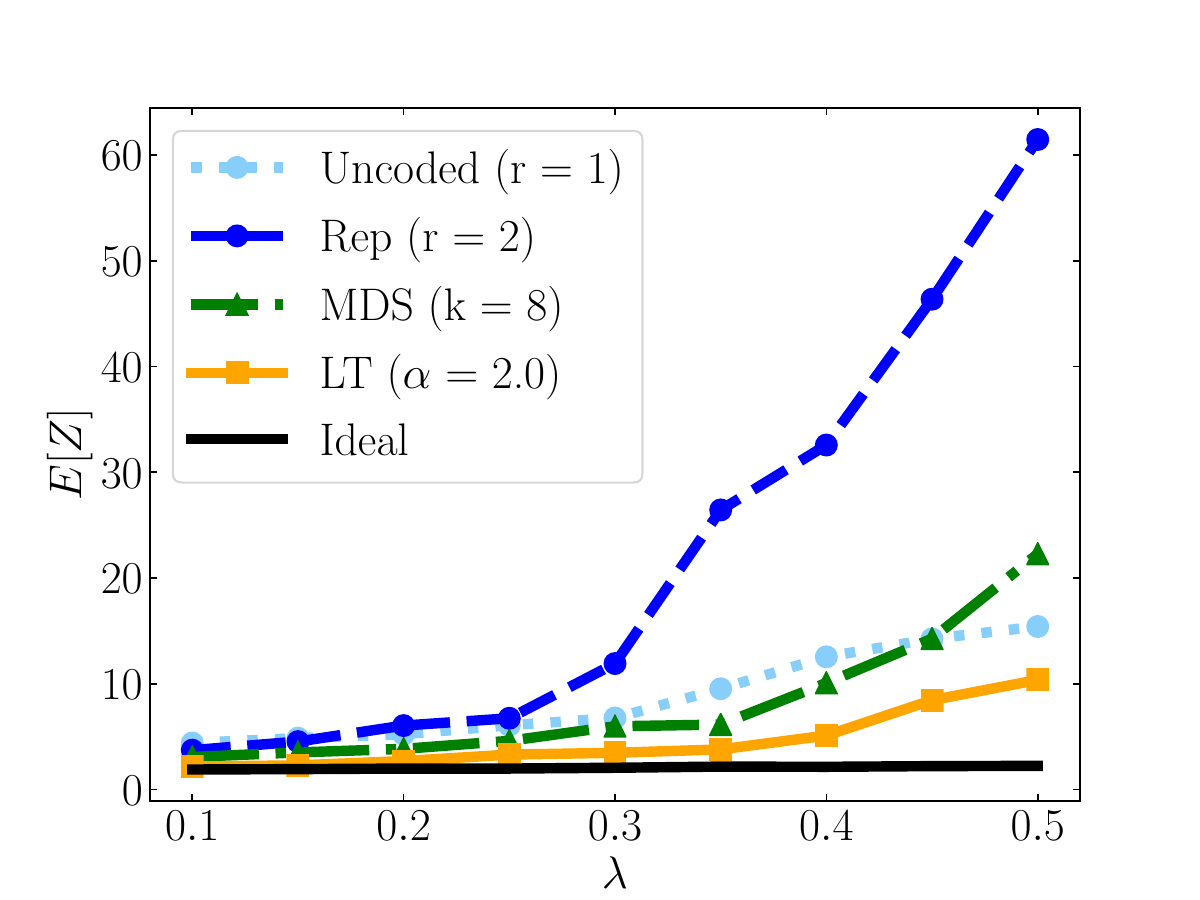}\label{fig:queueing_sim}}
  \caption{The tail probability of the latency is the highest for the replication schemes. MDS codes perform better in terms of latency but they perform a large number of redundant computations. The latency tail of LT codes is the minimum among all the schemes. Moreover the LT coded schemes performs significantly fewer redundant computations than MDS Codes or replication. When there are multiple jobs in the queue, the mean response time is least for the LT Coded setting under all values of arrival rate $\lambda$. All simulations are performed for a distributed matrix-vector multiplication task with $\rows = 10000$ matrix rows, $\proc = 10$ worker nodes, and delay model parameters $\exprate = 1.0, \shifttime = 0.001$.}
\label{fig:simulations}
\end{figure*}

The $r-Replication$ strategy involves replicating each of the $\proc/\numrep$ submatrices $\mat_1,\ldots,\mat_{\proc/\numrep}$ at $\numrep$ distinct workers and selecting the result of the fastest worker for each submatrix. 
\begin{lem}[Latency of the Replication Strategy]
\label{thm:replatency}
The latency of the $\numrep-$Replication strategy, $\runtime_{\text{rep}}$ is given by
\begin{align}
\runtime_{\text{rep}} &= \max_{1\leq i\leq \proc/\numrep}\min_{1\leq j\leq \numrep}X_{(i-1)\numrep+j} + \frac{\shifttime\rows\numrep}{\proc}.
\end{align}
\end{lem}
\begin{coro}
\label{coro:replatency}
The expected latency of the $\numrep-$Replication with $X_i \sim \exp(\exprate)$ for all workers $i=1,\ldots,\proc$ is
\begin{align}
\mathbb{E}[\runtime_{\text{rep}}] &= \frac{\shifttime\rows\numrep}{\proc} + \frac{1}{\mu}H_{\proc/\numrep} \simeq \frac{\shifttime\rows\numrep}{\proc} + \frac{1}{\mu}\log\frac{\proc}{\numrep}. \label{eqn:rep_latency}
\end{align}
\end{coro}
Once again we see that adding redundancy (increasing $r$) leads to an increase in the first term (more computation at each node) and decrease in the second term (less delay due to stragglers). Thus the extra computation at the workers may lead to an increase in latency even in this case. Moreover the log-factor in the second term causes $\runtime_{\text{Rep}}$ to always be larger than $\runtime_{\text{ideal}}$, just like $\runtime_{\text{MDS}}$.
\begin{lem}[Tail of Computations for Replication]
\label{thm:repcomptail}
The tail of the number of computations of the replication strategy, $\numcomp_{\text{rep}}$, with $\proc$ workers and the delay model of \Cref{eq:delaymodel} is bounded as
\begin{align}
\Pr(\numcomp_{\text{rep}}\geq\rows\numrep-\numcomp_0) \geq 1-\Pr\left(\sum_{i=1}^{\proc/\numrep}\sum_{j=1}^{r-1}(V_{j+1:\numrep}^{i}-V_{j:\numrep}^{i})\geq\frac{\shifttime C_0}{\numrep-1}-\frac{\shifttime\proc}{\numrep}\right),
\end{align}
where $V_{j}^{i} = X_{(i-1)\numrep+j}$ and $V_{j:\numrep}^{i}$ are the corresponding order statistics.
When $X_i \sim \exp(\exprate)\  \forall i=1,\ldots,\proc$ this reduces to
\begin{align}
\Pr(\numcomp_{\text{rep}}\geq\rows\numrep-\numcomp_0) \geq 1-\sum _{{i=0}}^{{\proc/\numrep-1}}{\frac  {1}{i!}}\exp({-\exprate\theta})(\exprate\theta)^{i},\\
\text{where }
\theta = \frac{\shifttime C_0}{(\numrep-1)^{2}}-\frac{\shifttime\proc}{\numrep(\numrep-1)}.
\end{align}.
\end{lem}
Thus with a high probability, the number of computations performed is very close to the worst-case number ($\rows\numrep$) i.e. when all the workers perform all the tasks assigned to them in time $\runtime_{\text{rep}}$.
\begin{rem}
\normalfont
While the benefits of using partial work from all workers can be obtained by using any random linear code on the rows of $\mat$, the key strength of LT codes is their low $\mathcal{O}(\rows\ln\rows)$ decoding complexity. Using an $(\encrows,\rows)$ MDS code on the rows of $\mat$ has $\mathcal{O}(\rows^3)$ decoding complexity which is unacceptable for large $\rows$. 
\end{rem}
We simulate the MDS, replication, and LT-coded schemes under our delay model \Cref{eq:delaymodel} for distributed matrix-vector multiplication with $\rows = 10000$ matrix rows, $\proc = 10$ workers and delay model parameters $\exprate = 1.0, \shifttime = 0.001$ (\Cref{fig:simulations}). We limit the amount of redundancy to $\strag = \encrows/\rows \leq 2.0$ since this is the amount of redundancy in the basic $2$-replication scheme. Observe that the LT coded strategy ($\strag = 2.0$) clearly outperforms MDS coding (with $\mdsnum = 8$) in that it not only exhibits near-ideal latency (\Cref{fig:latsim}) but also performs fewer total computations (\Cref{fig:compsim}) than MDS coding. Changing $\mdsnum$ does not improve the performance of MDS coding much. Specifically, increasing redundancy (reducing $\mdsnum$) in MDS coding leads to higher latency after a point, as illustrated in \Cref{fig:latcompsim} (and as expected from \Cref{thm:mdslatency}). On the other hand, the latency of LT coding converges to that of the Ideal scheme on increasing $\strag$, without any increase in computations. Additional simulations for $X_i \sim \text{Pareto}(1,3)$ given in \Cref{fig:simulations_par} in \Cref{sec:additional_simexp} also show similar improvements with LT coding.
\section{Queueing Analysis}
\label{sec:queueing}
In most real applications of matrix-vector multiplication in machine learning and data analytics, the matrix representing the model is fixed, while vectors representing the data that need to be multiplied with this matrix arrive as a real-time stream. Prior works on coded computing like \cite{lee2017speeding,dutta2016short} do not consider the effect of multiple incoming jobs which could lead to queueing delays at the master, in addition to straggling at workers. We analyze the latency with queueing for the proposed LT coded strategy as well as the MDS coded and Replication strategies. The LT coding results are presented below, while MDS and replication results are given in \Cref{sec:Proofs_Queueing}. 

Suppose that vectors $\vect_1,\vect_2,\ldots$ arrive according to a Poisson process with rate $\lambda$ and are broadcast by the master to the $\proc$ workers. Worker $i$ multiplies each vector it receives with the sub-matrix $\encmat_i$ stored in its memory (where $\encmat$ is generated according to the corresponding encoding strategy) and communicates the corresponding elements of $\encres=\encmat\vect$ to the master. Once the master has enough elements of $\encres$ to successfully decode $\res$, the remaining tasks at all the workers are cancelled. Then the mean response time $\E{Z}$ (waiting time in queue plus service time) of a matrix-vector multiplication job is as follows.
\begin{thm}[Latency of LT Coding with Queueing of Jobs]
\label{thm:ltmultiple}
For large $\encrows$ i.e. $\strag = \encrows/\rows \rightarrow \infty$, the mean response time of the LT coded scheme $\queuetime_{\text{LT}}$ when vectors $\vect$ are arriving at rate $\lambda$ according to a Poisson process is
\begin{align}
\mathbb{E}[\queuetime_{\text{LT}}] &= \mathbb{E}[\runtime_{\text{LT}}] + \frac{\lambda\mathbb{E}[(\runtime_{\text{LT}})^2]}{2(1-\lambda\mathbb{E}[\runtime_{\text{LT}}])},
\end{align}
where $\mathbb{E}[\runtime_{\text{LT}}]$ is bounded as described in \Cref{lem:ideallatency} and bounds on $\mathbb{E}[(\runtime_{\text{LT}})^2]$ are derived in \Cref{sec:Proofs_Queueing}.
\end{thm}
The proof is given in \Cref{sec:Proofs_Queueing}. The key idea used in the proof is that for large $\alpha$ this system becomes equivalent to an M/G/1 queue with service time $T_{\text{LT}}$. Then we simply apply the Pollaczek-Khinchine formula \cite{harchol2013_performance} for the mean response time of M/G/1 queues. When $\alpha$ is small the analysis becomes very difficult -- it is a generalization of the fork-join queueing system, whose response time is notoriously hard to analyze \cite{nelson_tantawi, kim_agarwal, varki_merc_chen}. This is an open question for future research.

For the MDS and replication strategies, we reduce the queueing system to a fork-join queueing system with redundancy, and then use previous results \cite{joshi2014delay, joshi2017efficient} to obtain bounds on the mean response time. The results are presented in \Cref{sec:Proofs_Queueing}.
\begin{rem}[Insights from MDS and Replication Queueing Analyses]
\normalfont
In the MDS and replication strategies, increasing redundancy (lower $\mdsnum$ and higher $\numrep$) reduces the number of workers that need to complete their tasks. However, the added redundancy increases the number of computations that each worker needs to perform due to which the waiting time for incoming jobs at the master increases, thus increasing the overall mean response time ($\queuetime_{\text{MDS}}$ and $\queuetime_{\text{rep}}$ respectively). On the other hand for the rateless coded (and ideal) strategies we just need to wait for $\randdecrows$ (or $\rows$) computations across all workers for each job. Moreover, for LT coding, adding redundancy (increasing $\encrows$) always reduces the service time for each job (\Cref{thm:probltlatency}) and thus the overall queueing delay $\queuetime_{\text{LT}}$ always decreases on adding redundancy.
\end{rem}
\Cref{fig:queueing_sim} shows simulation results of mean response time $\queuetime$ under our delay model with $X \sim \exp(1)$ and $\shifttime = 0.001$ for a distributed matrix-vector multiplication task with $\rows = 10000$ matrix rows using $\proc = 10$ worker nodes. The mean response time is averaged over $10$ trials with $100$ jobs in each trial. Jobs arrive according to a Poisson process with rate $\lambda \in (0.1, 0.6)$. The results illustrate that the benefits of our LT coded strategy over previous approaches are further enhanced when there is queueing of jobs.
\section{Experimental Results}
\label{sec:experiments}
\begin{figure*}[htb]
\centering
\subfloat[Average Latency (parallel)]{\includegraphics[width=0.33\linewidth]{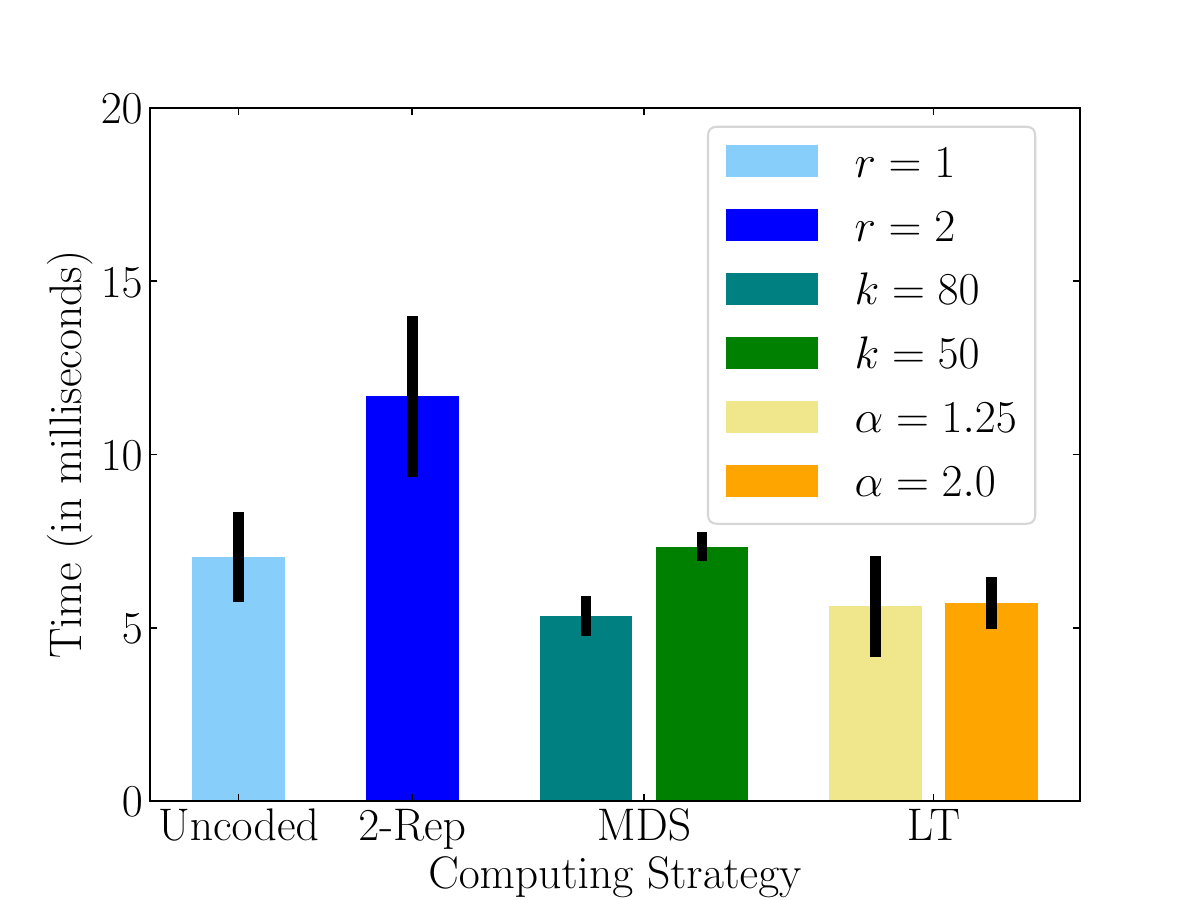}\label{fig:avg_parlat}}
\subfloat[Average Latency (distributed)]{\includegraphics[width=0.33\linewidth]{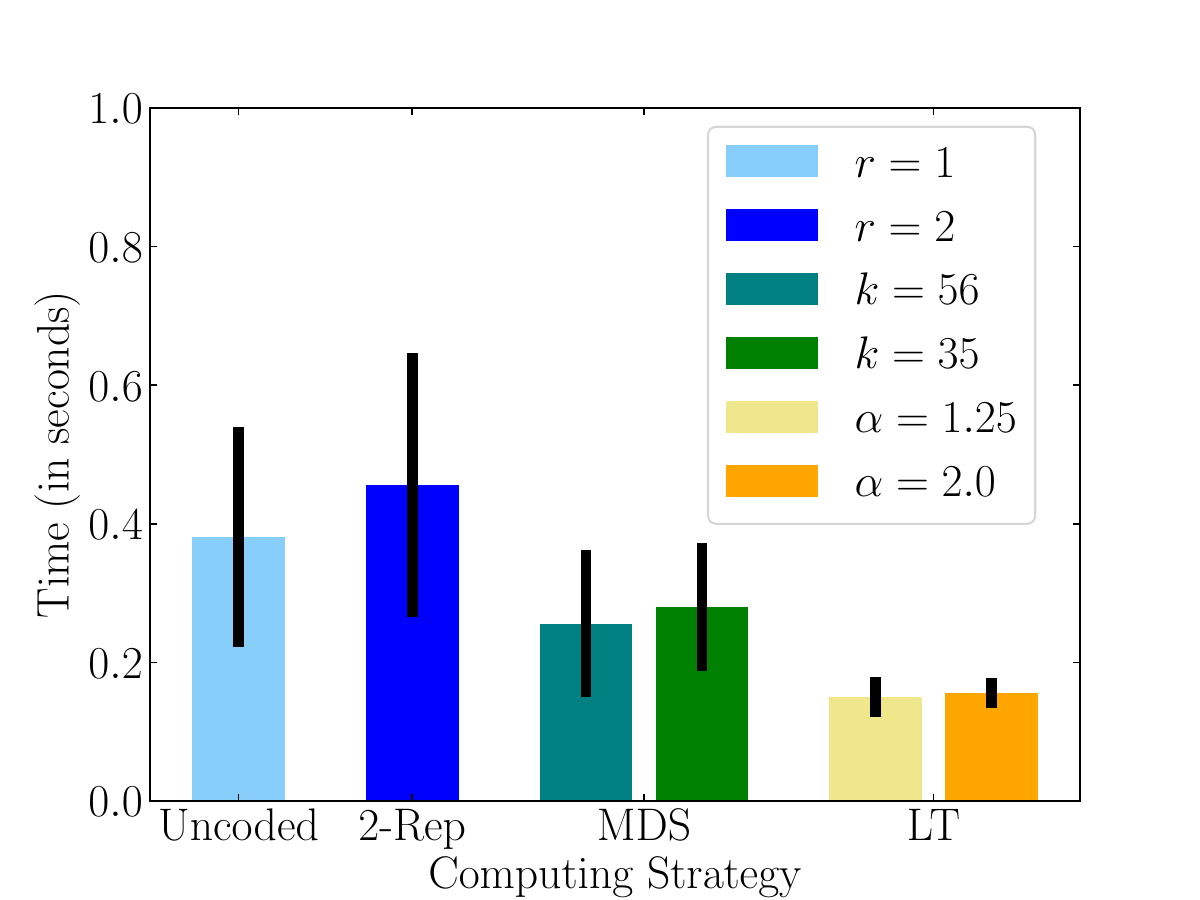}\label{fig:avg_distlat}}
\subfloat[Average Latency (serverless)]{\includegraphics[width=0.33\linewidth]{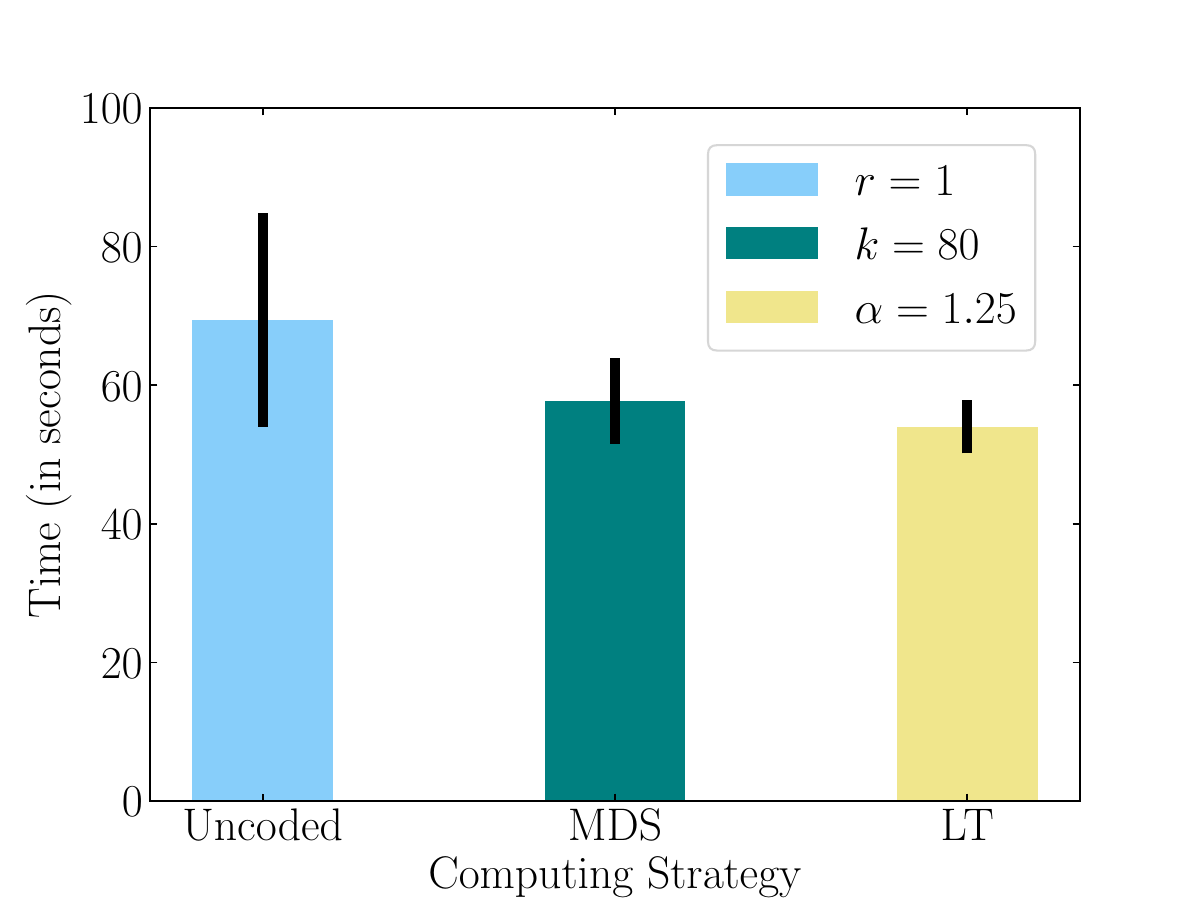}\label{fig:avg_lamlat}}\\
\subfloat[Average Comp. (parallel)]{\includegraphics[width=0.33\linewidth]{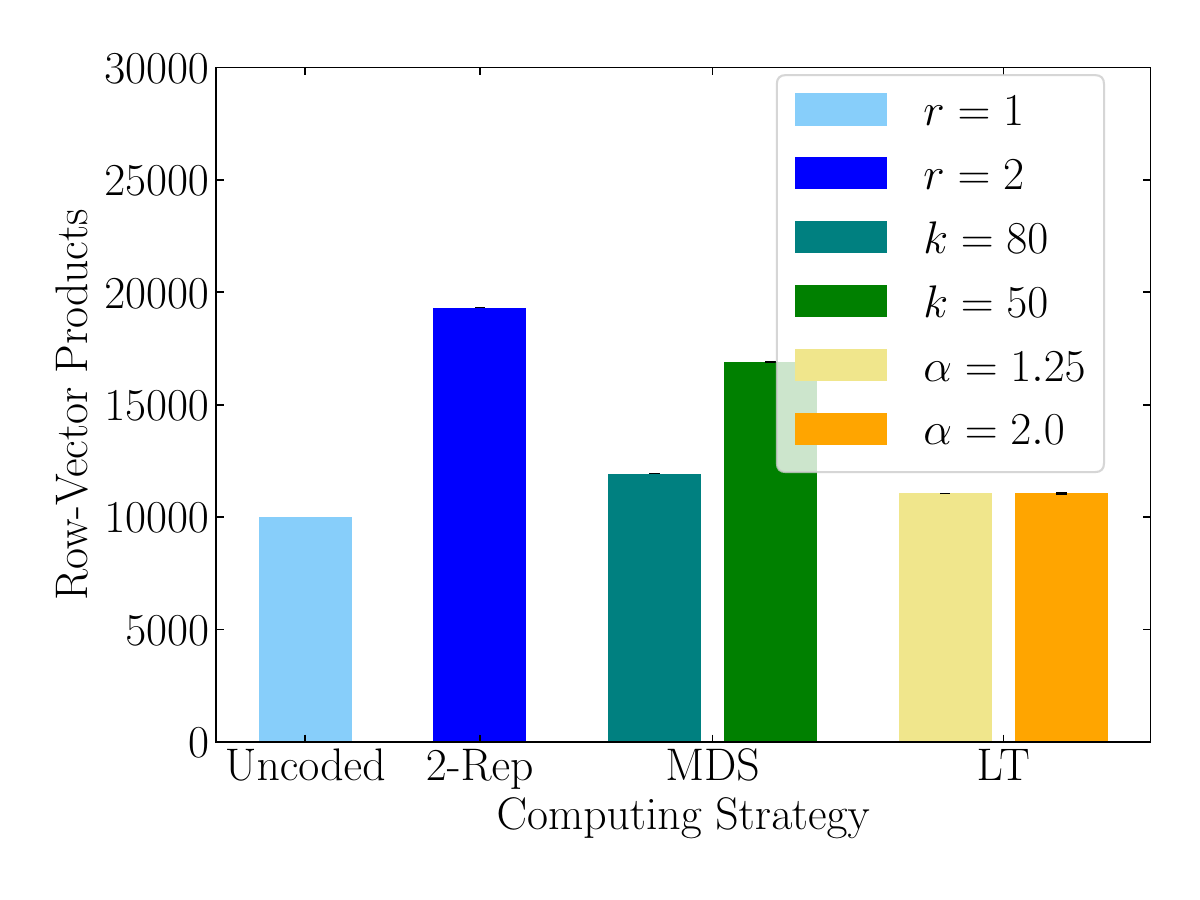}\label{fig:avg_parcomp}}
\subfloat[Average Comp. (distributed)]{\includegraphics[width=0.33\linewidth]{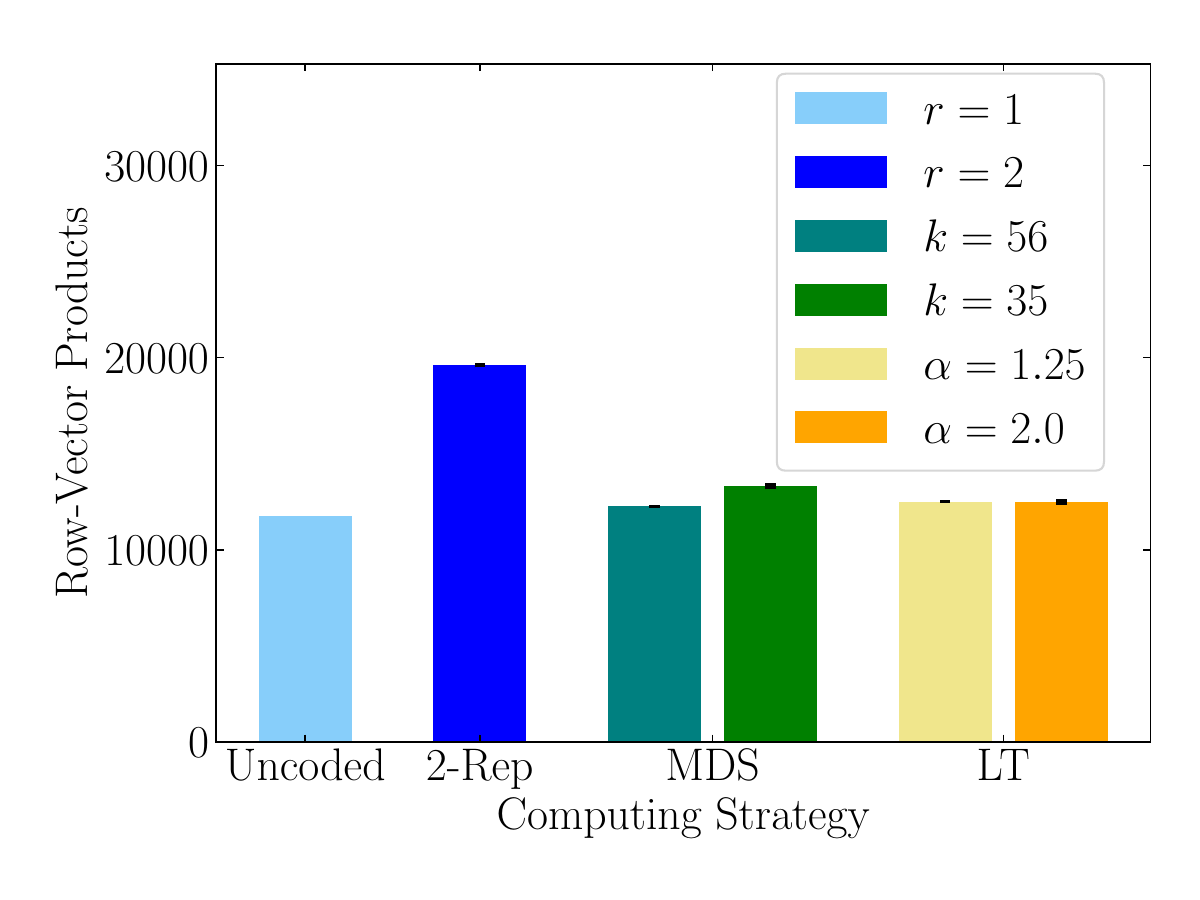}\label{fig:avg_distcomp}}
\subfloat[Average Comp. (serverless)]{\includegraphics[width=0.33\linewidth]{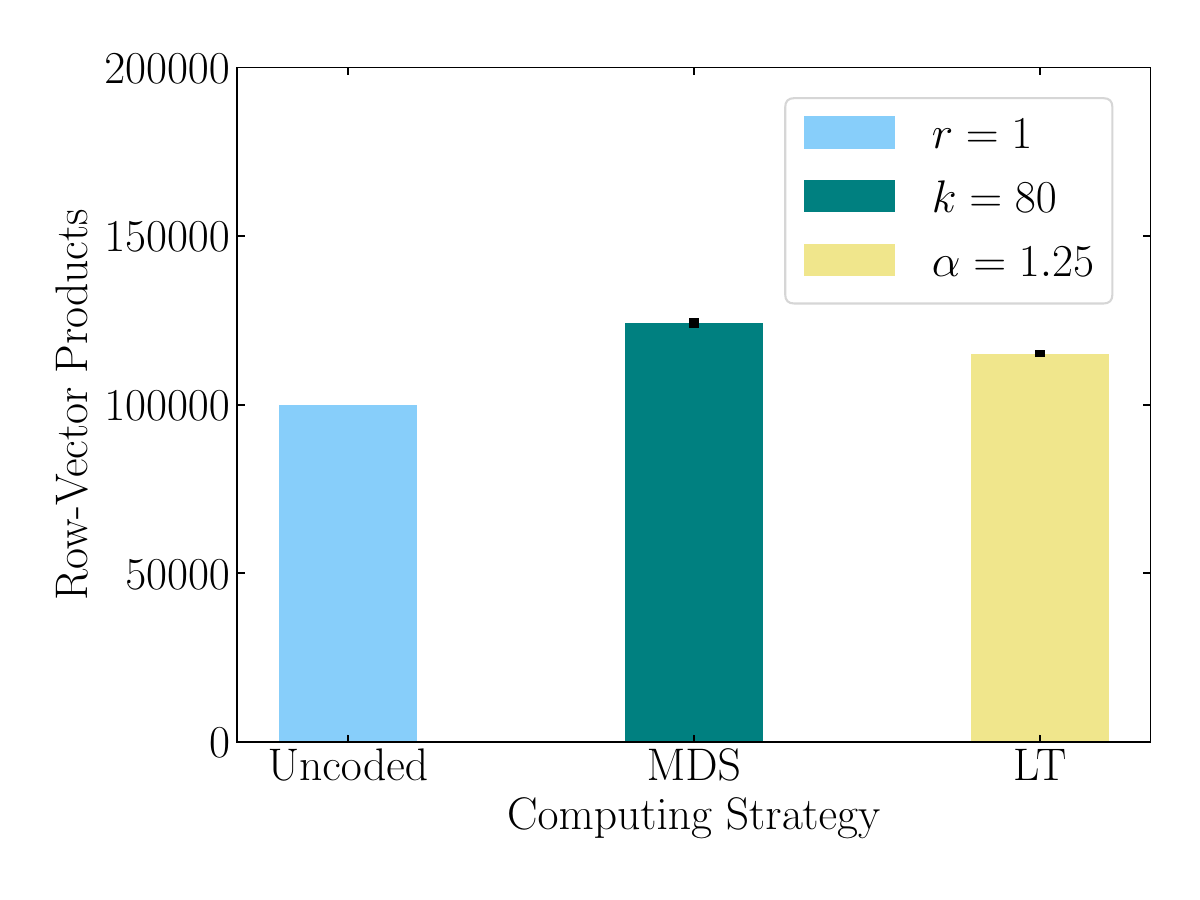}\label{fig:avg_lamcomp}}
\caption{Experiments on coded distributed matrix vector multiplication in parallel (Python Multiprocessing \cite{Multiprocessing}), distributed (AWS EC2 \cite{amazon_ec2}) and serverless (AWS Lambda \cite{amazon_lambda}) settings show that the LT Coded strategy has lower average latency than all other approaches ($1.2\times-$ to $3\times-$ improvement across scenarios) and performs fewer total computations than Replication or MDS Coding. Each error bar corresponds to 1 standard deviation.}
\label{fig:experiments}
\end{figure*}
We demonstrate the effectiveness of rateless codes in speeding up distributed matrix-vector multiplication in parallel, distributed and serverless enviornments. No artificial delays/background tasks were added to induce straggling in any experiments.
\subsection{Parallel Computing Experiments}
We consider multiplication of a $10000 \times 10000$ matrix $\mat$ of random integers with a $10000 \times 1$ vector $\vect$ of random integers on an iMac Desktop with 8 GB of RAM and a 3.6 GHz Intel i7 Processor. This computation is parallelized over 100 processes using Python's Multiprocessing Library \cite{Multiprocessing}. We compare the uncoded, 2-replication, MDS coding ($\mdsnum = 80,50$) and LT coding ($\strag=1.25,2.0$) approaches. For fair comparison, we consider instances of MDS and LT codes that have the same number of encoded rows $\encrows$. The encoded matrix $\encmat$ was divided equally among the $\proc=100$ processes and the experiment was repeated $10$ times with a different random $\vect$ each time. The processes multiply the rows of $\encmat$ with $\vect$ in parallel and we record the average latency (time required to collect enough row-vector products for successful decoding) and total computations. Results of average latency (\Cref{fig:avg_parlat}) show that LT coded and MDS coded approaches are clearly faster (about at least $1.2\times-$) than the Uncoded and 2-Replication approaches while \Cref{fig:avg_parcomp} shows that the LT coded approaches also perform fewer total computations than the MDS or 2-replication strategies thus leading to more efficient resource utilization. Note that while MDS coding with $\mdsnum=80$ has latency comparable to that of LT coding (both for $\strag=1.25$ and $\strag =2.0$), both latency and total computations with MDS coding increase on increasing $\mdsnum$ to $50$ due to the higher computational load at each node (as discussed in \Cref{sec:theoretical}). Recall that $\mdsnum$ corresponds to the number of "fast" workers in the system. In most real systems the number of "fast" workers is transient and thus, unpredictable. Our experiments show that MDS coding is highly sensitive to the choice of $\mdsnum$ with incorrect choices leading to \emph{higher} latency. LT Coding on the other hand is not only fast, but is also insensitive to the amount of redundancy ($\strag$) in that, the system designer can choose $\strag$ to be as large as permitted by memory constraints without a risk of loss in performance (unlike MDS).
\subsection{Distributed Computing Experiments}
We created a cluster of 70 t2.small workers on AWS EC2 \cite{amazon_ec2} using Kubernetes \cite{Kubernetes}. Each worker was allocated 1 GB of memory. Computations were performed using Dask \cite{Dask}, a popular framwork for parallel computing in Python. A $11760 \times 9216$ matrix $\mat$ was extracted from the STL-10 \cite{coates2011analysis} dataset. Once again we compared the uncoded, 2-replication, MDS coding ($\mdsnum = 56,35$) and LT coding ($\strag=1.25,2.0$) approaches. The encoded matrix $\encmat$ is divided equally among the $\proc=70$ workers and multiplied with 5 different vectors, each of length $9216$, also extracted from the STL-10 \cite{coates2011analysis} dataset. \Cref{fig:avg_distlat} shows the average latency of the different approaches. Both LT coded approaches are almost $2\times-$ faster than the MDS coded appraoches in this setting and almost $3\times-$ faster than the uncoded approaches. Each worker computes approximately $14$ row-vector products at a time before communicating the results to the master. This corresponds to approximately $10\%$ of the data stored in the workers memory thus balancing excessive communication (if results were communicated after every row-vector product computation) and memory limitations (if submatrices at the workers are too large to be communicated as a single chunk). \Cref{fig:avg_distcomp} shows that the LT coded strategies also perform fewer total computations than MDS or $2$-replication. Additional experiments in \Cref{sec:additional_simexp} show that LT coding also demonstrates greater resilience to node failures than Replication or MDS coding in this setting.
\subsection{Serverless Computing Experiments}
We also performed experiments in the serverless computing environment AWS Lambda \cite{amazon_lambda}. Serverless computing eschews the master-worker set-up in favor of only workers (resources) which read data from storage, perform computations on the data, and write it back to storage. Any further computations (like decoding) can be performed as and when desired by re-reading data from storage. As described in \cite{gupta2018oversketch}, there is typically significant variability (straggling) across workers in this setting and therefore we expect to obtain speedups through coding. We use Numpywren \cite{shankar2018numpywren} for performing linear algebra on AWS Lambda. We multiply a $100000 \times 10000$ matrix $\mat$ (approximately $10\times-$ larger than $\mat$ in the previous two experiments) with a $10000\times 1$ vector $\vect$ in this setting. We compare the uncoded, MDS-Coded ($\mdsnum = 80$) and LT-coded ($\strag=2.0$) approaches. As per the requirements of \cite{shankar2018numpywren}, the encoding is performed over blocks of $10$ rows instead of individual rows. Our results, averaged over 5 trials, are presented in \Cref{fig:avg_lamlat,fig:avg_lamcomp}. They clearly show that the LT coded approach is faster than previous approaches, and performs fewer computations than MDS coding. We note that the experimental nature of current serverless computing frameworks makes it challenging to perform fine-grained logging of task times and to use larger encoded matrices (larger $\alpha$). Thus we expect even better results once the limitations of current frameworks have been resolved.

\section{Concluding Remarks}
\label{sec:conclu}
We propose an erasure coding strategy based on \emph{rateless fountain codes} to speed up distributed matrix-vector multiplication in the presence of slow nodes (stragglers). For a matrix with $\rows$ rows, our strategy requires the nodes to \emph{collectively} finish slightly more than $\rows$ row-vector products. Thus, it seamlessly adapts to varying node speeds and achieves near-perfect load balancing. Moreover, it has a small overhead of redundant computations (asymptotically zero), and low decoding complexity. Theoretical analysis and experiments show that our approach strikes a better latency-computation trade-off than existing uncoded, replication and maximum-distance-separable (MDS) coding approaches. 

Going forward, we plan to extend our approach to other linear computations like sparse matrix-vector multiplication (SpMV), Matrix-Matrix multiplication, and Fourier Transforms. Previous work \cite{wang2018coded} has used fixed-rate variants of LT codes to speed-up SpMV; we expect even better performance by exploiting the rateless properties of fountain codes and utilizing partial work as described in this paper. Since erasure codes are inherently linear, extending coding techniques to speed-up distributed \emph{non-linear} computations such as neural network inference is difficult. Recently \cite{kosaian2019parity, kosaian2018learning} propose the use of neural networks to learn the encoder and decoder to handle non-linear computations. Coming up with a principled rateless coding approach in this setting remains an open problem.
\section{Acknowledgements}
The authors are grateful to Pulkit Grover, Sanghamitra Dutta, Yaoqing Yang, Haewon Jeong, Rashmi Vinayak, and Jack Kosaian for helpful discussions. Author Joshi also sincerely thanks Emina Soljanin, Alyson Fox, Fiona Knoll and Nadia Kazemi for fruitful initial
discussions during the Women in Data Science and Mathematics (WiSDM) Research Collaboration Workshop held at
Brown University in July 2017. This project was supported in part by the CMU Dean's fellowship, Qualcomm Innovation Fellowship, NSF CCF grant no. 1850029 and an Amazon Credits for Research Grant.
\bibliographystyle{ACM-Reference-Format}
\bibliography{Bibliography}
\appendix
\clearpage
\section{Properties of LT Codes}
\label{sec:LTProps}
\begin{figure}[htb]
  \centering
  \includegraphics[width=0.5\textwidth]{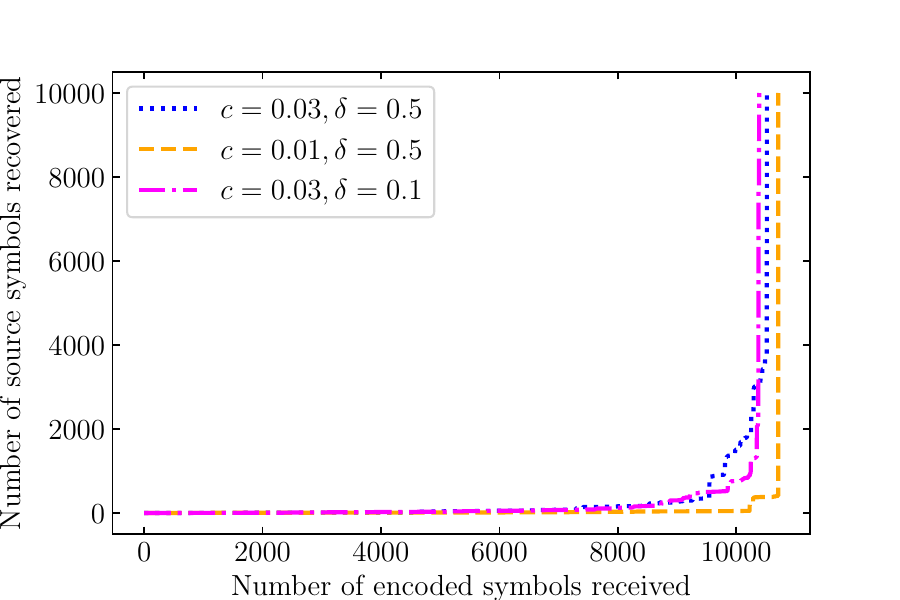}
  \caption{The number of decoded symbols is almost constant until $\rows = 10,000$ encoded symbols are received after which it increases rapidly.}
\label{fig:avalanche}
\end{figure}

\Cref{fig:avalanche} shows simulation results for the number of symbols decoded successfully for each encoded symbol received. For this we perform LT-Coded multiplication of a randomly generated $10,000 \times 10,000$ matrix with a $10,000 \times 1$ vector. The matrix $\mat$ is encoded using an LT code with parameters $\fcpc$ and $\fcpdel$ chosen according to the guidelines of \cite{mackay2003information}. We generate a single row of the encoded matrix $\encmat$ at a time which is then multiplied with the $10,000 \times 1$ size vector $\vect$ to give a single element of the encoded matrix vector product $\encres$. The process is repeated until we have enough symbols for successfully decoding the entire $10,000 \times 1$ size vector $\res$ using the peeling decoder. The plots of \Cref{fig:avalanche} correspond to different choices of $\fcpc$ and $\fcpdel$. In each case we observe an avalanche behavior wherein very few symbols are decoded up to a point ( approximately up to $10,000$ encoded symbols received) after which the decoding proceeds very rapidly to completion. This effectively illustrates the fact that the computation overhead of the proposed LT coded matrix vector multiplication strategy is very small ($\decrows =\rows(1+\epsilon)$). 
The theoretical encoding and decoding properties of LT codes are summarized in the following lemmas:
\begin{lem}[Theorem 13 in \cite{luby2002lt}]
\label{lem:LTcompthm}
For any constant $\fcpdel > 0$, the average degree of an encoded symbol is $\mathcal{O}( \log( \rows/\fcpdel))$ where $m$ is the number of source symbols. 
\end{lem}
\begin{coro}
\label{coro:enccomplexity}
Each encoding symbol can be generated using $\mathcal{O}( \log\rows)$ symbol operations on average. 
\end{coro}
\begin{lem}[Theorem 17 in \cite{luby2002lt}]
\label{lem:LTprobthm}
For any constant $\delta > 0$ and for a source block with $\rows$ source symbols, the LT decoder can recover all the source symbols from a set of $\randdecrows = \rows + \mathcal{O}( \sqrt{\rows}\log^{2}( \rows/\delta))$ with probability at least $1 - \delta$. 
\end{lem}
\begin{coro}
\label{coro:decthresh}
The expected decoding threshold $\mathbb{E}[\randdecrows]$ is given by $\mathbb{E}[\randdecrows] = \rows( 1+\epsilon)$ where $\epsilon \rightarrow 0$ as $\rows \rightarrow \infty$
\end{coro}
\begin{coro}
\label{coro:deccomplexity}
Since the average degree of an encoded symbol is $\mathcal{O}( \log( \rows/\fcpdel))$ the decoding requires $\mathcal{O}( \rows\log\rows)$ symbol operations on average.
\end{coro}

\section{On the Order Statistics of Exponential Random Variables}
\label{sec:ordexp}

We first state some standard results \cite{david2003order} on order statistics of exponential random variables to aid the understanding of the latency analysis presented subsequently. If $X_1$, $X_2$, \dots $X_{\proc}$ are exponential random variables with rate $\mu$, their $k^{th}$ order statistic is denoted by $X_{\mdsnum:\proc}$. Thus, $X_{1:\proc} = \min( X_1, X_2, \dots X_{\proc})$, and $X_{\proc:\proc} = \max( X_1, X_2, \dots X_{\proc})$. The expected value of $X_{\mdsnum:\proc}$ is given by
\begin{align}
\mathbb{E}[X_{\mdsnum:\proc}] &= \frac{1}{\mu}\left( \frac{1}{\proc}+ \dots +\frac{1}{\proc-\mdsnum+1} \right) = \frac{H_{\proc} - H_{\proc-\mdsnum}}{\mu} \label{eqn:harmonic},
\end{align}
where $H_{\proc}$ is the $\proc^{th}$ Harmonic number 
\begin{align}
H_\proc \triangleq \begin{cases} \sum_{i=1}^{\proc} \frac{1}{i} &\text{   for } p =1,2,\dots \\
 0 &\text{   for } p =0
\end{cases} \label{eqn:harmonic_def}
\end{align}
For large $\proc, H_\proc = \log\proc + \gamma$, where $\gamma$ is the Euler-Mascheroni constant and thus we can use the approximation $H_\proc \simeq \log\proc$ for large $\proc$.

Also the difference of consecutive order statistics of i.i.d exponential random variables is also exponentially distributed. Specifically, $U_{l} = X_{l+1:\proc}-X_{l:\proc} \sim \exp((\proc-l)\exprate)$ in this case.

\section{Proof of Delay Analysis Results}
\label{sec:Proofs_Delay}
\subsection{Ideal Load Balancing Strategy}
\begin{proof}[Proof of \Cref{thm:idealoptimal}]
The result follows from \Cref{lem:non_redundant} and \Cref{lem:redundant} for non-redundant and redundant task allocation policies respectively given below. Redundant policies refer to policies where the same task can be allocated to multiple workers (such as the $r-$ Replication policy).
\end{proof}
\begin{lem}
\label{lem:non_redundant}
For any distributed matrix-vector multiplication scheme following the delay model of \Cref{eq:delaymodel} without any redundancy in task allocation, the latency $\runtime$ is no less than $\runtime_{\text{ideal}}$.
\end{lem}
\begin{proof}[Proof of \Cref{lem:non_redundant}]
Consider any scheme other than the ideal load balancing scheme. Let $W_{i,i+1}$ be the time elapsed between completion of the $i^{\text{th}}$ and $(i+1)^{\text{th}}$ computation in this scheme and let $W_{i,i+1}^{\text{ideal}}$ be the time elapsed between completion of the $i^{\text{th}}$ and $(i+1)^{\text{th}}$ computation in the ideal scheme. By definition, the ideal load balancing is work-conserving, that is, no worker is idle while there are pending computations at other workers. Hence for the other scheme there must be some computation $\hat{i}$ after which at least one worker is idle even though there are pending computations at the other workers. Therefore $W_{\hat{i},\hat{i}+1}^{\text{ideal}} \leq W_{\hat{i},\hat{i}+1}$.
Moreover since the tasks are allocated by the master initially in our setting, it means that if a worker is idle after computation $\hat{i}$ is completed by the system, then it is idle for \emph{all} subsequent computations. Thus, $W_{i,i+1}^{\text{ideal}} \leq W_{i,i+1}$,  $\forall i\geq \hat{i}$. Therefore if $\runtime^{\text{ideal}}$ and $\runtime^{\text{other}}$ are the times taken by the ideal and any other scheme respectively to complete $\rows$ tasks,
\begin{align}
\runtime_{\text{ideal}} &= W_{0,1}^{\text{ideal}}+\ldots+W_{\hat{i},\hat{i}+1}^{\text{ideal}}+\ldots+W_{\rows-1,\rows}^{\text{ideal}},\\
\runtime &= W_{0,1}+\ldots+W_{\hat{i},\hat{i}+1}+\ldots+W_{\rows-1,\rows},
\end{align}
and $\runtime^{\text{ideal}} \leq \runtime$ for \emph{any} other task allocation strategy.
\end{proof}
\begin{lem}
\label{lem:redundant}
For any distributed matrix-vector multiplication scheme following the delay model of \Cref{eq:delaymodel} with potentially redundant task allocations, the latency $\runtime$ is no less than $\runtime_{\text{ideal}}$.
\end{lem}
\begin{proof}[Proof of \Cref{lem:redundant}]
Consider the a scheme with redundancy where task $j$ is allocated to $\numrep$ distinct workers. Let $V_1,\ldots,V_\numrep$ be the time instant at which each of the $\numrep$ workers start working on the task. If $W^j$ is the earliest time instant at which the task is received at the master then,
\begin{align}
    W^j &= \min_{1\leq i\leq \numrep}(V_{i} + \shifttime),\\
    &= \shifttime + \min_{1\leq i\leq \numrep}V_{i}.
\end{align}
Thus the time at which a task is completed is equal to the sum of the earliest time instant ($\min_{1\leq i\leq \numrep}V_{i}$) at which one of the $\numrep$ worker is available to process the task and the time ($\shifttime$) required by any worker to process the task. Since the ideal load balancing scheme greedily assigns tasks to the first of the $\proc$ workers that are available to process it and $\numrep \leq \proc$, therefore the ideal scheme is essentially equivalent to a scheme with \emph{maximum} redundancy and thus cannot be outperformed by any task allocation strategy with redundancy.
\end{proof}
\begin{proof}[Proof of \Cref{lem:ideallatency}]
As per our model, the time taken by worker $i$ to perform $\workcomp_i$ computations is given by
\begin{align}
\worktime_i = X_i + \shifttime\workcomp_i,\ \text{  for } i=1,\dots,\proc.
\end{align}
The latency $\runtime_{\text{ideal}}$ is the earliest time when $\sum_{i=1}^{p}\workcomp_i = \rows$, as illustrated in \Cref{fig:delayideal}. We note that, in this case it is not necessary that each worker has completed at least $1$ computation. Specifically, if $\runtime_{\text{ideal}} - X_i \leq \shifttime$ for any $i$ then it means that worker $i$ has not performed even a single computation in the time that the system as a whole has completed $\rows$ computations (owing to the large initial delay $X_i$). Therefore we define
\begin{align}
\mathcal{W}_{\text{ideal}}:=\{i:\runtime_{\text{ideal}} - X_i \geq \shifttime\}.
\end{align}
Here $\mathcal{W}_{\text{ideal}}$ is the set of workers for which $\workcomp_i > 0$. Thus
\begin{align}
\runtime_{\text{ideal}} &= \max_{i\in\mathcal{W}_{\text{ideal}}}\worktime_i = \max_{i\in\mathcal{W}_{\text{ideal}}}\left( X_i + \shifttime\workcomp_i \right), \label{eqn:lb_proof_1} \\
&\geq \min_{i \in \{ 1,\dots \proc \}} X_i +  \shifttime \max_{i\in\mathcal{W}_{\text{ideal}}}\workcomp_i \label{eqn:lb_proof_2}, \\
&\geq  X_{1:\proc} +  \frac{\shifttime\rows}{\proc} \label{eqn:lb_proof_4},
\end{align}
where to obtain \eqref{eqn:lb_proof_2}, we replace each $X_i$  in \eqref{eqn:lb_proof_1} by $\min_{i \in [1,\dots \proc]} X_i$ and then we can bring it outside the maximum. 
To obtain \eqref{eqn:lb_proof_4}, we observe that in order for the $\proc$ workers to collectively finish $\rows$ computations, the maximum number of computations completed by a worker has to be at least $\rows/\proc$.

To derive the upper bound, we note that
\begin{align}
\runtime_{\text{ideal}} &\leq X_i + \shifttime( \workcomp_i+1),\ \text{ for all } i = 1,\ldots,\proc
\end{align}
This is because at time $\runtime_{\text{ideal}}$ each of the workers $1,\ldots,\proc$, have completed $\workcomp_1,\ldots,\workcomp_p$ row-vector product tasks respectively, but they may have partially completed the next task. The $1$ added to each $\workcomp_i$ accounts for this edge effect, which is also illustrated in \Cref{fig:delay_model}.
Summing over all $i$ on both sides, we get
\begin{align}
\sum_{i=1}^{\proc}\runtime_{\text{ideal}}  &\leq \sum_{i=1}^{\proc} X_i + \sum_{i=1}^{\proc} \shifttime\left( \workcomp_i+1 \right),\\
\proc \runtime_{\text{ideal}} &\leq \sum_{i=1}^{\proc} X_i + \shifttime\left( \rows+\proc \right),\\
\runtime_{\text{ideal}} &\leq \frac{1}{\proc}\sum_{i=1}^{\proc} X_i + \frac{\shifttime\rows}{\proc} + \shifttime \label{eqn:ub_proof}
\end{align}
\end{proof}
\begin{proof}[Proof of \Cref{coro:ideallatency}]
If $X_i \sim \exp(\exprate)$ then taking expectation on both sides of \Cref{eqn:lb_proof_4} gives
\begin{align}
\mathbb{E}[\runtime_{\text{ideal}}] &\geq \mathbb{E}[X_{1:\proc}] + \frac{\shifttime\rows}{\proc}, \label{eqn:lb_proof_5} \\
&= \frac{1}{\proc \mu} +  \frac{\shifttime\rows}{\proc}. \label{eqn:lb_proof_6} 
\end{align}
where the lower bound in \eqref{eqn:lb_proof_6} follows from the result \eqref{eqn:harmonic} on order statistics of exponential random variables.
Likewise for the upper bound we can compute expectation on both sides of \Cref{eqn:ub_proof} to get,
\begin{align}
\mathbb{E}[\runtime_{\text{ideal}}]&\leq  \frac{1}{\proc}\sum_{i=1}^{\proc}\mathbb{E}[X_i] + \frac{\shifttime\rows}{\proc}+\shifttime\\
\mathbb{E}[\runtime_{\text{ideal}}] &\leq \frac{1}{\exprate} + \frac{\shifttime\rows}{\proc} + \shifttime.
\end{align}
\end{proof}
\subsection{Rateless Coded Strategy}
\begin{figure*}[t]
 \centering
 \subfloat[\label{fig:delayideal}]{\includegraphics[width=0.3\linewidth]{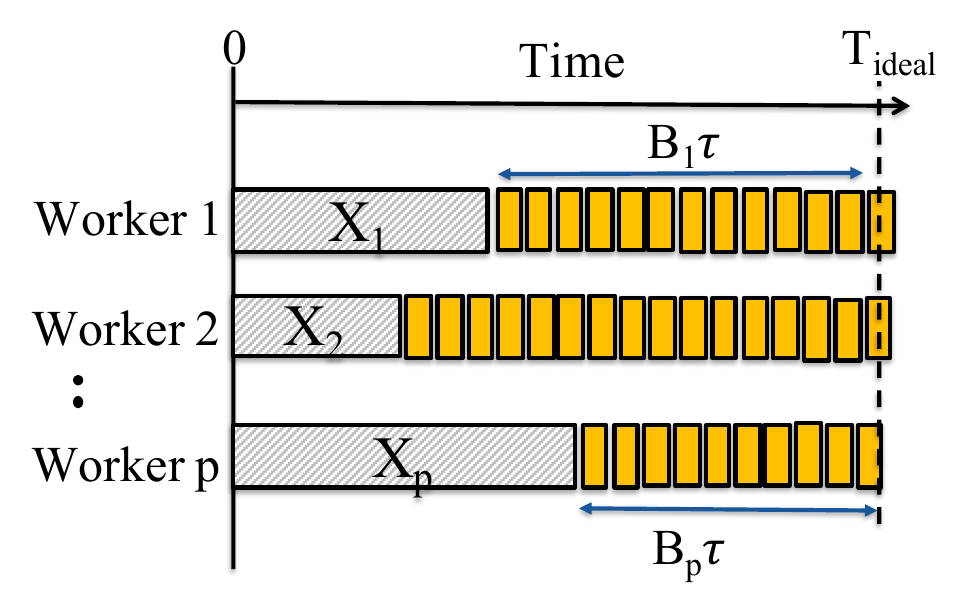}}
 \subfloat[\label{fig:delayfinite}]{\includegraphics[width=0.3\linewidth]{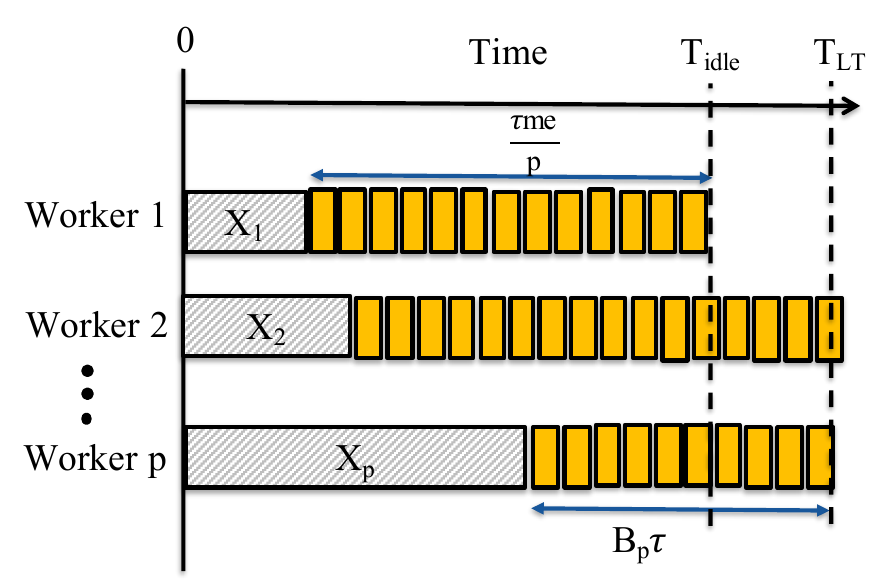}}
  \subfloat[\label{fig:delayE2}]{\includegraphics[width=0.3\linewidth]{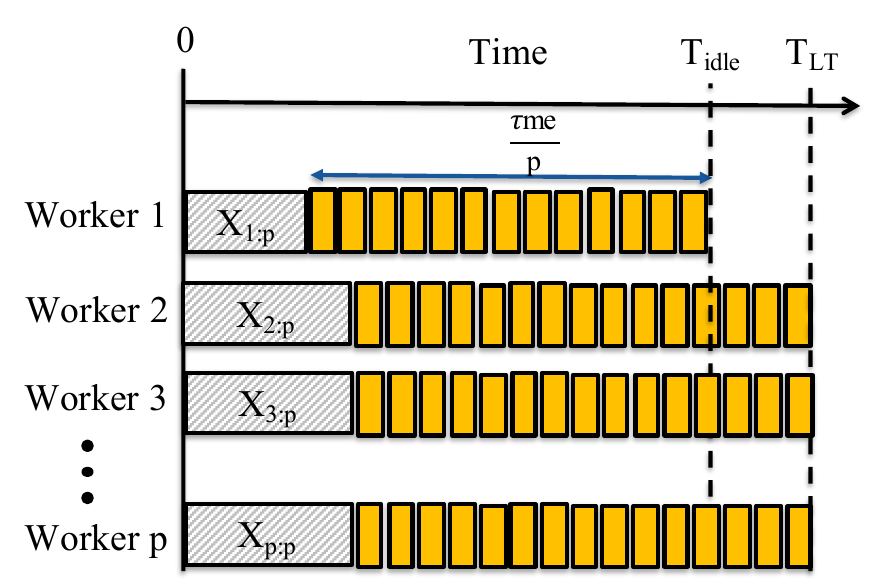}}
  \caption{(a) Worker $i$ has an random exponential initial delay $X_i$, after which it completes row-vector product tasks taking time $\tau$ per task. In the ideal case the latency $\runtime_{\text{ideal}}$ is simply the time to complete $\rows$ tasks in total. (b) A general scenario Worker 1 runs out of computations at $\runtime_{\text{idle}}$ before the system completes enough computations necessary for successful decoding. (c) The specific scenario (event $E_2$) where Worker $2$ starts so late that Worker $1$ runs out of computations even though workers $3,\ldots,\proc$ start at the same time as $2$.}
\end{figure*}
\begin{proof}[Proof of \Cref{thm:probltlatency}]
Recall that we make the following assumption for analysing the latency of the proposed rateless coded strategy.
\begin{assum}
The decoding threshold $\randdecrows$ (\Cref{defn:decrows}) of the LT coded strategy satisfies $\randdecrows \simeq \rows$. 
\end{assum}
We believe the above assumption is reasonable because the problem of distributed matrix vector multiplication arises only when $\rows$ (the number of rows of $\mat$) is large and the high probability bound of \Cref{lem:LTcomp} can be used to show that $\mathbb{E}[M'] = \rows(1+\epsilon)$, where $\epsilon \rightarrow 0$ as $\rows \rightarrow \infty$. 
To analyze the probability $\Pr(\runtime_{\text{LT}} > \runtime_{\text{ideal}})$ in light of the above assumption, let us first understand when $\runtime_{\text{LT}}$ will exceed $\runtime_{\text{ideal}}$. This situation will arise when the fastest node runs out of computations and becomes idle before $\randdecrows$ computations are collectively collectively completed by all the workers as illustrated in \Cref{fig:delayfinite}.

Recall that according to our delay model, the time $\worktime_i$ required by worker $i$ to perform $\workcomp_i$ computations is
\begin{equation}
\worktime_i = X_i + \shifttime\workcomp_i, \quad \text{for all } i=1,\dots, \proc
\end{equation}
where $X_i$ is an initial delay at worker $i$. Without loss of generality, we assume that nodes are ordered in increasing order of initial delays. Thus, $X_i = X_{i:\proc}$ for $i=1,\ldots,\proc$, where $X_{i:\proc}$ denotes the $i^{th}$ order statistic of $\proc$ i.i.d.\ random variables with distribution $F_X$. 

Now let us determine the events that cause $\runtime_{\text{LT}}$ to be greater than $\runtime_{\text{Ideal}}$ for a given set of realizations of the initial delays. Since each node is assigned $\encrows/\proc$ row-vector product tasks, the time $\runtime_{\text{idle}}$ at which the fastest node, node $1$, becomes idle is 
\begin{align}
\label{eq:idletime}
\runtime_{\text{idle}} = X_{1:\proc} + \frac{\shifttime\encrows}{\proc}
\end{align}
Suppose that the initial delays of the fastest $j$ workers are $X_{1:\proc}$, $X_{2:\proc}$,$\ldots$,$X_{j:\proc}$ respectively. Then let $\numcomp_j$ denote the number of computations collectively performed by the $\proc$ workers until time $\runtime_{\text{idle}}$ in the event that all the remaining $\proc-j$ workers start at the earliest possible instant and have initial delays equal to $X_{j:\proc}$.

Observe that $\runtime_{\text{LT}} > \runtime_{\text{Ideal}}$ if node $1$ becomes idle before all workers collectively perform $m$ computations. This situation (depicted in \Cref{fig:delayfinite}) arises if $X_{j:\proc}$ is so large for some node $j$ that $\numcomp_j < \rows$. We distill this event of node $1$ becoming idle before the result is recovered into sub-events $E_1$, $E_2$, \dots $E_{\proc}$, where we refer to $E_j$ as the event of idling due to node $j$, which is defined as follows.
\begin{defn}[Event $E_j$: Idling due to node $j$]
\label{defn:stragevent}
Given $X_{1:\proc}$, $X_{2:\proc}$,\ldots,$X_{j-1:\proc}$, $E_j$ occurs if $C_{j} < \rows$.
\end{defn}
The event $E_2$ is depicted in \Cref{fig:delayE2}. Therefore,
\begin{align}
\Pr(\runtime_{\text{LT}} > \runtime_{\text{ideal}}) &= \Pr(E_2 \cup \ldots \cup E_\proc) \\
&\leq \sum_{j=2}^{\proc} \Pr(E_j)\label{eq:probfail} \\
&= \sum_{j=2}^{\proc} \Pr(C_j < m)
\end{align}
where \eqref{eq:probfail} follows from the union bound. Note that the event of idling due to node $1$, $E_1$ is not defined since node $1$ is the fastest node by definition. 

Now we derive a lower bound on $C_j$ in terms of $X_{1:\proc},\ldots,X_{j:\proc}$ as follows.
\begin{align}
C_j &= \sum_{l=1}^{j-1} \left(\Bigr\lfloor \frac{T_{\text{idle}} - X_{l:p}}{\tau} \Bigr\rfloor \right)^{+} + (p-j+1) \left(\Bigr\lfloor \frac{T_{\text{idle}} - X_{j:p}}{\tau} \Bigr\rfloor\right)^{+} \label{eq:C_i_deriv_1}\\
&= \sum_{l=1}^{j-1} \left(\Bigr\lfloor \frac{\encrows}{\proc} - \frac{ (X_{l:p} - X_{i:p})}{\tau} \Bigr\rfloor \right)^{+} + (p-j+1) \left(\Bigr\lfloor \frac{\encrows}{\proc} - \frac{(X_{j:p} - X_{1:p})}{\tau} \Bigr\rfloor\right)^{+} \label{eq:C_i_deriv_2}\\
&\geq \encrows - \sum_{l=1}^{j-1} \left( \frac{ X_{l:p} - X_{i:p}}{\tau} \right)  - (p-j+1) \left( \frac{X_{j:p} - X_{1:p}}{\tau}\right) \label{eq:C_i_deriv_3} \\
&\geq \encrows - \frac{\proc-1}{\shifttime}\sum_{l=1}^{j-1}(X_{l+1:p}-X_{l:p}) \label{eq:C_i_deriv_4}\\
&= \encrows - \frac{\proc-1}{\shifttime}\sum_{l=1}^{j-1}U_{l} \label{eq:C_i_deriv_5}
\end{align}
where in \eqref{eq:C_i_deriv_3} we use the fact that for any $a,b \geq 0, [a - \lfloor b \rfloor]_{+} \geq (a-b)$.

As a consequence of the stochastic dominance implied by the above bound we have.
\begin{align}
\Pr(E_j) &= \Pr(\numcomp_j < \randdecrows)\\
&\leq \Pr\left(\sum_{l=1}^{j-1}U_{l:p} \geq \shifttime\frac{\encrows - \rows}{\proc - 1}\right) \label{eq:probEj}
\end{align}
Recall from \Cref{eq:probfail} that,
\begin{align}
\Pr(\runtime_{\text{LT}} > \runtime_{\text{ideal}}) &\leq \sum_{j=2}^{\proc} \Pr(E_j)\\
&\leq\sum_{j=2}^{\proc}\Pr\left(\sum_{l=1}^{j-1}U_{l} \geq \frac{\shifttime\rows(\strag - 1)}{\proc - 1}\right)
\end{align}
\end{proof}
\begin{proof}[Proof of \Cref{coro:probltlatency}]
If $X_i \sim \exp(\exprate)$ we can use the results from \Cref{sec:ordexp} to simplify \Cref{eq:probEj} further,
\begin{align}
\Pr(E_j)&\leq\Pr \left( \sum_{l=1}^{j-1}U_{l}\geq\frac{\shifttime\rows(\strag - 1)}{\proc - 1}\right)\\
&\leq\Pr \left((j-1)U_{j-1}\geq\frac{\shifttime\rows(\strag - 1)}{\proc - 1}\right)\label{eqn:ltstocdom2}\\
&=\Pr \left(U_{j-1}\geq\frac{\shifttime\rows(\strag - 1)}{(\proc - 1)(j-1)}\right)\\
&=\exp\left( -\frac{\exprate\shifttime\rows(\strag - 1)(\proc-j+1)}{(\proc - 1)(j-1)}\right)\label{eqn:ltlatencyfinal}
\end{align}
where \eqref{eqn:ltstocdom2} is obtained from the fact that $\Pr( U_{j-1}\geq u)\geq\Pr( U_{l}\geq u) \text{ for }l=1,\ldots,j-1$ for any $u$ since $U_{l}\sim\exp((\proc-l)\exprate)$. Lastly \eqref{eqn:ltlatencyfinal} is obtained from the expression for the tail distribution of an exponential random variable.\footnote{For a two-server system ($\proc = 2$), $\Pr(\runtime_{\text{LT}} > \runtime_{\text{ideasl}}) = \Pr(E_2) = \Pr(\numcomp_2 < \rows) \leq \Pr(X_{2:2}-X_{1:2} \geq \shifttime(\encrows - \rows))$ and we can see that increasing redundancy (increasing $\rows$) will decrease the probabilistic upper bound. Actually for any $\encrows > 2\rows$, $\Pr(\runtime_{\text{LT}} > \runtime_{\text{ideal}}) = 0$ since in this case the entire work can be done by the faster node but in this case our upper bound, which is a little loose, has a small positive value.}

Since $(\proc - j +1)/(j-1) > 1/(\proc - 1) \forall j=2,\ldots,\proc$, we obtain the final result
\begin{align}
\Pr(\runtime_{\text{LT}} > \runtime_{\text{ideal}}) &\leq \sum_{j=2}^{\proc} \Pr(E_j)\\
&\leq\sum_{j=2}^{\proc}\Pr\left(\sum_{l=1}^{j-1}U_{l} \geq \frac{\shifttime\rows(\strag - 1)}{\proc - 1}\right)\\
&\leq (\proc - 1)\exp \left(-\frac{\exprate\shifttime\rows(\strag-1)}{(\proc - 1)^2} \right)\\
&\leq \proc\exp \left(-\frac{\exprate\shifttime\rows(\strag-1)}{\proc^2}\right)
\end{align}
\end{proof}
\begin{proof}[Proof of \Cref{thm:ltlatency}]
We can write the expectation of $T_{\text{LT}}$ as follows.
\begin{align}
    \mathbb{E}[\runtime_{\text{LT}}] &= \Pr(\runtime_{\text{LT}} = \runtime_{\text{ideal}})\mathbb{E}[\runtime_{\text{LT}}\mid \runtime_{\text{LT}} = \runtime_{\text{ideal}}] + \Pr(\runtime_{\text{LT}} > \runtime_{\text{ideal}})\mathbb{E}[\runtime_{\text{LT}}\mid \runtime_{\text{LT}} > \runtime_{\text{ideal}}]\\
    \begin{split}
    &= \Pr(\runtime_{\text{LT}} = \runtime_{\text{ideal}})\mathbb{E}[\runtime_{\text{ideal}}\mid \runtime_{\text{LT}} = \runtime_{\text{ideal}}] + \Pr(\runtime_{\text{LT}} > \runtime_{\text{ideal}})\mathbb{E}[\runtime_{\text{ideal}}\mid \runtime_{\text{LT}} > \runtime_{\text{ideal}}] \\&+ \Pr(\runtime_{\text{LT}} > \runtime_{\text{ideal}})\mathbb{E}[\runtime_{\text{LT}} - \runtime_{\text{ideal}}\mid \runtime_{\text{LT}} > \runtime_{\text{ideal}}]
    \end{split}
    \\&= \mathbb{E}[\runtime_{\text{ideal}}] + \Pr(\runtime_{\text{LT}} > \runtime_{\text{ideal}})\mathbb{E}[\runtime_{\text{LT}} - \runtime_{\text{ideal}}\mid \runtime_{\text{LT}} > \runtime_{\text{ideal}}] \\
    &\leq \mathbb{E}[\runtime_{\text{ideal}}] + \Pr(\runtime_{\text{LT}} > \runtime_{\text{ideal}})\mathbb{E}[\runtime_{\text{LT}}\mid \runtime_{\text{LT}} > \runtime_{\text{ideal}}], \label{eq:lt_upper_1}
\end{align}
where \eqref{eq:lt_upper_1} follows from the fact that $\runtime_{\text{LT}} - \runtime_{\text{ideal}} < \runtime_{\text{LT}}$. Since we have already derived an upper bound for $\Pr(\runtime_{\text{LT}} > \runtime_{\text{ideal}})$ in \Cref{thm:probltlatency}, we will now derive an upper bound for $\mathbb{E}[\runtime_{\text{LT}}\mid \runtime_{\text{LT}} > \runtime_{\text{ideal}}]$. To do so, we first define $\mathcal{W}_{\text{LT}}$ as the set of workers that have \emph{not} completed all the $\strag\rows/\proc$ computations assigned to them i.e.\ $\mathcal{W}_{\text{LT}}:=\{i:\workcomp_i < \frac{\strag\rows}{\proc}\}$. The set $\mathcal{W}_{\text{LT}}$ does not include the workers which have completed all the $\strag\rows/\proc$ tasks assigned to them and are idle at $\runtime_{\text{LT}}$ since they have completed their tasks at some (unknown earlier) time and thus do not increase the upper bound for $\runtime_{\text{LT}}$. For workers in $\mathcal{W}_{\text{LT}}$,
\begin{align}
    \runtime_{\text{LT}} &\leq X_i +\shifttime(\workcomp_i + 1) \text{   for all }  i \in \mathcal{W}_{\text{LT}},\\
    &\leq \sum_{i=1}^{\proc}X_i + \frac{\shifttime\strag\rows}{\proc} + \shifttime, \label{eqn:T_lt_bnd_1} \\  
    &= X_{1:\proc} + X_{2:\proc}+\ldots+X_{\proc:\proc} + \frac{\shifttime\strag\rows}{\proc} + \shifttime \label{eqn:T_lt_bnd_2}\\
    &= \proc X_{1:\proc} + (\proc - 1)U_{1} +\ldots+U_{\proc-1}+ \frac{\shifttime\strag\rows}{\proc} + \shifttime, \label{eqn:T_lt_bnd_3}
\end{align}
where \eqref{eqn:T_lt_bnd_1} follows from the fact that $X_i \leq \sum_{i=1}^{\proc}X_{i}$ for any worker $i$. In \eqref{eqn:T_lt_bnd_2} we express the sum of $X_i$'s in terms of the order statistics $X_{l:\proc}$. In \eqref{eqn:T_lt_bnd_3}, $U_{l} = X_{l+1:\proc} - X_{l:\proc}$, $l=1,\ldots,\proc-1$.

We can then use \eqref{eqn:T_lt_bnd_3} to upper bound $\mathbb{E}[\runtime_{\text{LT}}\mid \runtime_{\text{LT}} > \runtime_{\text{ideal}}]$ as
\begin{align}
    \mathbb{E}[\runtime_{\text{LT}}\mid \runtime_{\text{LT}} > \runtime_{\text{ideal}}]
    & \leq \proc\mathbb{E}[X_{1:\proc}] + \mathbb{E}\left[\sum_{l=1}^{\proc - 1}(\proc - l)U_{l}\mid \runtime_{\text{LT}} > \runtime_{\text{ideal}} \right] + \frac{\shifttime\strag\rows}{\proc} + \shifttime, \label{eq:lt_upper_3}
\end{align}
where in the first term of \eqref{eq:lt_upper_3}, we do not condition by $\runtime_{\text{LT}} > \runtime_{\text{ideal}}$ since the initial delay  $X_{1:\proc}$ at the fastest worker $X_{1:\proc}$ is independent of the event $\runtime_{\text{LT}} > \runtime_{\text{ideal}}$. This is because as seen in the proof of \Cref{thm:probltlatency}, the event $\runtime_{\text{LT}} > \runtime_{\text{ideal}}$ depends on the values of $U_{1}, \ldots, U_{\proc-1}$ but not on $X_{1:\proc}$.

Now let us define $\bar{\mathbf{U}}_{l-1} = [U_{1}, \ldots, U_{l-1}]$ and $S_j = \sum_{l=1}^{j}(\proc - l)U_{l}$ for all $j=1,\ldots,\proc-1$. Using \Cref{lem:ltbound} (stated and proved below) we can simplify the second term in the right-hand-side of \Cref{eq:lt_upper_3} as,
\begin{align}
\mathbb{E}_{\bar{\mathbf{U}}_{\proc-1}}[S_j \mid \runtime_{\text{LT}} > \runtime_{\text{ideal}}]&= \mathbb{E}_{\bar{\mathbf{U}}_{\proc-2}}[\mathbb{E}_{U_{\proc - 1}}[U_{\proc -1} + S_{\proc-2}\mid \runtime_{\text{LT}} > \runtime_{\text{ideal}}, \bar{\mathbf{U}}_{\proc-2}]]\\
&\leq \mathbb{E}_{\bar{\mathbf{U}}_{\proc-3}}[R_{\proc-1} + \mathbb{E}_{U_{\proc - 2}}[2U_{\proc -2} + S_{\proc-3}\mid \runtime_{\text{LT}} > \runtime_{\text{ideal}}, \bar{\mathbf{U}}_{\proc-3}]]
\end{align}
where $R_{l}=\mathbb{E}[(\proc - l)U_{l}\mid U_{l} > \shifttime\strag\rows/\proc]$. 

Repeatedly applying \Cref{lem:ltbound} gives the final upper bound 
\begin{align}
    \mathbb{E}\left[\sum_{l=1}^{\proc - 1}(\proc - l)U_{l}\mid \runtime_{\text{LT}} > \runtime_{\text{ideal}} \right] \leq \sum_{l=1}^{\proc -1}R_{l} \label{eq:lt_upper_3_part}
\end{align}
Substituting \Cref{eq:lt_upper_3_part} in \Cref{eq:lt_upper_3} gives
\begin{align}
    \mathbb{E}[\runtime_{\text{LT}}\mid \runtime_{\text{LT}} > \runtime_{\text{ideal}}]
    &\leq \proc\mathbb{E}[X_{1:\proc}] + \sum_{l=1}^{\proc - 1}R_{l} +  \frac{\shifttime\strag\rows}{\proc} + \shifttime 
\end{align}
Thus finally we have
\begin{align}
    \mathbb{E}[\runtime_{\text{LT}}] - \mathbb{E}[\runtime_{\text{ideal}}] &\leq \Pr(\runtime_{\text{LT}} > \runtime_{\text{ideal}})\mathbb{E}[\runtime_{\text{LT}}\mid \runtime_{\text{LT}} > \runtime_{\text{ideal}}] \\
    &\leq \left(\sum_{j=2}^{\proc}\Pr\left(\sum_{l=1}^{j-1}U_{l:p} \geq \shifttime\frac{\encrows - \randdecrows}{\proc - 1}\right)\right)\left(\proc\mathbb{E}[X_{1:\proc}] + \sum_{l=1}^{\proc -1}R_{l} + \frac{\shifttime\strag\rows}{\proc} + \shifttime \right).
\end{align}
If $X_i \sim \exp(\exprate)$ we can use the results from \Cref{sec:ordexp} to simplify the above results further by using the bounds on $\Pr(\runtime_{\text{LT}} > \runtime_{\text{ideal}})$ for this case which is given by
\begin{align}
    \Pr(\runtime_{\text{LT}} > \runtime_{\text{ideal}}) \leq \proc\exp \left(-\frac{\exprate\shifttime\rows(\strag-1)}{\proc^2}\right),
\end{align}
and the fact that $\mathbb{E}[X_{1:\proc}] = 1/\proc\exprate$ and 
\begin{align}
    R_{l}&=\mathbb{E}[(\proc - l)U_{l}\mid U_{l} > \frac{\shifttime\strag\rows}{\proc}] \nonumber \\
    &= \frac{\shifttime\strag\rows}{\proc}(\proc - l) + (\proc - l)\frac{1}{(\proc - l)\exprate} \nonumber \\
    &= \frac{\shifttime\strag\rows}{\proc}(\proc - l) + \frac{1}{\exprate},
\end{align}
since $U_{l}\sim \exp((\proc - l)\exprate)$ which implies that 
\begin{align}
    \proc\mathbb{E}[X_{1:\proc}] + \sum_{l=1}^{\proc - 1}R_l &= \proc\frac{1}{\proc\exprate} + \frac{\shifttime\strag\rows}{\proc}\frac{\proc(\proc - 1)}{2} + \frac{\proc - 1}{\exprate},\\
    &\leq \frac{\shifttime\strag\rows\proc}{2} + \frac{\proc}{\exprate}.
\end{align}
Thus,
\begin{align}
    \mathbb{E}[\runtime_{\text{LT}}] - \mathbb{E}[\runtime_{\text{ideal}}] &\leq \Pr(\runtime_{\text{LT}} > \runtime_{\text{ideal}})\mathbb{E}[\runtime_{\text{LT}}\mid \runtime_{\text{LT}} > \runtime_{\text{ideal}}] \\
    &\leq \proc\exp\left(-\frac{\exprate\shifttime\rows(\strag-1)}{\proc^2}\right)\left(\frac{\shifttime\strag\rows\proc}{2} + \frac{\proc}{\exprate} + \frac{\shifttime\strag\rows}{\proc} + \shifttime\right)\\
    &\leq \left(\shifttime\strag\rows\proc^2 + \frac{\proc^2}{\exprate} + \shifttime\proc \right)\exp \left(-\frac{\exprate\shifttime\rows(\strag-1)}{\proc^2}\right)
\end{align}
where the last expression is true for $\proc\geq 2$ which is always the case in distributed settings.
\end{proof}
\begin{lem}
\label{lem:ltbound}
Given the values of $U_{1}, \ldots, U_{l-1}$ we have the following upper bound
\begin{align}
     \mathbb{E}[(\proc - l)U_{l} \mid \runtime_{\text{LT}} > \runtime_{\text{ideal}}, \bar{U}_{l-1}] \leq \mathbb{E}[(\proc - l)U_{l}\mid U_{l} > \frac{\shifttime\strag\rows}{\proc}]
\end{align}
where $\bar{\mathbf{U}}_{l-1} = [U_{1}, \ldots, U_{l-1}]$.
\end{lem}
\begin{proof}
Observe that given the values of $U_{1}, \ldots, U_{l-1}$, the condition $U_{l} > \shifttime\strag\rows/\proc$ is always sufficient to guarantee that $\runtime_{\text{LT}} > \runtime_{\text{ideal}}$ even if that may not be guaranteed by the values of $U_{1:\proc}, \ldots, U_{l-1:\proc}$. This is because $U_{l}  = X_{l+1:\proc} - X_{l:\proc}$ and thus $U_{l} > \shifttime\strag\rows/\proc$ implies that worker $l+1$ starts computing only \emph{after} worker $l$ has completed all $\strag\rows/\proc$ computations assigned to it. For eg. If $U_{2} > \shifttime\strag\rows/\proc$ then the second fastest worker (and all subsequent workers) will only start computing after the fastest worker has completed all of its $\strag\rows/\proc$ computations. Based on this we can show that,
\begin{align}
    &\mathbb{E}[(\proc - l)U_{l} \mid \runtime_{\text{LT}} > \runtime_{\text{ideal}}, \bar{U}_{l-1}] \nonumber \\ 
    &= \Pr(U_{l} < \frac{\shifttime\strag\rows}{\proc})\mathbb{E}[(\proc - l)U_{l} \mid \runtime_{\text{LT}} > \runtime_{\text{ideal}}, U_{l} < \frac{\shifttime\strag\rows}{\proc}, \bar{U}_{l-1}] +\nonumber \\
    &\,\,\Pr(U_{l} > \frac{\shifttime\strag\rows}{\proc})\mathbb{E}[(\proc - l)U_{l} \mid \runtime_{\text{LT}} > \runtime_{\text{ideal}}, U_{l} > \frac{\shifttime\strag\rows}{\proc},\bar{U}_{l-1}]
\end{align}
Since $\mathbb{E}[(\proc - l)U_{l} \mid \runtime_{\text{LT}} > \runtime_{\text{ideal}}, U_{l} < \frac{\shifttime\strag\rows}{\proc}, \bar{U}_{l-1}] < \mathbb{E}[(\proc - l)U_{l} \mid \runtime_{\text{LT}} > \runtime_{\text{ideal}}, U_{l} > \frac{\shifttime\strag\rows}{\proc},\bar{U}_{l-1}]$, it follows that
\begin{align}
    &\mathbb{E}[(\proc - l)U_{l} \mid \runtime_{\text{LT}} > \runtime_{\text{ideal}}, \bar{U}_{l-1}] \nonumber \\ 
    & \quad \leq \mathbb{E}[(\proc - l)U_{l} \mid \runtime_{\text{LT}} > \runtime_{\text{ideal}}, U_{l} > \frac{\shifttime\strag\rows}{\proc},\bar{U}_{l-1}]
\end{align}
However since $U_{l} > \frac{\shifttime\strag\rows}{\proc}$ is sufficient to guarantee $\runtime_{\text{LT}} > \runtime_{\text{ideal}}$ and $(\proc - l)U_{l}$ does not depend on $\bar{U}_{l-1} = [U_{1}, \ldots, U_{l-1}]$, the above expression reduces to 
\begin{align}
     \mathbb{E}[(\proc - l)U_{l} \mid \runtime_{\text{LT}} > \runtime_{\text{ideal}}, \bar{U}_{l-1}] \leq \mathbb{E}[(\proc - l)U_{l}\mid U_{l} > \frac{\shifttime\strag\rows}{\proc}]
\end{align}
\end{proof}
\subsection{MDS Coded Strategy}
\begin{proof}[Proof of \Cref{thm:mdslatency}]
The latency in the MDS-coded case is $\runtime_{\text{MDS}} = \worktime_{\mdsnum:\proc}$, where $\worktime_{\mdsnum:\proc}$ is the $\mdsnum^{\text{th}}$ order statistic of the individual worker latencies $\worktime_1,\worktime_2,\ldots,\worktime_\proc$ since we only wait for the fastest $\mdsnum$ workers to finish the task assigned to them. In this case, each of the fastest $\mdsnum$ workers performs $\frac{\rows}{\mdsnum}$ computations and thus the overall latency is given by
\begin{equation}
\runtime_{\text{MDS}} = \worktime_{\mdsnum:\proc} = X_{\mdsnum:\proc} + \shifttime\frac{\rows}{\mdsnum}
\end{equation}
\end{proof}
\begin{proof}[Proof of \Cref{coro:mdslatency}]
If $X_i \sim \exp(\exprate)$, the expected overall latency is given by
\begin{align}
\mathbb{E}[\runtime_{\text{MDS}}] &= \mathbb{E}[X_{\mdsnum:\proc}] + \shifttime\frac{\rows}{\mdsnum},\\
&= \frac{\shifttime \rows}{\mdsnum} + \frac{1}{\mu}\left( H_\proc - H_{\proc-\mdsnum} \right), \label{eqn:mds_3} \\
&\simeq \frac{\shifttime \rows}{\mdsnum} + \frac{1}{\mu} \log\frac{\proc}{\proc - \mdsnum} \label{eqn:mds_4}.
\end{align}
where \eqref{eqn:mds_3} and \eqref{eqn:mds_4} follow from the exponential order statistics results in \eqref{eqn:harmonic} and \eqref{eqn:harmonic_def}.
\end{proof}
\begin{proof}[Proof of \Cref{thm:mdscomptail}]
As per our model, we represent the number of computations at worker $i$ by the random variable $\workcomp_i$ . We also use the random variable $\numcomp_{\text{MDS}}$ to denote the total number of computations performed by all $\proc$ workers until $\runtime_{\text{MDS}}$, which is the time when the master collects enough computations to be able to recover the matrix-vector product $\res=\mat\vect$. Thus
\begin{align}
\numcomp_{\text{MDS}}&=\workcomp_1+\workcomp_2+\ldots+\workcomp_\proc\\
&=\workcomp_{1:\proc}+\workcomp_{2:\proc}+\ldots+\workcomp_{\proc:\proc},
\end{align}
where the second expression is simply the right-hand side of the first expression written in terms of the corresponding order statistics.
We note that under our model the time spent by worker $i$ in performing $\workcomp_i$ computations is $\worktime_i=X_i+\shifttime\workcomp_i$ where $X_i$ denotes setup/initial delay and $\shifttime$ is a constant denoting the time taken to perform a single computation. Thus $\workcomp_{1:\proc}$ corresponds to the worker that performs the least number of computations which is also the worker with the largest value of setup time i.e $X_{\proc:\proc}$ since all workers stop computing at the same time ( $\runtime_{\text{MDS}}$). Thus for a given $C_0$, the tail of the total number of computations performed in the MDS Coded strategy is given by
\begin{align}
\Pr\left( \numcomp_{\text{MDS}}\leq\frac{\rows\proc}{\mdsnum}-C_0 \right)&=\Pr\left( \sum_{i=1}^{\proc}\workcomp_{i:\proc}\leq\frac{\rows\proc}{\mdsnum}-C_0 \right)\\
&=\Pr\left( \sum_{i=1}^{\proc-\mdsnum}\workcomp_{i:\proc}+\frac{\rows}{\mdsnum}\times\mdsnum\leq\frac{\rows\proc}{\mdsnum}-C_0 \right)\label{eqn:topk}\\
&=\Pr\left( \sum_{i=1}^{\proc-\mdsnum}\workcomp_{i:\proc}\leq\frac{\rows\left( \proc-\mdsnum \right)}{\mdsnum}-C_0 \right)\\
&\leq\Pr\left( \left( \proc-\mdsnum \right)\workcomp_{1:\proc}\leq\frac{\rows\left( \proc-\mdsnum \right)}{\mdsnum}-C_0 \right)\label{eqn:mdsstocdom1}\\
&=\Pr\left( \workcomp_{1:\proc}\leq\frac{\rows}{\mdsnum}-\frac{C_0}{\proc-\mdsnum} \right)
\end{align}
where \eqref{eqn:topk} follows from the fact that the fastest $\mdsnum$ workers correspond to $\workcomp_{\proc-\mdsnum+1:\proc},\workcomp_{\proc-\mdsnum+2:\proc},\ldots,\workcomp_{\proc:\proc}$ and must perform all the tasks assigned to them i.e. $\rows/\mdsnum$ computations each, while \eqref{eqn:mdsstocdom1} follows from the fact that $\workcomp_{2:\proc},\ldots,\workcomp_{\proc:\proc}$ are always larger than $\workcomp_{1:\proc}$ by definition.

At this point we note that the worker which performs $\workcomp_{1:\proc}$ computations has setup time $X_{\proc:\proc}$. There can be two possibilities -- either $\runtime_{\text{MDS}} > X_{\proc:\proc}$, or $\runtime_{\text{MDS}} \leq X_{\proc:\proc}$. If $\runtime_{\text{MDS}} > X_{\proc:\proc}$ then
\begin{align}
\runtime_{\text{MDS}} \leq X_{\proc:\proc}+\shifttime\left( \workcomp_{1:\proc}+1 \right)
\label{eqn:Tmdsbound}
\end{align}
where the added $1$ accounts for the edge effect of partial computations at the nodes.
If $\runtime_{\text{MDS}} \leq X_{\proc:\proc}$ then also the upper bound \eqref{eqn:Tmdsbound} holds. Thus overall (by rearranging terms in \eqref{eqn:Tmdsbound}) we obtain,
\begin{align}
\workcomp_{1:\proc}\geq\frac{\runtime_{\text{MDS}}-X_{\proc:\proc}}{\shifttime}-1.
\end{align}
Thus we can write
\begin{align}
\Pr\left( \numcomp_{\text{MDS}}\leq\frac{\rows\proc}{\mdsnum}-C_0 \right)&\leq\Pr\left( \frac{\runtime_{\text{MDS}}-X_{\proc:\proc}}{\shifttime}-1\leq\frac{\rows}{\mdsnum}-\frac{C_0}{\proc-\mdsnum} \right)\\
&=\Pr\left( X_{\proc:\proc}-X_{\mdsnum:\proc}\geq\frac{\shifttime C_0}{\proc-\mdsnum}-\shifttime \right)\label{eqn:mdsrearrange}\\
&=\Pr\left( \sum_{l=\mdsnum}^{\proc-1}\left( X_{l+1:\proc}-X_{l:\proc} \right)\geq\frac{\shifttime C_0}{\proc-\mdsnum}-\shifttime \right),
\end{align}
where \eqref{eqn:mdsrearrange} follows from the fact that $\runtime_{\text{MDS}}=X_{\mdsnum:\proc}+\shifttime\rows/\mdsnum$. 
If $X_i \sim \exp(\exprate)$ we can use the result on the difference of consecutive order statistics of exponential random variables from \Cref{sec:ordexp} to simplify the above expression further,
\begin{align}
\Pr\left( \numcomp_{\text{MDS}}\leq\frac{\rows\proc}{\mdsnum}-C_0 \right)&\leq\Pr \left( \sum_{l=\mdsnum}^{\proc-1}U_{l}\geq\frac{\shifttime C_0}{\proc-\mdsnum}-\shifttime  \right)\\
&\leq\Pr \left(  \left( \proc-\mdsnum \right)U_{\proc-1}\geq\frac{\shifttime C_0}{\proc-\mdsnum}-\shifttime  \right)\label{eqn:mdsstocdom2}\\
&=\Pr \left(  U_{\proc-1}\geq\frac{\shifttime C_0}{\left( \proc-\mdsnum \right)^2}-\frac{\shifttime}{\proc-\mdsnum}  \right)\\
&=\exp\left( -\exprate\left( \frac{\shifttime C_0}{\left( \proc-\mdsnum \right)^2}-\frac{\shifttime}{\proc-\mdsnum} \right) \right)\label{eqn:mdscompfinal}
\end{align}
where \eqref{eqn:mdsstocdom2} is obtained from the fact that $\Pr( U_{\proc-1}\geq u)\geq\Pr( U_{l}\geq u) \text{ for }l=\mdsnum,\ldots,\proc-1$ for any $u$ since $U_{l}\sim\exp((\proc - l)\exprate)$. Lastly \eqref{eqn:mdscompfinal} is obtained from the expression for the tail distribution of an exponential random variable.
\end{proof}
\subsection{Replication Strategy}
\begin{proof}[Proof of \Cref{thm:replatency}]
In the $\numrep-$replication strategy each submatrix $\mat_{i}$ is replicated at $\numrep$ workers and we wait for the fastest of these $\numrep$ workers. Without loss of generality, we assume that submatrix $\mat_{1}$ is stored at workers $1,2,\ldots,\numrep$, submatrix $\mat_{2}$ is stored at workers $\numrep+1,\numrep+2,\ldots,2*\numrep$ and so on. More generally submatrix $\mat_{i}$ is stored at workers $( i-1)\numrep+1,\ldots,i\numrep$. Thus the time taken to compute the product $\mat_{i}\vect$ is given by
\begin{align}
V_i&=\min\left( \worktime_{( i-1)\numrep+1},\worktime_{( i-1)\numrep+2},\ldots,\worktime_{i\numrep} \right)\\
&=\min\left(X_{( i-1)\numrep+1}+\shifttime\workcomp_{( i-1)\numrep+1},\ldots,X_{i\numrep}+\shifttime\workcomp_{i\numrep} \right)\\
&=\min\left(X_{( i-1)\numrep+1},\ldots,X_{i\numrep} \right)+\frac{\shifttime\rows\numrep}{\proc}\\
&=W_i+\frac{\shifttime\rows\numrep}{\proc}\label{eqn:rep_0}.
\end{align}
This is because the fastest of the $\numrep$ workers that store $\mat_{i}$ corresponds to $\min( X_{( i-1)\numrep+1},\ldots,X_{i\numrep})$ and this worker must perform $\frac{\rows\numrep}{\proc}$ computations to compute the product $\mat_{i}\vect$.

The latency $\runtime_{\text{rep}}$ is the time at which the product $\mat_{i}\vect$ is computed for all $i=1,\ldots,\proc/\numrep$ since $\mat$ is split into $\proc/\numrep$ submatrices. Thus
\begin{align}
\runtime_{\text{rep}} &= \max\left( V_1,V_2,\ldots,V_{\proc/\numrep} \right), \label{eqn:rep_1}\\
&=\max\left( W_1,W_2,\ldots,W_{\proc/\numrep} \right) + \frac{\shifttime\rows\numrep}{\proc}\label{eqn:rep_2},
\end{align}
\end{proof}
\begin{proof}[Proof of \Cref{coro:replatency}]
If $X_j \sim \exp(\exprate)$ then observe that $W_i=\min( X_{( i-1)\numrep+1},\ldots,X_{i\numrep})$ is an $\exp( \numrep\exprate)$ random variable since it is the minimum of $\numrep$ $\exp( \exprate)$ random variables. Thus taking expectation on both sides of \Cref{{eqn:rep_2}},
\begin{align}
\mathbb{E}[\runtime_{\text{rep}}] &= \frac{\shifttime\rows\numrep}{\proc}+\mathbb{E}[\max\left( W_1,W_2,\ldots,W_{\proc/\numrep} \right)], \\
& = \frac{\shifttime\rows\numrep}{\proc} + \frac{1}{\numrep\mu}H_{\proc/\numrep}, \label{eqn:rep_4} \\
&\simeq \frac{\shifttime\rows\numrep}{\proc} + \frac{1}{\numrep\mu}\log\frac{\proc}{\numrep}  \label{eqn:rep_5},
\end{align}
where \eqref{eqn:rep_4} and \eqref{eqn:rep_5} follow from \eqref{eqn:harmonic} and \eqref{eqn:harmonic_def}.
\end{proof}
\begin{proof}[Proof of \Cref{thm:repcomptail}]
As per our model, we represent the number of computations at worker $i$ by the random variable $\workcomp_i$. We also use the random variable $\numcomp_{\text{rep}}$ to denote the total number of computations performed by all $\proc$ workers until $\runtime_{\text{rep}}$, which is the time when the master collects enough computations to be able to recover the matrix-vector product $\res=\mat\vect$. Thus
\begin{align}
\numcomp_{\text{rep}}&=\workcomp_1+\workcomp_2+\ldots+\workcomp_\proc\\
&=\sum_{i=1}^{\proc/\numrep}\sum_{j=1}^{\numrep}\workcomp_{(i-1)\numrep+j},
\end{align}
where the term inside the summation in the second expression represents the number of computations performed by each worker that store a copy of the submatrix $\mat_{i}$ (for a given $i$). In what follows, we use the shorthand notation $D_{j}^{i} = \workcomp_{(i-1)\numrep+j}$ and use $D_{j:\numrep}^{i}$ to denote the order statistics of $D_{1}^{i},\ldots,D_{\numrep}^{i}$. Rewriting the above expression in terms of the order statistics we get,
\begin{align}
\numcomp_{\text{rep}}&=\sum_{i=1}^{\proc/\numrep}\sum_{j=1}^{\numrep}D_{j:\numrep}^{i},
\end{align}
and the tail bound,
\begin{align}
\Pr(\numcomp_{\text{rep}}\leq\rows\numrep-C_0)&=\Pr\left(\sum_{i=1}^{\proc/\numrep}\sum_{j=1}^{\numrep}D_{j:\numrep}^{i}\leq\rows\numrep-C_0\right)\\
&=\Pr\left(\sum_{i=1}^{\proc/\numrep}\sum_{j=1}^{\numrep-1}D_{j:\numrep}^{i}\leq\rows(\numrep-1)-C_0\right)\label{eqn:grouptop}\\
&\leq\Pr\left((\numrep-1)\sum_{i=1}^{\proc/\numrep}D_{1:\numrep}^{i}\leq\rows(\numrep-1)-C_0\right)\label{eqn:repstocdom1}\\
&=\Pr\left(\sum_{i=1}^{\proc/\numrep}D_{1:\numrep}^{i}\leq\rows-\frac{C_0}{\numrep-1}\right)
\end{align}
where \eqref{eqn:grouptop} follows from the fact that for any given submatrix $\mat_i, i=1,\ldots,\proc/\numrep$, the fastest worker that stores a copy of that submatrix, which corresponds to $D_{\numrep:\numrep}^{i}$ (fastest worker performs the most computations) must perform all the tasks assigned to it i.e. $\rows\numrep/\proc$ computations each, while \eqref{eqn:repstocdom1} follows from the fact that $D_{2:\numrep}^{i},\ldots,D_{\numrep:\numrep}^{i}$ are always larger than $D_{1:\numrep}^{i}$ by definition.

At this point we introduce the shorthand notation $V_{j}^{i} = X_{(i-1)\numrep+j}$ for the setup time of the worker that stores the $j^{\text{th}}$ copy of submatrix $\mat_{i}$ and note that the worker which performs $D_{1:\numrep}^{i}$ computations has setup time $V_{\numrep:\numrep}^{i}$ ($V_{j:\numrep}^{i}$ are the order statistics of $V_{1}^{i},\ldots,V_{\numrep}^{i}$). There can be two possibilities -- either $\runtime_{\text{rep}} > V_{\numrep:\numrep}^{i}$, or $\runtime_{\text{rep}} \leq V_{\numrep:\numrep}^{i}$. If $\runtime_{\text{rep}} > V_{\numrep:\numrep}^{i}$ then
\begin{align}
\runtime_{\text{rep}} \leq V_{\numrep:\numrep}^{i}+\shifttime(D_{1:\numrep}^{i}+1)
\label{eqn:Trepbound}
\end{align}
where the added $1$ accounts for the edge effect of partial computations at the nodes.
If $\runtime_{\text{rep}} \leq V_{\numrep:\numrep}^{i}$ then also the upper bound \eqref{eqn:Trepbound} holds. Thus overall (by rearranging terms in \eqref{eqn:Trepbound}) we obtain,
\begin{align}
D_{1:\numrep}^{i}\geq\frac{\runtime_{\text{rep}}-V_{\numrep:\numrep}^{i}}{\shifttime}-1
\end{align}
Thus we can write
\begin{align}
\Pr(\numcomp_{\text{rep}}\leq\rows\numrep-C_0)&\leq\Pr\left(\sum_{i=1}^{\proc/\numrep}\left(\frac{\runtime_{\text{rep}}-V_{\numrep:\numrep}^{i}}{\shifttime}-1\right)\leq\rows-\frac{C_0}{\numrep-1}\right)\\
&=\Pr\left(\sum_{i=1}^{\proc/\numrep}(V_{\numrep:\numrep}^{i}-W_{\text{rep}})\geq\frac{\shifttime C_0}{\numrep-1}-\frac{\shifttime\proc}{\numrep}\right)\label{eqn:reprearrange}
\end{align}
where $W_{\text{rep}} = \max_{i}V_{1:\numrep}^{i} = \max_{1\leq i\leq \proc/\numrep}\min_{1\leq j\leq \numrep}X_{(i-1)\numrep+j}$ and \eqref{eqn:reprearrange} follows from the fact that $\runtime_{\text{rep}}=\max_{1\leq i\leq \proc/\numrep}\min_{1\leq j\leq \numrep}X_{(i-1)\numrep+j}+\shifttime\rows\numrep/\proc$. 
From our definition of $W_{\text{rep}}$ we see that,
\begin{align}
V_{\numrep:\numrep}^{i}-W_{\text{rep}} \leq V_{\numrep:\numrep}^{i}-V_{1:\numrep}^{i}
\end{align}
and the consequent stochastic dominance can be used to get an upper bound on \eqref{eqn:reprearrange} as,
\begin{align}
\Pr(\numcomp_{\text{rep}}\leq\rows\numrep-C_0)&\leq\Pr\left(\sum_{i=1}^{\proc/\numrep}(V_{\numrep:\numrep}^{i}-V_{1:\numrep}^{i})\geq\frac{\shifttime C_0}{\numrep-1}-\frac{\shifttime\proc}{\numrep}\right)\\
&=\Pr\left(\sum_{i=1}^{\proc/\numrep}\sum_{j=1}^{r-1}(V_{j+1:\numrep}^{i}-V_{j:\numrep}^{i})\geq\frac{\shifttime C_0}{\numrep-1}-\frac{\shifttime\proc}{\numrep}\right)
\end{align}
If $X_i \sim \exp(\exprate)$ we can use the result from \Cref{sec:ordexp} on the difference of consecutive order statistics of exponential random variables to simplify the above expression further (since $V_{j}^{i} = X_{(i-1)\numrep+j}$ are also exponentially distributed and thus $U_{j}^{i} = (V_{j+1:\numrep}^{i}-V_{j:\numrep}^{i})\sim\exp( ( \numrep-j)\exprate)$),
\begin{align}
\Pr(\numcomp_{\text{rep}}\leq\rows\numrep-C_0)&\leq\Pr\left(\sum_{i=1}^{\proc/\numrep}\sum_{j=1}^{r-1}U_{j}^{i}\geq\frac{\shifttime C_0}{\numrep-1}-\frac{\shifttime\proc}{\numrep}\right)\\
&\leq\Pr\left((\numrep-1)\sum_{i=1}^{\proc/\numrep}U_{\numrep-1}^{i}\geq\frac{\shifttime C_0}{\numrep-1}-\frac{\shifttime\proc}{\numrep}\right)\label{eqn:repstocdom2}\\
&=\Pr \left(\sum_{i=1}^{\proc/\numrep}U_{\numrep-1}^{i}\geq\frac{\shifttime C_0}{(\numrep-1)^{2}}-\frac{\shifttime\proc}{\numrep(\numrep-1)}\right)\\
&=\sum _{{i=0}}^{{\proc/\numrep-1}}{\frac  {1}{i!}}\exp({-\exprate\theta})(\exprate\theta)^{i}.
&\label{eqn:repcompfinal}
\end{align}
where \eqref{eqn:repstocdom2} is obtained from the fact that $\Pr( U_{\numrep-1}^{i}\geq u)\geq\Pr( U_{j}^{i}\geq u) \text{ for }j = 1,\ldots,\numrep-2$ for any $u$ since $U_{j}\sim\exp((\numrep - j)\exprate)$. Lastly \eqref{eqn:repcompfinal} is obtained from the expression for the tail distribution of an Erlang random variable which is the sum of $\proc/\numrep$ exponential random variables with rate $\exprate$ and
\begin{align}
\theta = \frac{\shifttime C_0}{(\numrep-1)^{2}}-\frac{\shifttime\proc}{\numrep(\numrep-1)}
\end{align}
\end{proof}

\section{Theoretical Results for Queueing Analysis}
\label{sec:Proofs_Queueing}
\begin{proof}[Proof of \Cref{thm:ltmultiple}]
For large $\encrows$ i.e. $\strag = \encrows/\rows \rightarrow \infty$, under the proposed rateless coded strategy we just need to wait for $\randdecrows$ computations to be performed by the workers in total. Hence the $\proc$ workers can be treated as a single server with service time equal to $\runtime_{\text{LT}} $. According to the Pollaczek Khinchine formula \cite{harchol2013_performance} for Poisson arrivals with rate $\lambda$, the expected total processing time (time in queue + service time) for a job in such a queue is given by 
\begin{align}
\mathbb{E}[\queuetime_{\text{LT}}] &= \mathbb{E}[\runtime_{\text{LT}} ] + \frac{\lambda\mathbb{E}[(\runtime_{\text{LT}} )^2]}{2(1-\lambda\mathbb{E}[\runtime_{\text{LT}} ])}
\end{align}
Recall that for $\strag = \encrows/\rows \rightarrow \infty$ the LT and Ideal schemes are identical by definition. Therefore we can use the bounds derived for $\runtime_{\text{ideal}}$ in \Cref{lem:ideallatency} to bound $\runtime_{\text{LT}}$ as,
\begin{align}
 \runtime_{\text{LT}}  &\leq \shifttime\left( \frac{\randdecrows}{\proc}+1\right) + \frac{1}{\proc}\sum_{i=1}^{\proc} X_i \\
\runtime_{\text{LT}}  &\geq \shifttime \frac{\randdecrows}{\proc} + X_{1:\proc}
\end{align}
 To analyze the second moment $\mathbb{E}[(\runtime_{\text{LT}} )^2]$, let $\shifttime' = \shifttime/\proc$ and let $\bar{X} = (1/\proc)\sum_{i=1}^{\proc} X_i$. Then,
\begin{align}
( \runtime_{\text{LT}} )^2 &\leq (\shifttime + \shifttime'\randdecrows)^2 + \bar{X}^2 + 2(\shifttime + \shifttime'\randdecrows)\bar{X} \\
(\runtime_{\text{LT}} )^2 &\geq (\shifttime'\randdecrows)^2 + \bar{X}^2 + 2\shifttime'\randdecrows\bar{X}
\end{align}
Taking expectations and noting that $\randdecrows$ and $\bar{X}$ are independent,
\begin{align}
\mathbb{E}[(\runtime_{\text{LT}} )^2] &\leq \shifttime^2 + 2\shifttime\shifttime'\decrows +(\shifttime')^2\mathbb{E}[(\randdecrows)^2] + \mathbb{E}[\bar{X}^2] + \nonumber \\
& \hspace{1cm} 2(\shifttime + \shifttime'\decrows)\mathbb{E}[\bar{X}] \\
\mathbb{E}[(\runtime_{\text{LT}} )^2] &\geq (\shifttime')^2\mathbb{E}[(\randdecrows)^2] + \mathbb{E}[\bar{X}^2] + 2\shifttime'\decrows\mathbb{E}[\bar{X}]
\end{align}
\end{proof}

\begin{lem}[Multiple jobs with MDS Coding]
\label{lem:mdsmultiple}
The expected latency of the MDS coded scheme $\queuetime_{\text{MDS}}$ when a stream of vectors $\vect_1,\vect_2,\ldots$ need to be multiplied with the same matrix $\mat$ (assuming Poisson arrivals with rate $\lambda$ for the vectors) is given by
\begin{align}
\mathbb{E}[\queuetime_{\text{MDS}}] &\leq \mathbb{E}[\worktime_{\mdsnum:\proc}] + \frac{\lambda((\mathbb{E}[\worktime_{\mdsnum:\proc}])^2 + \text{Var}[\worktime_{\mdsnum:\proc}])}{2(1-\lambda\mathbb{E}[\worktime_{\mdsnum:\proc}])}\\
\mathbb{E}[\queuetime_{\text{MDS}}] &\geq \mathbb{E}[\worktime_{\mdsnum:\proc}] + \frac{\lambda((\mathbb{E}[\worktime_{1:\proc}])^2 + \text{Var}[\worktime_{1:\proc}])}{2(1-\lambda\mathbb{E}[\worktime_{1:\proc}])}
\end{align}
where $\worktime_i = X_i + \shifttime\rows/\mdsnum$ is the service time at worker $i$, $i=1,\ldots,\proc$ for the MDS coded case.
\end{lem}
\begin{proof}[Proof of \Cref{lem:mdsmultiple}]
When a stream of incoming vectors $\vect_1,\vect_2,\ldots$ need to be multiplied with the matrix $\mat$ over $\proc$ workers, the resulting system is a $(\proc,\mdsnum)$ fork-join queue since the task of computing matrix vector products of the form $\mat\vect$ is forked to the $\proc$ workers and we need to wait for $\mdsnum$ workers to complete the tasks assigned to them. The expression for latency $\queuetime_{\text{MDS}}$ follows from Theorem 4 of \cite{joshi2017efficient} which gives bounds on the latency of $(\proc,\mdsnum)$ fork-join queues assuming Poisson arrivals.
\end{proof}
\begin{lem}[Multiple jobs with Replication]
\label{lem:repmultiple}
The expected latency of the replication scheme $\queuetime_{\text{rep}}$ when a stream of vectors $\vect_1,\vect_2,\ldots$ need to be multiplied with the same matrix $\mat$ (assuming Poisson arrivals with rate $\lambda$ for the vectors) is given by
\begin{align}
\mathbb{E}[\queuetime_{\text{rep}}] &\leq \mathbb{E}[V_{\proc/\numrep:\proc/\numrep}] + \frac{\lambda((\mathbb{E}[V_{\proc/\numrep:\proc/\numrep}])^2 + \text{Var}[V_{\proc/\numrep:\proc/\numrep}])}{2(1-\lambda\mathbb{E}[V_{\proc/\numrep:\proc/\numrep}])}\\
\mathbb{E}[\queuetime_{\text{rep}}] &\geq \mathbb{E}[V_{\proc/\numrep:\proc/\numrep}] + \frac{\lambda((\mathbb{E}[V_{1:\proc/\numrep}])^2 + \text{Var}[V_{1:\proc/\numrep}])}{2(1-\lambda\mathbb{E}[V_{1:\proc/\numrep}])}
\end{align}
where $W_i = \min_{i}X_{(i-1)\numrep+j} + \shifttime\rows\numrep/\proc$, $i=1,\ldots,\proc/\numrep$.
\end{lem}
\begin{proof}[Proof of \Cref{lem:repmultiple}]
When a stream of incoming vectors $\vect_1,\vect_2,\ldots$ need to be multiplied with the matrix $\mat$ over $\proc$ workers, the resulting system is a $(\proc/\numrep,\proc/\numrep)$ fork-join queue. This is because each submatrix $\mat_{1},\ldots,\mat_{\proc/\numrep}$ is replicated at $\numrep$ workers and we need to wait for the fastest worker for each submatrix. Thus $i^{\text{th}}$ group of $\numrep$ workers has an effective service time of $V_i$ as defined in \Cref{eqn:rep_0}. Since we need to wait for all $\proc/\numrep$ groups of $\numrep$ workers in this fashion it is equivalent to a $(\proc/\numrep,\proc/\numrep)$ fork-join queue where the $i^{\text{th}}$ node has service time $V_{i}$. Once again the expression for latency $\queuetime_{\text{rep}}$ follows from Theorem 4 of \cite{joshi2017efficient} assuming Poisson arrivals.
\end{proof}
\section{Additional Theoretical Results}
\label{sec:additional_theory}

\begin{figure*}[t]
  \centering
  \subfloat[Latency Tail]{\includegraphics[width=0.33\linewidth]{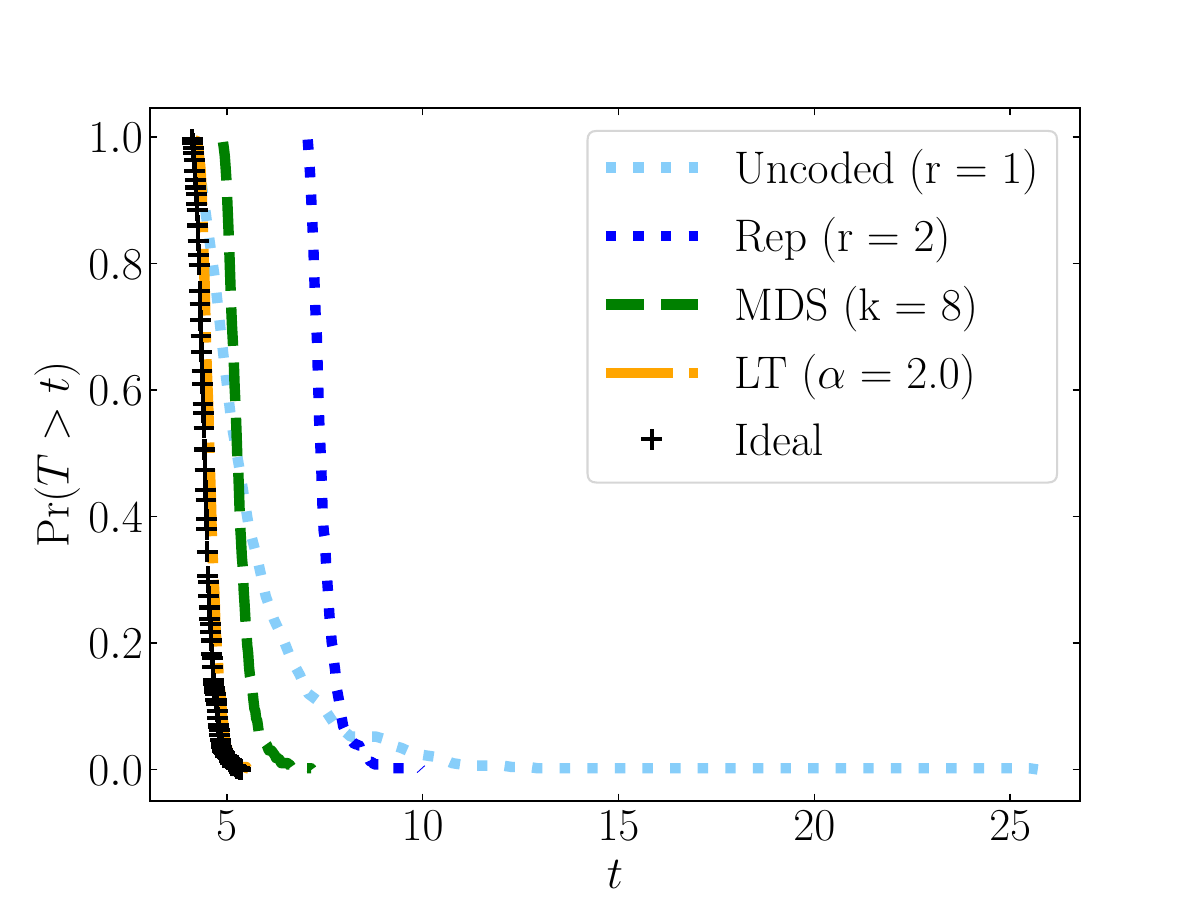}\label{fig:latsim_par}}
  \subfloat[Computation Tail]{\includegraphics[width=0.33\linewidth]{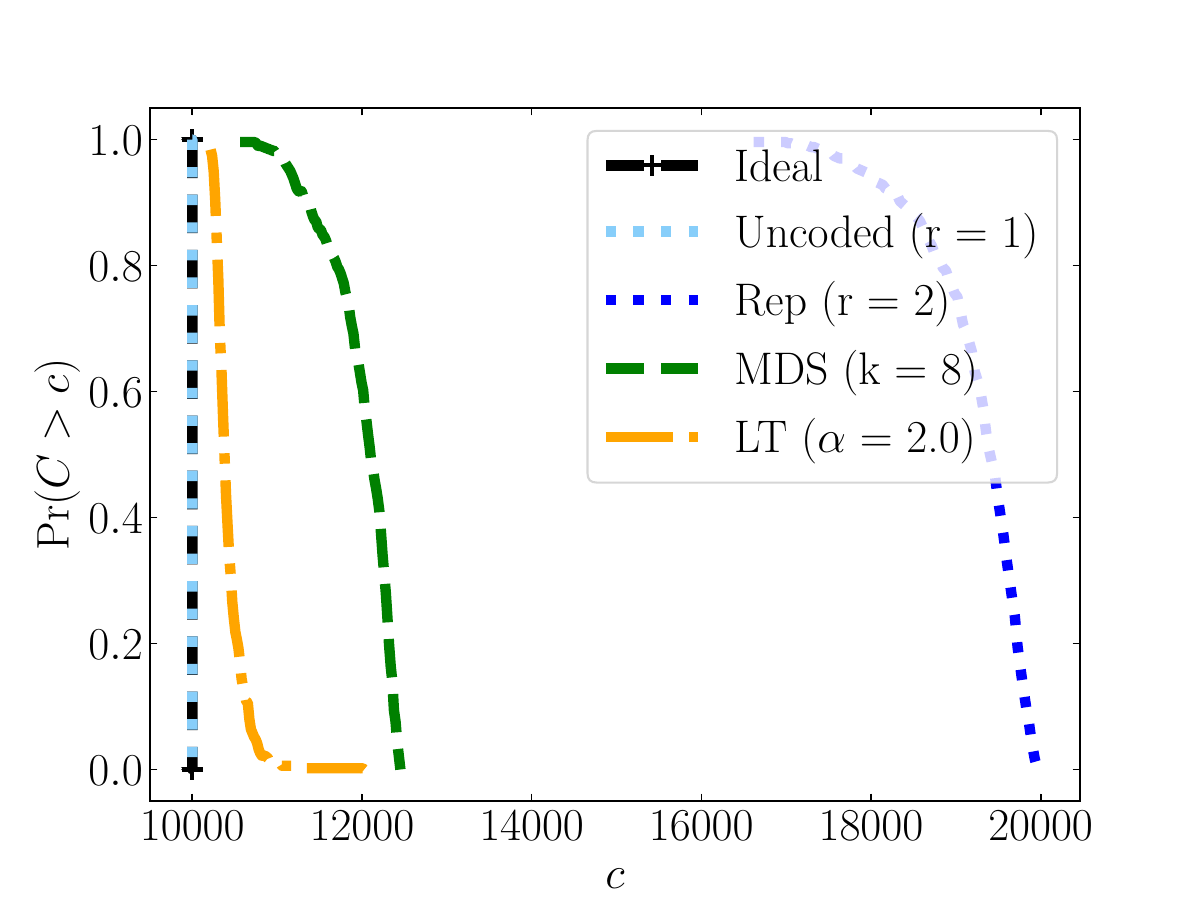}\label{fig:compsim_par}}
  \subfloat[Mean Response Time]{\includegraphics[width=0.33\linewidth]{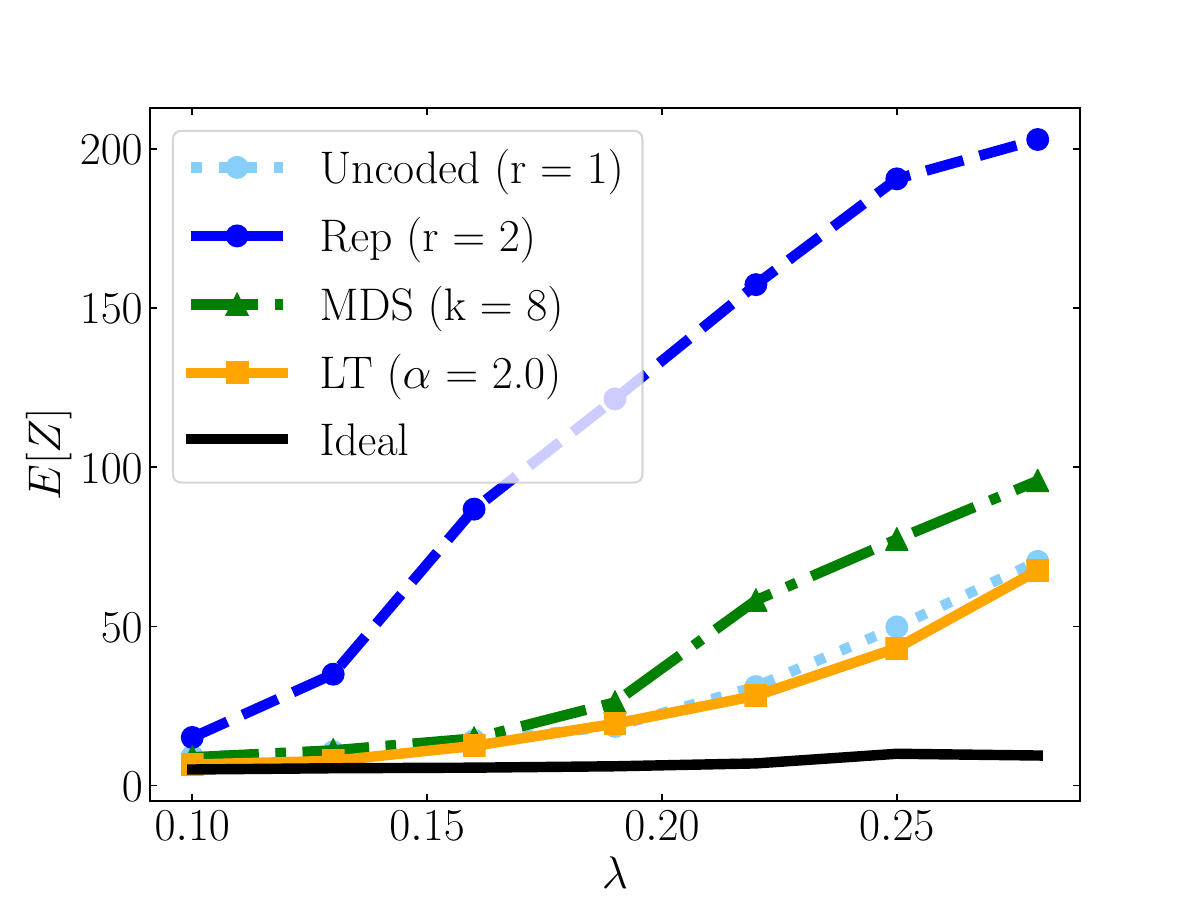}\label{fig:queueing_sim_par}}
  \caption{The tail probability of the latency is the highest for the replication schemes. MDS codes perform better in terms of latency but they perform a large number of redundant computations. The latency tail of LT codes is the minimum among all the schemes. Moreover the LT coded schemes performs significantly fewer redundant computations than MDS Codes or replication. When there are multiple jobs in the queue, the mean response time is least for the LT Coded setting under all values of arrival rate $\lambda$. All simulations are performed with $\rows = 10000$ matrix rows, $\proc = 10$ worker nodes, $\shifttime = 0.001$, and a Pareto (1,3) distribution on the initial delays $X_i$, $i=1,\ldots,\proc$}
\label{fig:simulations_par}
\end{figure*}

The following theorem compares the latency of the MDS coded, and ideal load balancing strategies. It indicates that $\runtime_{\text{MDS}}$ approaches $\runtime_{\text{ideal}}$ only when the fastest $\mdsnum$ workers start computing at approximately the same time while the stragglers do not start until much later. In this rare situation neglecting partial computations by stragglers does not adversely impact latency performance. However in all other cases discarding work done by stragglers causes $\runtime_{\text{MDS}}$ to be larger than $\runtime_{\text{ideal}}$.
\begin{thm}[MDS v/s Ideal]
\label{thm:probmdslatency}
The latency of the MDS coded strategy, $\runtime_{\text{MDS}}$ is larger than the latency of the ideal strategy $\runtime_{\text{ideal}}$ with a high probability. Specifically $\Pr(\runtime_{\text{MDS}} > \runtime_{\text{ideal}}) = 1 - \delta_{\text{MDS}}$ where
\begin{align}
\delta_{\text{MDS}} = \Pr( X_{\mdsnum:\proc} - X_{1:\proc} \leq \shifttime, X_{\mdsnum+1:\proc} - X_{\mdsnum:\proc} > \shifttime\left(\frac{\rows}{\mdsnum} - 1\right))
\end{align}
\end{thm}
\begin{proof}
To compare the latency of the MDS coded strategy to that of the ideal scheme we note that the latency $\runtime_{\text{ideal}}$ of the ideal scheme, is the earliest time when $\sum_{i=1}^{p}\workcomp_i = \rows$, as illustrated in \Cref{fig:delay_model}. We note that, in this case it is not necessary that each worker has completed at least $1$ computation. Specifically, if $\runtime_{\text{ideal}} - X_i \leq \shifttime$ for any $i$ then it means that worker $i$ has not performed even a single computation in the time that the system as a whole has completed $\rows$ computations ( owing to the large initial delay $X_i$). Therefore we define
\begin{align}
\mathcal{W}_{\text{ideal}}:=\{i:\runtime_{\text{ideal}} - X_i \geq \shifttime\}
\end{align}
Here $\mathcal{W}_{\text{ideal}}$ is the set of workers for which $\workcomp_i > 0$ in time up to $\runtime_{\text{ideal}}$. 
We also note that
\begin{align}
\label{eq:idealupper}
\runtime_{\text{ideal}} &< X_i + \shifttime( \workcomp_i+1),\ \text{ for all } i = 1,\ldots,\proc
\end{align}
This is because at time $\runtime_{\text{ideal}}$ each of the workers $1,\ldots,\proc$, have completed $\workcomp_1,\ldots,\workcomp_\proc$ row-vector product tasks respectively, but they may have partially completed the next task. The $1$ added to each $\workcomp_i$ accounts for this edge effect, which is also illustrated in \Cref{fig:delay_model}.

We will compare $\runtime_{\text{MDS}}$ and $\runtime_{\text{ideal}}$ for the following three cases assuming the same realizations of initial delay $X_i$, $i = 1,\ldots,\proc$ for both schemes (we will also assume without loss of generality that $X_1<X_2<\ldots<X_\proc$ i.e $X_i = X_{i:\proc} \forall i$):
\begin{itemize}
\item Case 1: If node $\mdsnum \not\in \mathcal{W}_{\text{ideal}}$, $\workcomp_k = 0$ in the ideal scheme. Thus if we use the ideal scheme the latency is $\runtime_{\text{ideal}} < X_{\mdsnum:\proc} + \shifttime$ (from \Cref{eq:idealupper}) whereas if we use the MDS coded scheme the latency is $\runtime_{\text{MDS}} = X_{\mdsnum:\proc} + \shifttime\rows/\mdsnum > \runtime_{\text{ideal}}$. 
%
\item Case 2: If node $\mdsnum \in \mathcal{W}_{\text{ideal}}$ but $X_\mdsnum \neq \max_{i\in\mathcal{W}_{\text{ideal}}}X_i$ then nodes $1,\ldots,\mdsnum-1$ must also lie in $\mathcal{W}_{\text{ideal}}$ (since $X_1<X_2<\ldots<X_\mdsnum$) and at least one of nodes $\mdsnum+1,\ldots,\proc$ must also lie in $\mathcal{W}_{\text{ideal}}$ (since $X_\mdsnum \neq \max_{i\in\mathcal{W}_{\text{ideal}}}X_i$. If we use the ideal scheme and assume that node $\mdsnum$ performs $\rows/\mdsnum$ computations then nodes $1,\ldots,\mdsnum-1$ must perform $\rows/\mdsnum$ or more computations each (since since $X_1<X_2<\ldots<X_\mdsnum$). Together nodes $1,\ldots,\proc$ perform at least  $\mdsnum \times \rows/\mdsnum = \rows$ computations and since at least one of nodes $\mdsnum+1,\ldots,\proc$ must also lie in $\mathcal{W}_{\text{ideal}}$, the total number of computations performed is greater than $\rows$ which is not possible since the number of computations performed in the ideal scheme is $\rows$ by definition. Thus node $\mdsnum$ can perform at most $(\rows/\mdsnum - 1)$ computations and thus from $\Cref{eq:idealupper}$ we have that $\runtime_{\text{ideal}} \leq X_{\mdsnum:\proc} + \shifttime(\rows/\mdsnum - 1) < \runtime_{\text{MDS}}$. 
%
\item Case 3: If node $\mdsnum \in \mathcal{W}_{\text{ideal}}$ and $X_\mdsnum = \max_{i\in\mathcal{W}_{\text{ideal}}}X_i$ then there are exactly $\mdsnum$ workers in $\mathcal{W}_{\text{ideal}}$. If $X_\mdsnum - X_1 > \shifttime$ then node $1$ performs at least $1$ more computation than node $\mdsnum$ since it takes time $\shifttime$ for a node to perform a computation. In that case node $\mdsnum$ performs fewer than $\rows/\mdsnum$ computations since total number of computations performed by nodes $1,\ldots,\mdsnum$ must be $\rows$. Thus in this case $\runtime_{\text{ideal}} < X_\mdsnum + \shifttime\rows/\mdsnum = \runtime_{\text{MDS}}$ (from \Cref{eq:idealupper}). Likewise for nodes $2,\ldots,\mdsnum-1$ as well. Thus $\runtime_{\text{ideal}} = \runtime_{\text{MDS}}$ only when $X_{1},\ldots,X_{\mdsnum-1} \geq X_{\mdsnum} - \shifttime$ 
\end{itemize}
Thus overall it is only in Case 3 above under specific circumstances that we can have $\runtime_{\text{ideal}} = \runtime_{\text{MDS}}$. In all other cases $\runtime_{\text{ideal}} < \runtime_{\text{MDS}}$. Hence probabilistically $\Pr(\runtime_{\text{MDS}} > \runtime_{\text{ideal}}) = 1 - \delta_{\text{MDS}}$ where
\begin{align}
\delta_{\text{MDS}} = \Pr( X_{\mdsnum:\proc} - X_{1:\proc} \leq \shifttime, X_{\mdsnum+1:\proc} - X_{\mdsnum:\proc} > \shifttime\left(\frac{\rows}{\mdsnum} - 1\right))
\end{align}
where the first condition is equivalent to $X_{1:\proc},\ldots,X_{\mdsnum-1:\proc} \leq X_{\mdsnum:\proc}$ by the definition of the order statistics and the second condition ensures that $X_{\mdsnum:\proc} = \max_{i\in\mathcal{W}_{\text{ideal}}}X_i$ (nodes $\mdsnum+1,\ldots,\proc$ do not start until node $\mdsnum$ has completed $\rows/\mdsnum$ computations).
\end{proof}

The following theorem shows that except in rare cases $\runtime_{\text{Rep}}$ is larger than $\runtime_{\text{ideal}}$.
\begin{thm}[Replication v/s Ideal]
\label{thm:probreplatency}
The latency of the replication coded strategy, $\runtime_{\text{rep}}$ is larger than the latency of the ideal strategy $\runtime_{\text{ideal}}$ with a high probability. Specifically $\Pr(\runtime_{\text{rep}} > \runtime_{\text{ideal}}) = 1 - \delta_{\text{rep}}$ where
\begin{align}
\delta_{\text{rep}} = \Pr\left(\min_{1\leq i\leq \proc/\numrep}X_{2:r}^{(i)} - \max_{1\leq i\leq \proc/\numrep}X_{1:r}^{(i)} > \shifttime\left(\frac{\rows\numrep}{\proc} - 1\right)\right)
\end{align}
and $X_{j}^{(i)} = X_{(i-1)\numrep+j}$.
\end{thm}

\begin{proof}
In the $\numrep-$replication scheme the matrix $\mat$ is split into $\proc/\numrep$ submatrices $\mat_{1},\ldots\,\mat_{\proc/\numrep}$. Each submatrix is replicated at $\numrep$ distinct workers and we wait for the fastest worker for each submatrix. There are $\proc/\numrep$ such groups of workers, with all workers in group $i$ computing the product $\mat_i\vect$.

Following similar arguments to the MDS coded case, we can conclude that $\runtime_{\text{rep}} = \runtime_{\text{ideal}}$ only when the workers in each group other than the fastest worker do not start computing until the fastest workers of \emph{all} groups have finished their computations. The fastest worker group $i$ is the one that takes time $V_i$ (in \Cref{eqn:rep_0}) to compute $\mat_{i}\vect$.  

This is because in any other scenario the workers in the replications scheme that are not the fastest in their respective groups perform redundant computations and hence some work is wasted as compared to the ideal scheme where some computations could have been transferred from the fastest workers to the slower workers in each group thus reducing the total time taken to compute $\rows$ row-vector products. However if the slower workers in the groups do not start their computations until after \emph{all} the fastest workers across \emph{all} groups complete their computations then the replication and ideal schemes are essentially identical. This is because in this case even in the ideal scheme no work could have been transferred from the fastest workers in each group to the slow workers. 

The following condition quantifies the above explanation,
\begin{align}
\min_{1\leq i\leq \proc/\numrep}X_{2:r}^{(i)} - \max_{1\leq i\leq \proc/\numrep}X_{1:r}^{(i)} > \shifttime\left(\frac{\rows\numrep}{\proc} - 1\right)
\end{align}
This is because $X_{1:r}^{(i)}+\shifttime(\rows\numrep/\proc - 1)$ is the time at which the fastest worker in group $i$ (with initial delay corresponding to $X_{1:r}^{(i)}$) completes $\rows\numrep/\proc - 1$ computations and starts working on the last $((\rows\numrep/\proc)^{\text{th}})$ computation. Thus even if the second slowest worker of the group (corresponding to $X_{2:r}^{(i)}$) and any other worker(s) starts after this time they cannot perform any computations before the fastest worker completes all the computations assigned to it. Thus no computations are transferred to the slower workers even in the ideal scheme in this case.

Here $\max_{1\leq i\leq \proc/\numrep}X_{1:r}^{(i)} + \shifttime\left(\frac{\rows\numrep}{\proc} - 1\right)$ is the time taken by the slowest of the fastest workers across all groups to complete $\rows\numrep/\proc - 1$ computations and $\min_{1\leq i\leq \proc/\numrep}X_{2:r}^{(i)}$ is the initial delay of the worker with the least initial delay among the second fastest workers of all groups.

Therefore $\Pr(\runtime_{\text{rep}} > \runtime_{\text{ideal}}) = 1 - \delta_{\text{rep}}$ where
\begin{align}
\delta_{\text{rep}} = \Pr\left(\min_{1\leq i\leq \proc/\numrep}X_{2:r}^{(i)} - \max_{1\leq i\leq \proc/\numrep}X_{1:r}^{(i)} > \shifttime\left(\frac{\rows\numrep}{\proc} - 1\right)\right)
\end{align}
and $X_{j}^{(i)} = X_{(i-1)\numrep+j}$
\end{proof}
\section{Additional Simulations and Experiments}
\label{sec:additional_simexp}
\textbf{Additional Simulations: } We simulate the MDS, replication, and LT-coded schemes under our delay model (\Cref{eq:delaymodel}) with initial delays $X_i$ distributed according to a Pareto (1,3) distribution. The results are summarised in \Cref{fig:simulations_par}. Once again we observe that LT coding ($\strag = 2.0$) clearly outperforms MDS coding ($\mdsnum = 8$) both in terms of latency (\Cref{fig:latsim_par}), number of redundant computations (\Cref{fig:compsim_par}), and mean response time averaged over $10$ trials with $100$ jobs per trial and Poisson $(\lambda)$ arrivals (\Cref{fig:queueing_sim_par}).

\begin{figure}[t]
\centering
  \includegraphics[width=0.4\textwidth]{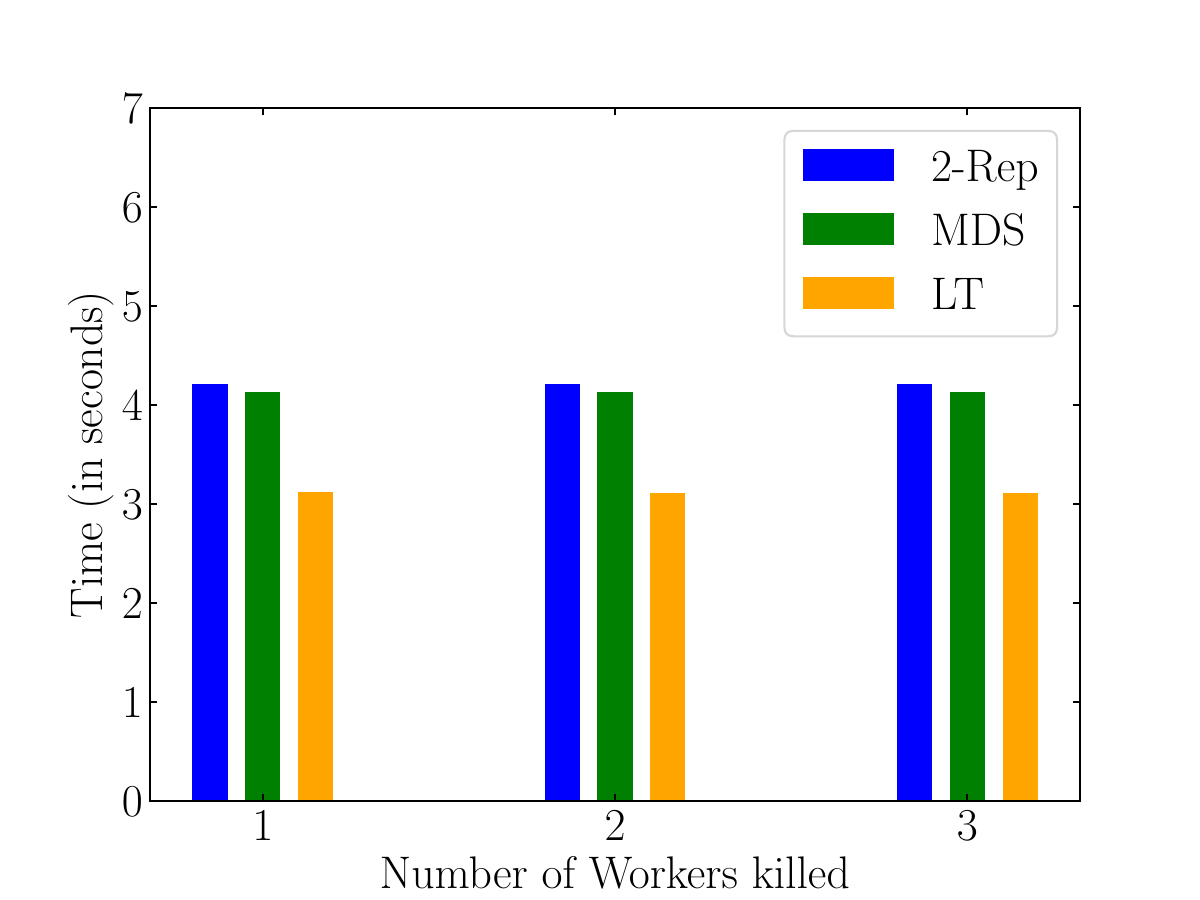}
  \caption{Worker $i$ has a random initial delay $X_i$, after which it completes row-vector product tasks (denoted by the small rectangles), taking time $\tau$ per task. The latency $\runtime$ is the time until enough tasks have been completed for the product $\res = \mat\vect$ to be recovered. \label{fig:worker_kill}}
\end{figure}

\noindent \textbf{Additional Experiments: } We study the effect of worker failures in the different coded computing strategies using an EC2 \cite{amazon_ec2} cluster 10 t2.micro workers. Note that redundancy/coding is essential if workers fail as the naive uncoded approach cannot handle the resulting loss of data. $\mat$ is a $10000 \times 10000$ identity matrix and is encoded using replication $(\numrep=2)$, MDS coding $(\mdsnum=5)$, and LT coding ($\strag=2.0$). Results in \Cref{fig:worker_kill} clearly show that the LT coded approach is more robust multiple node failures than the other approaches.

\end{document}